\documentclass[11pt,a4paper]{article}

\usepackage{xcolor}
\usepackage[colorlinks=true, linkcolor=black!50!blue, urlcolor=blue, citecolor=blue, anchorcolor=blue]{hyperref}
\usepackage[font=small,labelfont=bf,margin=0mm,labelsep=period,tableposition=top]{caption}
\usepackage[a4paper,top=3cm,bottom=2.5cm,left=2.5cm,right=2.5cm,bindingoffset=0mm]{geometry}
\usepackage{amsmath}
\usepackage{amsfonts}
\usepackage{amssymb}
\usepackage{authblk}
\usepackage{dsfont}
\usepackage{pifont}
\usepackage{booktabs}
\usepackage{tabularx}
\usepackage{siunitx}
\usepackage{graphicx}
\usepackage{subcaption}
\usepackage{epstopdf}
\usepackage{epsfig}
\usepackage{framed}
\usepackage{makeidx}
\usepackage{simplewick}
\usepackage{placeins}
\usepackage{bbold}
\usepackage{braket}
\graphicspath{{./figs/}}

\newcommand{\nreps}{N_{\rm rep}}

\newcommand{\ie}{{\it i.e.}}
\newcommand{\eg}{{\it e.g.}}

\newcommand{\ndat}{N_{\mathrm{dat}}}
\newcommand{\ngrid}{N_{\mathrm{grid}}}
\newcommand{\nflav}{N_{f}}

\newcommand{\cov}{\mathrm{Cov}}
\newcommand{\FKtab}{(\mathrm{FK})}
\newcommand{\FKtabT}{(\mathrm{FK})^T}

\newcommand{\ddt}{\frac{d}{dt}}
\newcommand{\fin}{f^{\rm in}}
\newcommand{\finperp}{f^{\rm in \perp}}
\newcommand{\finpar}{f^{\rm in \parallel}}

\newcommand{\fimm}{f_{{\rm Im}(M)}}
\newcommand{\fker}{f_{\ker(M)}}

\newcommand{\bpmat}{\begin{pmatrix}}
\newcommand{\epmat}{\end{pmatrix}}

\begin{document}
\newgeometry{top=1.5cm,bottom=1.5cm,left=1.5cm,right=1.5cm,bindingoffset=0mm}
\vspace{-2.0cm}
\begin{flushright}
\end{flushright}
\vspace{0.3cm}

\begin{center}
  {\Large \bf Quantitative Understanding of PDF Fits and their Uncertainties}
  
  \vspace{1.0cm}

\vspace{1.1cm}

  Amedeo Chiefa, Luigi Del Debbio and Richard Kenway

  \vspace{0.2cm}

  {\it \small
    The Higgs Centre for Theoretical Physics, \\ 
    School of Physics and Astronomy, The University of Edinburgh,\\
    Peter Guthrie Tait Road, Edinburgh EH9 3FD, United Kingdom\\[0.1cm]
  }
  \vspace{0.7cm}
\end{center}

\begin{abstract}
  Parton Distribution Functions (PDFs) play a central role in describing
  experimental data at colliders and provide insight into the 
  structure of nucleons. As the LHC enters an era of high-precision measurements, a robust
  PDF determination with a reliable uncertainty quantification has become
  mandatory in order to match the experimental precision. The NNPDF
  collaboration has pioneered the use of Machine Learning (ML) techniques for PDF
  determinations, using Neural Networks (NNs) to parametrise the unknown PDFs 
  in a flexible and unbiased way. The NNs are then trained on experimental data
  by means of stochastic gradient descent algorithms. The statistical robustness 
  of the results is validated by extensive closure tests using synthetic data.
  In this work, we develop a theoretical framework based on the
  Neural Tangent Kernel (NTK) to analyse the training dynamics of neural
  networks. This approach allows us to derive, under precise assumptions, an
  analytical description of the neural network evolution during training,
  enabling a quantitative understanding of the training process. 
  Having an analytical handle on the training dynamics allows us to clarify the
  role of the NN architecture and the impact of the experimental data in a 
  transparent way. Similarly, we are able to describe the evolution of the covariance of
  the NN output during training, providing a quantitative description of how
  uncertainties are propagated from the data to the fitted function.
  Interestingly, the methodology developed in this work can be used to understand the 
  minimization of a loss function for any kind of parametrization, thereby 
  providing a unified framework to compare different PDF determinations, like,
  \eg, fits based on a particular functional form. While our results are {\em not}\ 
  a substitute for PDF fitting, they do provide a powerful diagnostic tool to
  assess the robustness of current fitting methodologies.
  Beyond its relevance for particle physics phenomenology, our analysis of 
  PDF determinations provides a testbed to apply theoretical ideas about the 
  learning process developed in the ML community. As seen in applications from 
  other domains, we find that our results deviate from the simple picture 
  of the \textit{lazy training} regime discussed in the ML literature.
\end{abstract}

\newpage

\tableofcontents
\clearpage

\section{Introduction}
\label{sec:intro}

Parton Distribution Functions (PDFs) are a central ingredient in describing
experimental data at hadron colliders and in gaining insights into the internal
structure of the proton. The high-precision era of particle physics that we are
now witnessing calls for equally precise theoretical predictions. Since PDFs are
a key ingredient in these predictions, the need for robust PDF determinations
with reliable uncertainty quantification has become increasingly important for
both Standard Model measurements and searches for new physics.

PDFs are typically extracted from global analyses of experimental and lattice data. 
Their determination is a classic example of an \textit{inverse problem}, as it
involves inferring a continuous function from a finite set of data points. This
process is inherently ill-defined, and the limited amount of experimental
information prevents us from obtaining a unique solution to the problem. The
solution will inevitably depend on the assumptions made and on the prior
knowledge introduced to regularise the problem, either explicitly stated or
implicitly embedded in the fitting framework.

The complex nature of inverse problems has prompted the development of
sophisticated statistical methods and
tools to tackle them. In general, PDF determinations can be broadly classified
into two main categories, depending on whether a specific functional form is
assumed for the PDFs or whether a non-parametric approach is adopted. Although
the former approach has been widely used in the literature, non-parametric
approaches based on Bayesian inference have been successfully applied to the
problem of PDF determination, albeit in a limited
scenario~\cite{DelDebbio:2021whr,Candido:2024hjt,Medrano:2025cmg}.
Bayesian-based approaches are promising tools that ensure a rigorous framework
where prior information and assumptions are spelled out explicitly. Yet, a
global PDF determination based on these methods has not yet been attempted, and the 
impact of the prior needs to be carefully studied in these frameworks. 

On the other hand, state-of-the-art PDF determinations rely on parametric
approaches, where a specific functional form is assumed for the PDFs at a given
initial scale $Q_0$. These functions are typically chosen to be flexible enough
to capture the main features of the PDFs, while their internal parameters are
optimised to reproduce the experimental data. Several
groups~\cite{NNPDF:2021njg,Ablat:2024hbm,Bailey:2020ooq,Alekhin:2017kpj} have
set the standard for PDF determinations through continuous refinement of their
global fits as new data and theoretical advances become available, with an
increasing emphasis on uncertainty quantification. Although these determinations
have been shown to perform well on a wide range of new experimental
data~\cite{Chiefa:2025loi}, the different methodological frameworks adopted by
the various groups lead to PDF sets whose differences are yet to be fully
understood~\cite{Harland-Lang:2024kvt,PDF4LHCWorkingGroup:2022cjn}. These
differences become significantly visible when considering parameter
determinations that are particularly sensitive to the choice of the PDF set,
both on the central values and, most importantly, on the associated uncertainties
(see Refs.~\cite{ATLAS:2023lhg,CMS:2024ony,ATLAS:2024erm,CMS:2024lrd} for some recent
examples).

In this work, we build upon the work of
Refs.~\cite{DelDebbio:2021whr,Candido:2024hjt}, which aims at providing a
sound statistical framework for PDF determination, with all
underlying assumptions clearly stated. We focus on the NNPDF
methodology~\cite{NNPDF:2021njg}, which pioneered the use of ML
tools in the context of PDF determinations and has been validated through
extensive studies over the
years~\cite{DelDebbio:2021whr,Barontini:2025lnl,Cruz-Martinez:2021rgy}. It
combines a Monte Carlo sampling of the experimental data and a feed-forward
neural network parametrization of the PDFs. We adopt a simplified 
framework to analyse the training process,
aiming at providing a quantitative description of its key aspects, and making
transparent the assumptions that are often implicitly embedded in the fitting
procedure.

We demonstrate that the training dynamics of a neural network can be fully
reformulated in functional space, leading to an interpretable description of the
learning process. We show that the training dynamics is dictated by the Neural
Tangent Kernel (NTK)~\cite{jacot2018neural}, which encodes and factorises the
dependence on the architecture and the parameters of the neural network. Similar
approaches leveraging NTK properties have been explored in other
contexts~\cite{tovey2025collective}. In fact, the spectral properties of the NTK
provide a powerful lens through which we can understand the learning process:
only the directions that are orthogonal to the kernel of the NTK are actually
learned in the training process. At initialisation, the NTK is characterised by
a wide spectrum of eigenvalues, with only a few large eigenvalues being
significantly different from zero. Even though the NNs span a very broad
functional space, the training explores a much smaller subspace. During the
training process, the hierarchy in the NTK spectrum is preserved, but
eigenvalues that were initially subleading, or zero, grow in magnitude. Since
the only directions that contribute to the learning process are those associated
to the non-zero eigenvalues, with the actual value of the eigenvalue setting the
learning speed along the corresponding eigenvector direction, the growth of some
eigenvalues indicates that new features in the functional space emerge during
training and that the network thus becomes capable of representing more complex
functions. The space of functions explored in the training process is therefore
dynamically determined during the training itself, exploiting the flexibility of
the parametrization in order to explore multiple functional forms and select the
preferred one based on inference from data rather than a priori decisions.

Another key result of this work is that, after an initial transient phase where
the NTK evolves significantly, the training process enters a second regime where
the NTK becomes approximately constant. This regime is often referred to as
\textit{lazy training} in the Machine Learning
literature~\cite{jacot2018neural}, and it has important
implications for the training dynamics. In this regime, we show that the
training process can be described analytically, allowing us to obtain a single
and clean closed-form expression for the output of the network at any training
time $t$. The main result of this analysis is summarised in Eq.~\eqref{eq:AnalyticSol}
which we report here,
\[
    f_{t}
        = U(t) f_{0} + V(t) Y\, , 
\]
where $f_{t}$ is the network output at training time $t$, $f_{0}$
is the initial output at $t=0$, $Y$ are the training data,
and $U(t)$ and $V(t)$ are time-dependent matrices that depend
on the NTK and are computed explicitly in Sect.~\ref{sec:LazyTraining}.
It is interesting to remark that this expression decomposes into two
contributions: one that depends on the initial condition and another that
depends on the data, thus making explicit the role of prior information and of
experimental measurements in shaping the final result. This analytical expression
also allows us to compute the evolution of the covariance of the network output
during training, providing a quantitative description of how uncertainties
are propagated from the data to the fitted function.
Although applicable
only when the NTK reaches stability, this analytical description is a
powerful tool to bridge the gap between the parametric regression approach
adopted in NNPDF and other methods for solving inverse problems that are receiving growing
attention in the community.

Being derived in a simplified setting -- considering a single PDF flavor
combination with DIS data and vanilla gradient descent optimization -- we
present this study as an exploration of foundational aspects, primarily 
focussed on the theoretical issues; 
further investigations are in progress in order to extend these ideas
to the full complexity of modern global PDF fits. We particularly emphasise that
the present analysis is not limited to neural networks, but can be extended to
any functional parametrization that undergoes a gradient-based training
process. It will be interesting to explore the properties of the NTK together
with its spectral structure in more realistic PDF fits, translating the differences
between various fitting methodologies in terms of the NTK. We leave these
studies to future work.

The remainder of this paper is organized as follows. In Section~\ref{sec:Init}
the inverse problem of PDF determination is briefly reviewed in the simplified
case of theoretical predictions that depend linearly on the PDFs. We then review
some fundamental statistical aspects of the Neural Networks at initialisation,
which will be relevant in the rest of the paper. The training dynamics is then
discussed in Section~\ref{sec:Training}, where the learning process of the
neural network is reformulated in functional space by means of the NTK. The
implications of the \textit{lazy training} regime are used in
Section~\ref{sec:LazyTraining} to derive an analytical description of the
training process. Finally, we present our conclusions and outlook in
Section~\ref{sec:conclusions}.

\section{Neural Networks and PDFs}
\label{sec:Init}

In the following, we prepare the ground for the study of the training dynamics
of neural networks used in the NNPDF framework. We start by briefly presenting
the inverse problem of PDF determination using data depending linearly on the PDFs,
setting the notation and introducing the statistical
vocabulary used in the rest of this study. We then discuss some
statistical aspects of the neural networks at initialisation, which will help us
understand the implications in the training process. These properties, derived
in the large-width limit~\cite{lee2019wide,jacot2018neural}, are analysed for the
specific architecture used in the NNPDF methodology. An exhaustive and detailed
review of wide-network properties is beyond the scope of this work, and the
reader is encouraged to refer to Ref.~\cite{Roberts:2021fes} for a comprehensive
review.

\subsection{The 1-dimensional regression problem of PDFs}
\label{subsec:inverse_problem}

The extraction of PDFs from experimental data is a classic example of an inverse
problem, namely the reconstruction of a function $f(x)$ from a finite set of
data points $Y_I$, where the index $I=1, \ldots, \ndat$.\footnote{When omitting
the data index $I$, we will always assume $Y \in \mathbb{R}^{\ndat}$.} In
particular, for this study, we will focus on DIS data, which depend linearly on
the function $f(x)$. The theoretical prediction for the data point $Y_I$ is
given by
\begin{equation}
    \label{eq:TheoryPred}
    T_I[f] = \sum_{i=1}^{\nflav} \int dx\, C_{Ii}(x) f_{i}(x)\, ,
\end{equation}
where $C_{Ii}(x)$ is a coefficient function, known to some given order in
perturbation theory, $i = 1, \ldots, \nflav$, labels the parton flavor, 
and $f_i(x)$ is the PDF (or set of PDFs) that we want to determine.

Attempting to determine a function $f$ in an infinite dimensional space of
solutions using a finite set of data is inherently ill-posed. The solution
inevitably depends on assumptions and prior knowledge -- conscious or not --
introduced to regularise the problem. Different methodologies, based either on
non-parametric methods or parametric regression, have been proposed to address
these challenges, yielding increasingly precise PDFs. Yet, despite
the longstanding effort to provide robust uncertainty quantification and
establish the relationships between different methodologies and their solutions,
some discrepancies remain unresolved, see, \eg, \cite{PDF4LHCWorkingGroup:2022cjn}. 
Understanding such differences between the
various approaches is thus crucial for precision physics.

Following the ideas highlighted in
Refs.~\cite{DelDebbio:2021whr,Candido:2024hjt}, the solution of the inverse
problem is conveniently phrased in a Bayesian framework. The functions $f_i$ are
promoted to stochastic processes; for any grid of points $x_{\alpha}$,
$\alpha=1, \ldots, \ngrid$, the vector $f_{i\alpha}=f_{i}(x_{\alpha})$ is a
vector of $\nflav\times\ngrid$ stochastic variables, for which we introduce a
{\em prior}\ distribution $p(f)$.\footnote{Following the same convention used for the
data, when omitting the grid index $\alpha$, and/or the flavor index $i$, we
will always refer to a vector $f \in \mathbb{R}^{\nflav\times\ngrid}$.} In this
perspective, any fitting procedure is interpreted as a recipe that yields the
{\em posterior}\ distribution $\tilde{p}(f) = p(f | Y)$.
In this study, following the NNPDF methodology, probability distributions are represented by
ensembles of i.i.d. neural network replicas. So, for instance, the prior
distribution $p(f)$ is described by an ensemble
\begin{equation}
    \label{eq:RepDef}
    \left\{f^{(k)} \in \mathbb{R}^{\nflav\times\ngrid}; k=1, \ldots, \nreps\right\}\, ,
\end{equation}
drawn from the distribution $p$, so that
\begin{equation}
    \label{eq:ReplicaEnsemble}
    \mathbb{E}_{p}[O(f)] = \frac{1}{\nreps} \sum_{k=1}^{\nreps} O(f^{(k)})\, ,
\end{equation}
for any observable $O$ that is built from the PDFs.

The prior distribution $p(f)$ is defined by initializing a set of neural networks (NNs) 
replicas using a Glorot normal initializer~\cite{glorot2010understanding}. The result of this
initialisation is discussed below in Sec.~\ref{sec:NNinit}.

In order to account for the experimental uncertainties and propagate them to the
fitted PDFs, the NNPDF collaboration uses Monte Carlo replicas~\cite{Costantini:2024wby}. For each
replica, labelled by the index $k$, a new set of data $Y^{(k)}$ is generated from an $\ndat$ dimensional
Gaussian distribution centred at the experimental central value $Y$, with the
covariance given by the experimental covariance matrix $C_Y$,
\begin{equation}
    \label{eq:ExpReplicaDistr}
    Y^{(k)} \sim \mathcal{N}\left(Y, C_Y\right)\, .
\end{equation}
Each replica $f^{(k)}$ is trained on its corresponding data set $Y^{(k)}$. We
denote the replicas at training time $t$ as $f^{(k)}_{t} \in
\mathbb{R}^{\nflav\times\ngrid}$. Stopping the training at time $T$, the
posterior probability distribution is represented by the set of 
{\em trained}\ replicas
$\left\{f^{(k)}_{T}\in \mathbb{R}^{\nflav\times\ngrid}; k=1, \ldots,
\nreps\right\}$, so that averages over the posterior distribution are computed
as
\begin{equation}
    \label{eq:PostEnsemble}
    \mathbb{E}_{\tilde{p}}[O(f)] = \frac{1}{\nreps} \sum_{k=1}^{\nreps}
        O\left(f^{(k)}_{T}\right)\, .
\end{equation}
All knowledge about the solution of the inverse problem, $f$, is encoded in the
posterior $\tilde{p}$ and is expressed as expectation values of observables $O$
using Eq.~\eqref{eq:PostEnsemble}. Let us stress once again that the expectation values
with respect to the prior and posterior distributions are both obtained by taking 
averages over replicas. The expectation value with respect to the prior is the average over
replicas at initialization. The expectation value with respect to the posterior is the average
over the replicas at training time $T$. 

Training may yield different posteriors depending on the initial network
configuration. To understand this dependence, we pause to examine the
statistical properties of network ensembles at initialization. This analysis
provides a quantitative insight into how prior knowledge embedded in the initialization
interacts with, and evolves throughout, the training process, as we show in
Sec.~\ref{sec:LazyTraining}.

\subsection{Neural Networks at Initialisation}
\label{sec:NNinit}

When initializing a neural network, the weights and biases -- which we denote
collectively as the {\em parameters}\ of the network -- are drawn from some
probability distribution. In the NNPDF formalism, the set of network parameters
at initialisation for each replica is an instance of i.i.d. stochastic
variables. More importantly, the probability distribution of the network
parameters induces a probability distribution for the output of the neural
networks at initialisation. It is well known that the probability distribution
of these outputs becomes approximately Gaussian when the size of the hidden
layers is increased~~\cite{Roberts:2021fes}. We call this limit the {\em large-network} limit.

As detailed in Ref.~\cite{NNPDF:2021njg}, the NNs used for the NNPDF fit have a
2-25-20-8 architecture, a $\tanh$ activation function, and are initialized using
a Glorot normal distribution~\cite{glorot2010understanding}. The preactivation
function of a neuron is denoted as $\phi^{(\ell)}_{i,\alpha} =
\phi^{(\ell)}_i(x_\alpha)$, where $\ell = 0, \ldots, L$, denotes the layer of the neuron, 
and, for each $\ell$, $i=1, \ldots, n_{\ell}$
identifies the neuron within the layer.\footnote{We refer to $i$ as the {\em
neuron}\ index.} Furthermore, $x_{\alpha}$ is the input to the NN, \ie, a point in the 
interval $[0,1]$. A grid of
$\ngrid=50$ points in $x$ is used to compute observables in the NNPDF formalism and in
this work we focus on the values of $f$ at those values of $x_\alpha$, where 
the index $\alpha = 1, \ldots, \ngrid$ labels the points on the grid. For
completeness, we list the values of $x_\alpha$ in Tab.~\ref{tab:Xvals}.

\begin{table}[ht]
    \centering
    \begin{tabular}{|c|c|c|c|c|c|c|c|c|c|}
    \hline
    $\alpha$ & $x_\alpha$ & $\alpha$ & $x_\alpha$ & $\alpha$ & $x_\alpha$ & $\alpha$ & $x_\alpha$ & $\alpha$ & $x_\alpha$ \\
    \hline
    $1$  & $2.00 \times 10^{-7}$ & $11$ & $1.29 \times 10^{-5}$ & $21$ & $8.31 \times 10^{-4}$ & $31$ & $0.0434$ & $41$ & $0.422$ \\
    $2$  & $3.03 \times 10^{-7}$ & $12$ & $1.96 \times 10^{-5}$ & $22$ & $1.26 \times 10^{-3}$ & $32$ & $0.0605$ & $42$ & $0.480$ \\
    $3$  & $4.60 \times 10^{-7}$ & $13$ & $2.97 \times 10^{-5}$ & $23$ & $1.90 \times 10^{-3}$ & $33$ & $0.0823$ & $43$ & $0.540$ \\
    $4$  & $6.98 \times 10^{-7}$ & $14$ & $4.51 \times 10^{-5}$ & $24$ & $2.87 \times 10^{-3}$ & $34$ & $0.109$ & $44$ & $0.601$ \\
    $5$  & $1.06 \times 10^{-6}$ & $15$ & $6.84 \times 10^{-5}$ & $25$ & $4.33 \times 10^{-3}$ & $35$ & $0.141$ & $45$ & $0.665$ \\
    $6$  & $1.61 \times 10^{-6}$ & $16$ & $1.04 \times 10^{-4}$ & $26$ & $6.50 \times 10^{-3}$ & $36$ & $0.178$ & $46$ & $0.730$ \\
    $7$  & $2.44 \times 10^{-6}$ & $17$ & $1.57 \times 10^{-4}$ & $27$ & $9.70 \times 10^{-3}$ & $37$ & $0.220$ & $47$ & $0.796$ \\
    $8$  & $3.70 \times 10^{-6}$ & $18$ & $2.39 \times 10^{-4}$ & $28$ & $0.0144$ & $38$ & $0.265$ & $48$ & $0.863$ \\
    $9$  & $5.61 \times 10^{-6}$ & $19$ & $3.62 \times 10^{-4}$ & $29$ & $0.0211$ & $39$ & $0.314$ & $49$ & $0.931$ \\
    $10$ & $8.52 \times 10^{-6}$ & $20$ & $5.49 \times 10^{-4}$ & $30$ & $0.0305$ & $40$ & $0.367$ & $50$ & $1.00$ \\
    \hline
\end{tabular}

    \caption{Values of $x_\alpha$ used in the NNPDF grids for the computation of
    observables. The points are equally spaced on a logarithmic scale
    for $\alpha = 1, \ldots, 27$, and linearly spaced for $\alpha > 27$.
    \label{tab:Xvals}}
\end{table}

The output of the neuron is identified by the pair $(\ell,i)$ is
$\rho^{(\ell)}_{i\alpha} = \tanh\left(\phi^{(\ell)}_{i\alpha}\right)$.
The parameters of the NN are the weights $w^{(\ell)}_{ij}$ and the biases $b^{(\ell)}_i$, which are
collectively denoted as $\theta_\mu$, where $\mu = 1, \ldots, P$, and the total number of parameters
is
\begin{equation}
    \label{eq:TotPar}
    P = \sum_{\ell=1}^{L} \left(n_{\ell} n_{\ell-1} + n_\ell\right)\, .
\end{equation}
The preactivation function in layer $(\ell+1)$ is a weighted average of the outputs of the neurons on 
the previous layer, namely
\begin{align}
    \label{eq:RecursionNN}
    \phi^{(\ell+1)}_{i\alpha} = \sum_{j=1}^{n_\ell} w^{(\ell+1)}_{ij} \rho^{(\ell)}_{j\alpha} + b^{(\ell+1)}_{i}\, .
\end{align}
The PDFs in the so-called evolution basis are parametrized by the preactivation
functions of the output layer $L$, $x_\alpha f_i(x_\alpha)=
\phi^{(L)}_{i,\alpha}$, where the neuron index on the last layer, $i=1, \ldots,
8$, labels the flavors.\footnote{For simplicity, we ignore the preprocessing
function $x^{1-\delta_i} (1-x)^{\eta_i}$ that is currently used in the NNPDF
fits. While the preprocessing may be useful in speeding the training it does not
affect the current discussion.} The input layer is identified by $\ell=0$ and
the activation function for that specific layer is the identity, so that
\begin{equation}
    \label{eq:InitLayerPhi}
    \rho^{(0)}_{i,\alpha} = \phi^{(0)}_{i,\alpha} = x_{i,\alpha} =
    \begin{cases}
        x_\alpha\, , \quad &\text{for}\ i=1\, ;\\
        \log\left(x_\alpha\right)\, , \quad &\text{for}\ i=2\, .
    \end{cases}
\end{equation}
In the following we refer to the preactivation functions as {\em fields}.
In Eq.~\eqref{eq:InitLayerPhi}, each input $x$ is augmented with its
logarithm, so that the vector $(x, \log x)$ is fed into the first hidden layer.
This transformation, used in the NNPDF fits and referred to as ``scaled
inputs'', gives the network direct access to both linear and logarithmic scales
of the input. Although the $x$-values are already sampled on a mixed
logarithmic/linear grid (see Tab.~\ref{tab:Xvals}), with a logarithmic spacing
in the small-$x$ region and a linear spacing in the large-$x$ region, feeding
$\log x$ explicitly allows the network to resolve PDF features across the full
$x$-range more efficiently.

\begin{figure}[t!]
  \centering
  \includegraphics[width=0.95\textwidth]{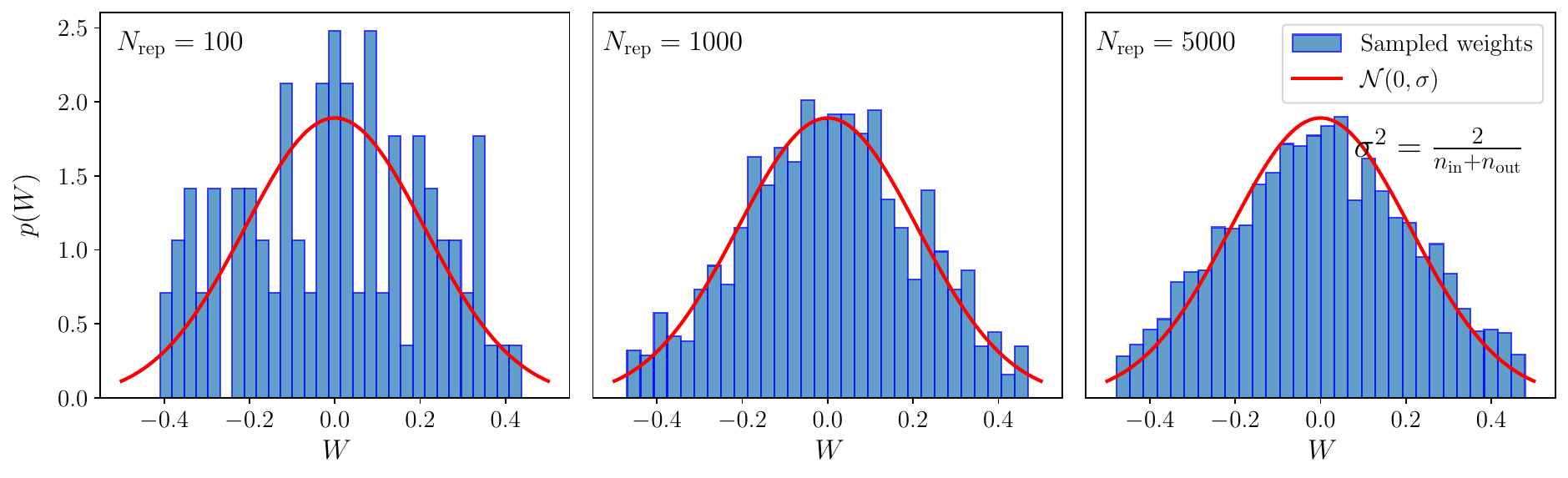}
  \caption{Sampled distribution of a selected weight as a function of the
  number of replicas. The red line represents the underlying Gaussian distribution
  from which the weights are drawn. As the number of replicas is increased the 
  distribution of the weight converges to the expected Gaussian.}
  \label{fig:weight_distribution}
\end{figure}
The Glorot normal initialiser draws each weight and bias of the NN from independent Gaussian
distributions, denoted $p_w$ and $p_b$ respectively, centred at zero and with variances
rescaled by the number of nodes in adjacent layers,
\begin{equation}
    \label{eq:RescaledGlorotVariances}
    \frac{C^{(\ell)}_{w}}{n_{\ell-1} + n_{\ell}}\, ,
    \quad \frac{C^{(\ell)}_{b}}{n_{\ell-1} + n_{\ell}}\, .
\end{equation}
Following the NNPDF prescription, we have $C^{(\ell)}_{w}=C^{(\ell)}_{b}=1$.
Figure~\ref{fig:weight_distribution} shows the binned distribution of one of the
weights in the network as a function of the number of replicas. Together with the
histogram, the underlying Gaussian, as dictated by the Glorot normal
initialisation, is also shown. The figure illustrates how the distribution of
the weights converges to the expected Gaussian as the number of replicas increases.

The probability distribution of the NN parameters induces a probability distribution for the
preactivations; the probability distribution of the fields in layer $\ell$, for given values of 
the field in the layer $\ell-1$ is
\begin{align}
    \label{eq:PreactAtInit}
    p\left(\phi^{(\ell)} | \phi^{(\ell-1)}\right)
      &= \int \mathcal{D}w\, p_w(w)\,
        \mathcal{D}b\, p_b(b)\, \prod_{i,\alpha}
        \delta\left(
          \phi^{(\ell)}_{i\alpha} - \sum_{j} w^{(\ell)}_{ij}
          \rho\left(\phi^{(\ell-1)}_{j\alpha}\right)
          - b^{(\ell)}_i
          \right)\, .
\end{align}
For clarity of writing, we will omit the condition in the probability distribution, and write 
simply $p(\phi^{(\ell)})$. 
Note that, here and in what follows, $p(\phi^{(\ell)})$ denotes the joint probability for all the
$n_{\ell}\times\ngrid$ components of $\phi^{(\ell)}$,
\begin{align}
    \label{eq:ExplIndices}
    p\left(\phi^{(\ell)}\right) = p\left(\phi^{(\ell)}_{1,\alpha_1}, \phi^{(\ell)}_{2,\alpha_1}, \ldots,
        \phi^{(\ell)}_{n_\ell,\alpha_1}, \phi^{(\ell)}_{1,\alpha_2}, \ldots, \phi^{(\ell)}_{n_\ell,\alpha_2},
        \ldots,
        \phi^{(\ell)}_{n_\ell,\ngrid}\right)\, .
\end{align}
This duality between parameter-space and function-space provides a powerful framework to study
the behaviour of an ensemble of NNs, and in particular the symmetry properties of the distribution
$p(\phi^{(\ell)})$ (see, \eg, Ref.~\cite{Maiti:2021fpy}). Working in parameter space, \ie, computing the
expectation values of correlators of fields as integrals over the NN parameters, one can readily
show that
\begin{align}
    \label{eq:NeurRotInv}
    \mathbb{E}\left[
        R_{i_1j_1} \phi^{(\ell)}_{j_1 \alpha_1} \ldots
        R_{i_nj_n} \phi^{(\ell)}_{j_n \alpha_n}
    \right] =
    \mathbb{E}\left[
        \phi^{(\ell)}_{i_1 \alpha_1} \ldots
        \phi^{(\ell)}_{i_n \alpha_n}
    \right]\, ,
\end{align}
where $R$ is an orthogonal matrix in $\text{SO}(n_{\ell})$. Eq.\eqref{eq:NeurRotInv} implies
that the probability distribution in Eq.~\eqref{eq:PreactAtInit} is also invariant under rotations,
and therefore it can only be a function of $\text{SO}(n_{\ell})$ invariants. Therefore
\begin{align}
    \label{eq:PriorAction}
    p\left(\phi^{(\ell)}\right) =
        \frac{1}{Z^{(\ell)}} \exp\left(-S\left[\phi^{(\ell)}_{\alpha_1}
            \cdot \phi^{(\ell)}_{\alpha_2}\right]\right)\, ,
\end{align}
where
\begin{align}
    \label{eq:PhiInvariant}
    \phi^{(\ell)}_{\alpha_1}
            \cdot \phi^{(\ell)}_{\alpha_2} =
    \sum_{i=1}^{n_\ell} \phi^{(\ell)}_{i \alpha_1} \phi^{(\ell)}_{i \alpha_2}\, .
\end{align}
The action can be expanded in powers of the invariant bilinear,
\begin{align}
    \label{eq:ExpandAction}
    S\left[\phi^{(\ell)}_{\alpha_1}
            \cdot \phi^{(\ell)}_{\alpha_2}\right] =
        \frac12 \gamma^{(\ell)}_{\alpha_1\alpha_2}
            \phi^{(\ell)}_{\alpha_1} \cdot \phi^{(\ell)}_{\alpha_2} +
            \frac{1}{8 n_{\ell-1}} \gamma^{(\ell)}_{\alpha_1\alpha_2,\alpha_3\alpha_4}
            \phi^{(\ell)}_{\alpha_1} \cdot \phi^{(\ell)}_{\alpha_2} \,
            \phi^{(\ell)}_{\alpha_3} \cdot \phi^{(\ell)}_{\alpha_4} + O(1/n_{\ell-1}^2)\, ,
\end{align}
so that the probability distribution is fully determined by the couplings 
$\gamma^{(\ell)}$.\footnote{ We have denoted {\em all}\ couplings by
  $\gamma^{{(\ell)}}$. Different couplings are identified by the number of
  indices, so that $\gamma^{(\ell)}_{\alpha_1\alpha_2}$ is a two-point coupling,
  $\gamma^{(\ell)}_{\alpha_1\alpha_2,\alpha_3\alpha_4}$ is a four-point coupling,
  etc. } 
In Eq.~\eqref{eq:ExpandAction}, we have factored out inverse powers of
$n_{\ell-1}$ for each coupling. With this convention, and with the scaling of the
parameters variances in Eq.~\eqref{eq:RescaledGlorotVariances}, the couplings in
the action are all $O(1)$ in the limit where $n_\ell\to\infty$. As a
consequence, the probability distribution at initialisation is a
multidimensional Gaussian at leading order -- \ie, $\mathcal{O}(1)$ -- in
$1/n_\ell$, with quartic corrections that are $O(1/n_\ell)$, while higher powers
of the invariant bilinear are suppressed by higher powers of the width of the
layer. This power counting defines an effective field theory, where deviations
from Gaussianity can be computed in perturbation theory to any given order in
$1/n_\ell$, see, \eg\, Ref.~\cite{Roberts:2021fes,Chiefa:2026TBA} for a detailed presentation
of these ideas. While the actual calculations become rapidly cumbersome, the
conceptual framework is straightforward.

\begin{figure}[!htb]
    \centering
    \includegraphics[scale=0.58]{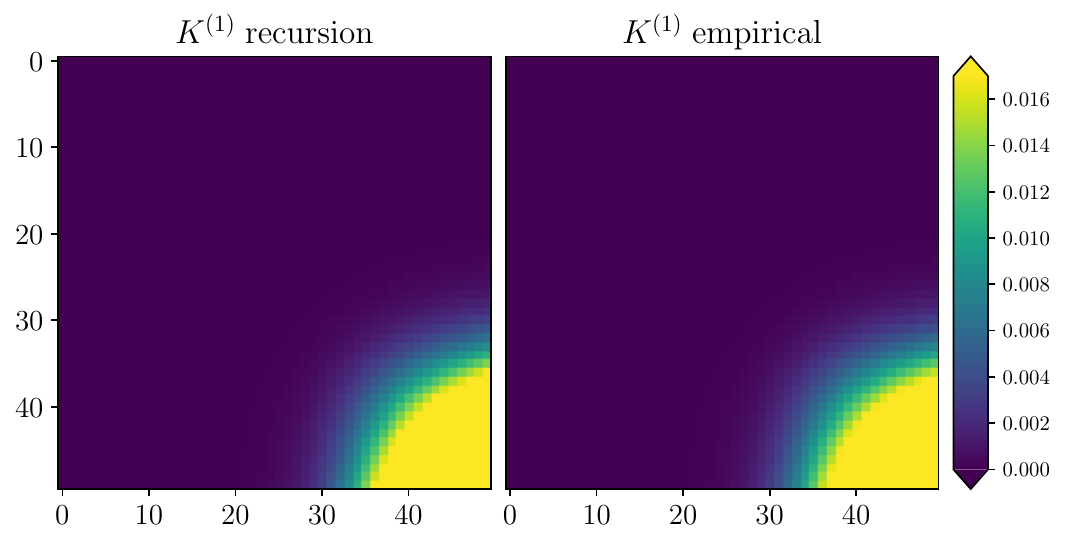}
    \includegraphics[scale=0.58]{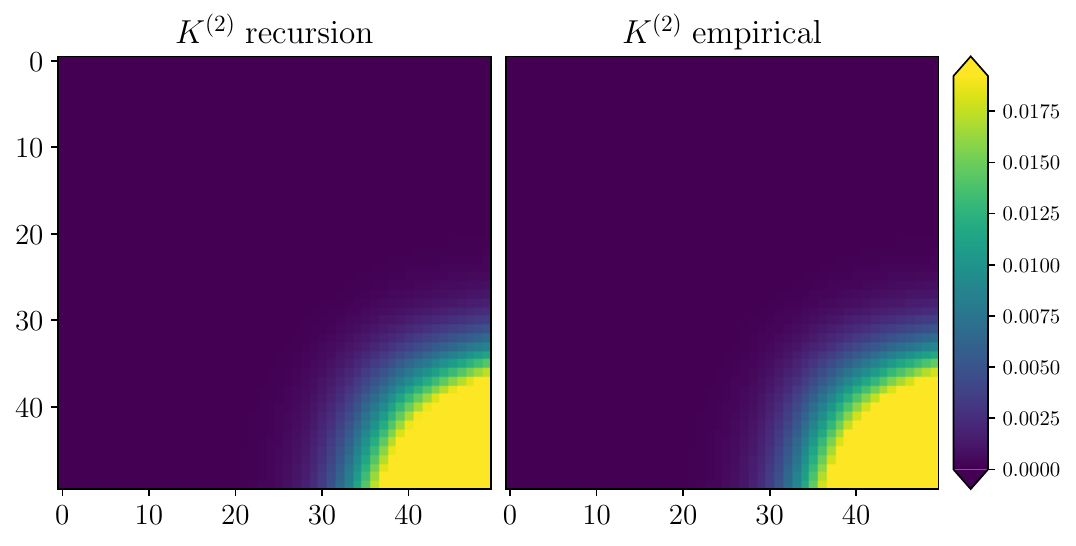}
    \includegraphics[scale=0.58]{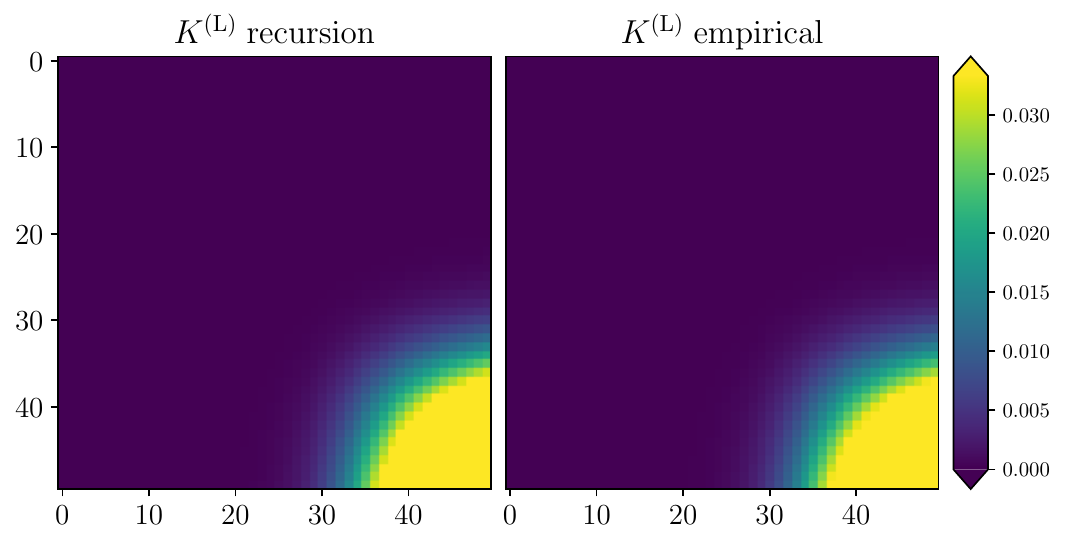}
    \caption{The empirical (left) and analytical (right) covariance matrices of
    the first, second and output layers of the NNPDF architecture (top to
    bottom). The covariance in the left panel is computed ``bootstrapping'' over
    an ensemble of replicas, initialised using the Glorot normal distribution.
    The covariance in the right panel is obtained by solving
    Eq.~\eqref{eq:RecursionForK} numerically. In order to reduce the bootstrap
    errors in the empirical covariance, an ensemble of 1000 replicas has been
    used for this figure.}
    \label{fig:KRecursion}
\end{figure}

\begin{figure}[!htb]
    \centering
    \includegraphics[width=0.86\textwidth]{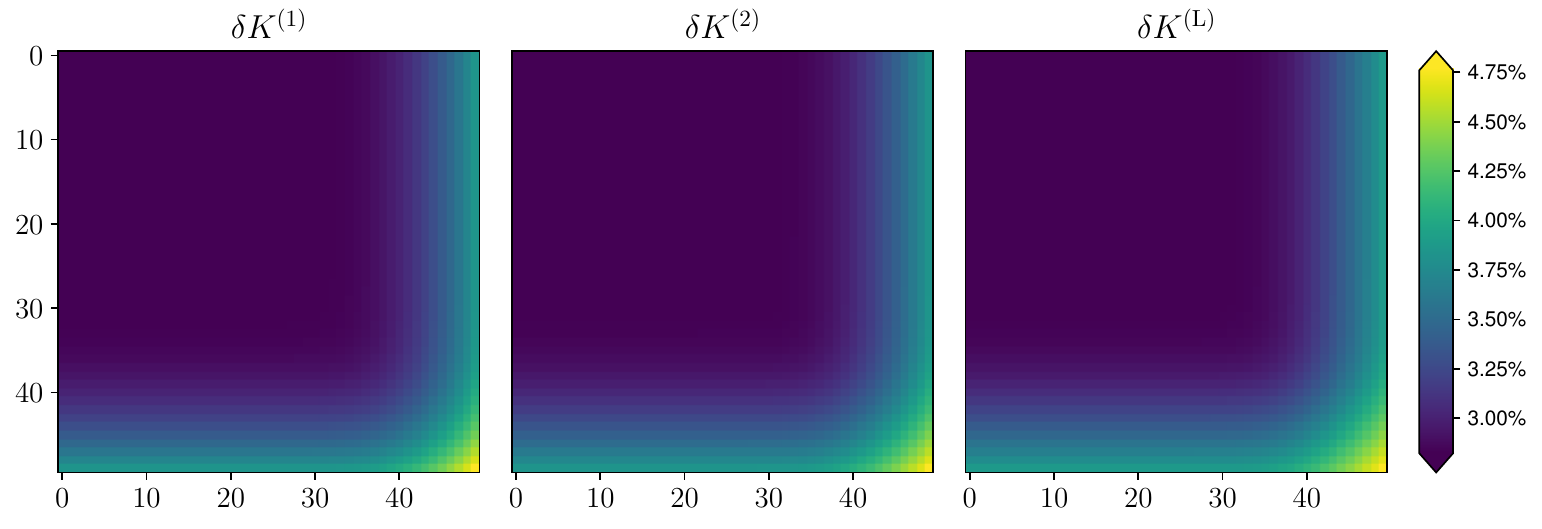}
    \caption{Relative difference between the empirical kernel,
    computed from an ensemble of networks at initialisation, and the recursive
    kernel obtained by iterating Eq.~\eqref{eq:RecursionForK} for the three layers
    of the NNPDF architecture. An ensemble of 1000 replicas has been used to
    reduce the bootstrap errors in the empirical covariance.}
    \label{fig:delta_K}
\end{figure}

At leading order, the second and fourth cumulant are respectively
\begin{align}
    &\langle \phi^{(\ell)}_{i_1,\alpha_1} \phi^{(\ell)}_{i_2,\alpha_2}\rangle
      = \delta_{i_1 i_2} K^{(\ell)}_{\alpha_1\alpha_2} + O(1/n_{\ell-1})\, , \\
    &\langle \phi^{(\ell)}_{i_1,\alpha_1} \phi^{(\ell)}_{i_2,\alpha_2}
      \phi^{(\ell)}_{i_3,\alpha_3} \phi^{(\ell)}_{i_4,\alpha_4}\rangle_c
      = O(1/n_{\ell-1})\, ,
\end{align}
where\footnote{
    The notation here refers to the matrix element $(\alpha_1,\alpha_2)$ of the inverse matrix of $\gamma^{(\ell)}$, and {\em not}\ to the inverse
    of the matrix element $\gamma^{(\ell)}_{\alpha_1\alpha_2}$.
}
\begin{equation}
    \label{eq:DefineKmat}
    K^{(\ell)}_{\alpha_1\alpha_2} = \left(\gamma^{(\ell)}\right)^{-1}_{\alpha_1\alpha_2}\, .
\end{equation}
The ``evolution'' of the couplings as we go deep in the NN, \ie, the dependence of the couplings on
$\ell$, is governed by Renormalization Group (RG) equations, which preserve the power counting in
powers of $1/n_{\ell}$. At leading order,
\begin{align}
    K^{(\ell+1)}_{\alpha_1\alpha_2} &=
      \left.
      C_b^{(\ell+1)} + C_w^{(\ell+1)} \frac{n_\ell}{n_\ell+n_{\ell+1}}\frac{1}{n_\ell}
      \langle \vec{\rho}^{\,(\ell)}_{\alpha_1} \cdot
      \vec{\rho}^{\,(\ell)}_{\alpha_2} \rangle
      \right|_{O(1)} \\
      \label{eq:RecursionForK}
      &= C_b^{(\ell+1)} + C_w^{(\ell+1)} \frac{n_\ell}{n_\ell+n_{\ell+1}}\frac{1}{n_\ell}
      \langle \vec{\rho}^{\,(\ell)}_{\alpha_1} \cdot
      \vec{\rho}^{\,(\ell)}_{\alpha_2} \rangle_{K^{(\ell)}}\, ,
\end{align}
where
\begin{align*}
    \frac{1}{n_\ell}
      \langle \vec{\rho}^{\,(\ell)}_{\alpha_1} \cdot
      \vec{\rho}^{\,(\ell)}_{\alpha_2} \rangle_{K^{(\ell)}} =
    \int \mathcal{D}\phi\,
      \frac{e^{-\frac12 \left(K^{(\ell)}\right)^{-1}_{\beta_1\beta_2}
        \phi_{\beta_1} \phi_{\beta_2}}}
        {\left|2\pi K^{(\ell)}\right|^{1/2}}\,
        \rho(\phi_{\alpha_1}) \rho(\phi_{\alpha_2})\, , 
\end{align*}
and
\begin{align}
    \label{eq:FunctIntDef}
    \mathcal{D}\phi = \prod_{\alpha=1}^{\ngrid} d\phi_\alpha\, .
\end{align}
Note that the integration variables in Eq.~\eqref{eq:FunctIntDef} do not have a
neuron index and the integrals are $\ngrid$ dimensional integrals.
Eq.~\eqref{eq:RecursionForK} is iterated for the NNPDF architecture, yielding
$K^{(\ell)}$ for arbitrary $\ell$, \ie, the covariance at initialisation for
various depths. These are compared with the empirical covariance computed from
an ensemble replicas in Fig.~\ref{fig:KRecursion} for the first two hidden layers
and the output layer. Furthermore, the relative difference between the empirical
covariance and the theoretical prediction is shown in Fig.~\ref{fig:delta_K}. In
order to reduce the bootstrap errors in the empirical covariance, an ensemble
with $\nreps=1000$ has been used for these figures. The agreement between the
theoretical prediction and the empirical computation is excellent, confirming
the validity of the large-network expansion even for networks of moderate size,
as those used in the NNPDF fits.

\begin{figure}
  \centering
  \includegraphics[width=0.95\textwidth]{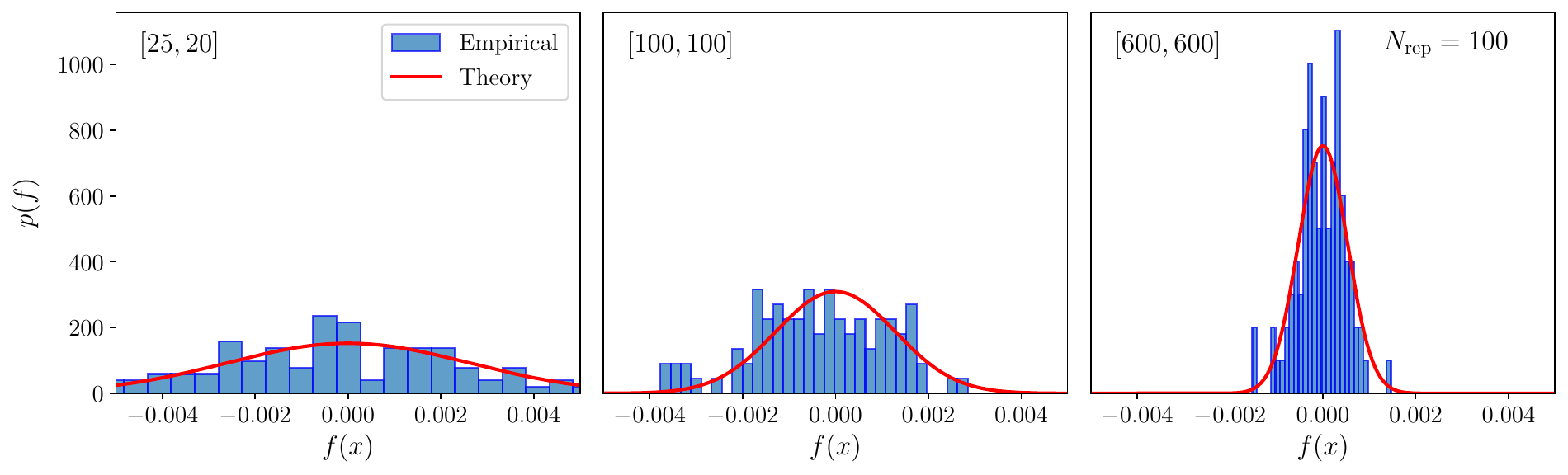}
  \includegraphics[width=0.95\textwidth]{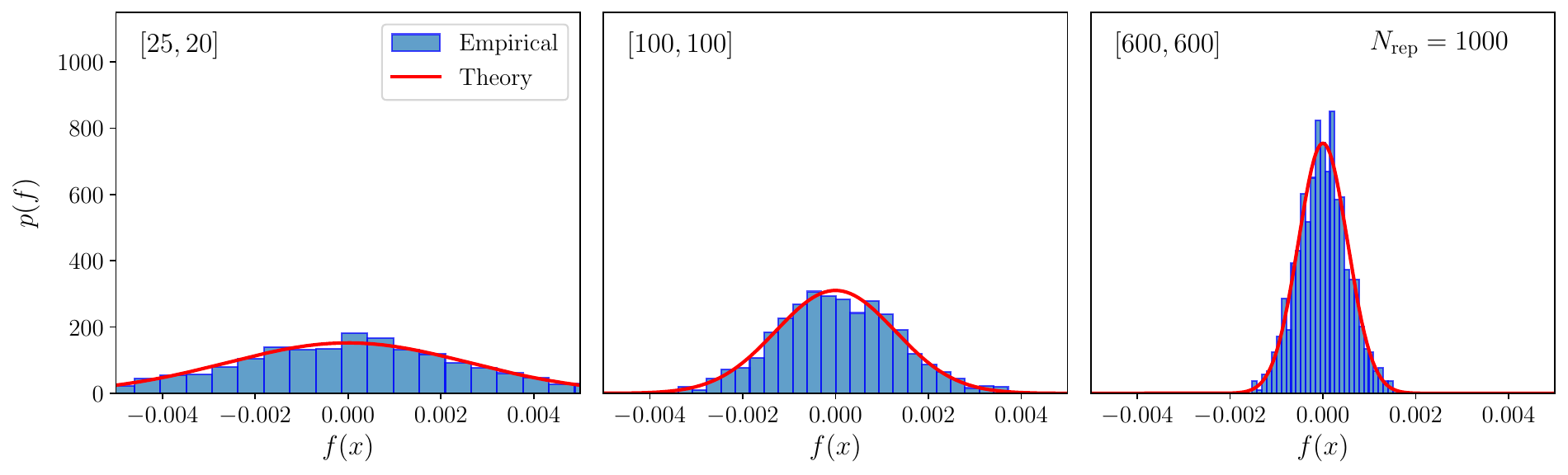}
  \caption{Sampled distribution of the output $xT_3$ at $x=0.0065$ for two
  different ensemble sizes, $\nreps=100$ (top) and $\nreps=1000$ (bottom). Each
  column shows the distribution for a different network architecture, the latter
  displayed in the top left corner of each panel. The red line the represents
  the predicted Gaussian distribution as dictated by the kernel recursion
  formula in Eq.~\eqref{eq:RecursionForK}.
  \label{fig:OutputDist}}
\end{figure}

\begin{figure}[!t]
    \centering
    \includegraphics[width=0.45\textwidth]{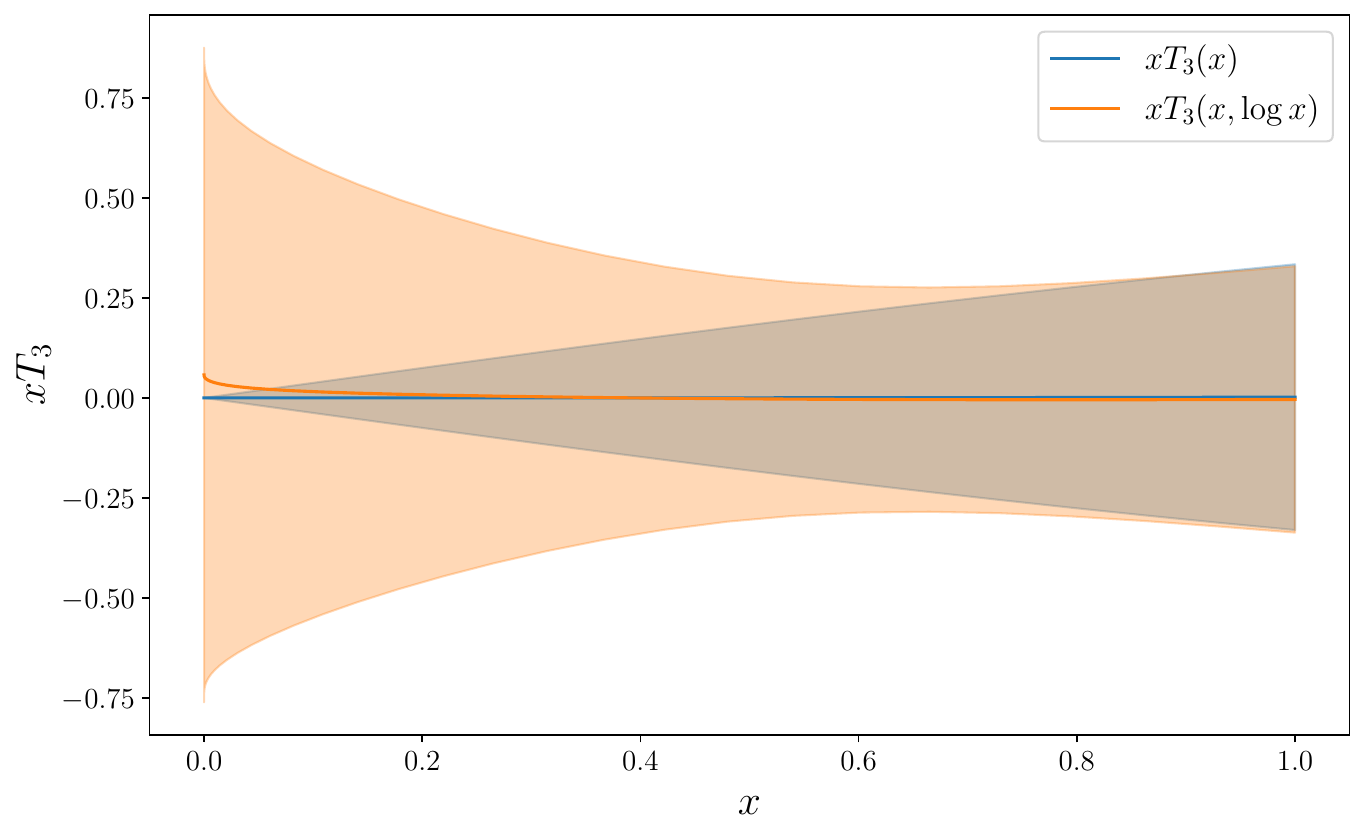}
    \includegraphics[width=0.45\textwidth]{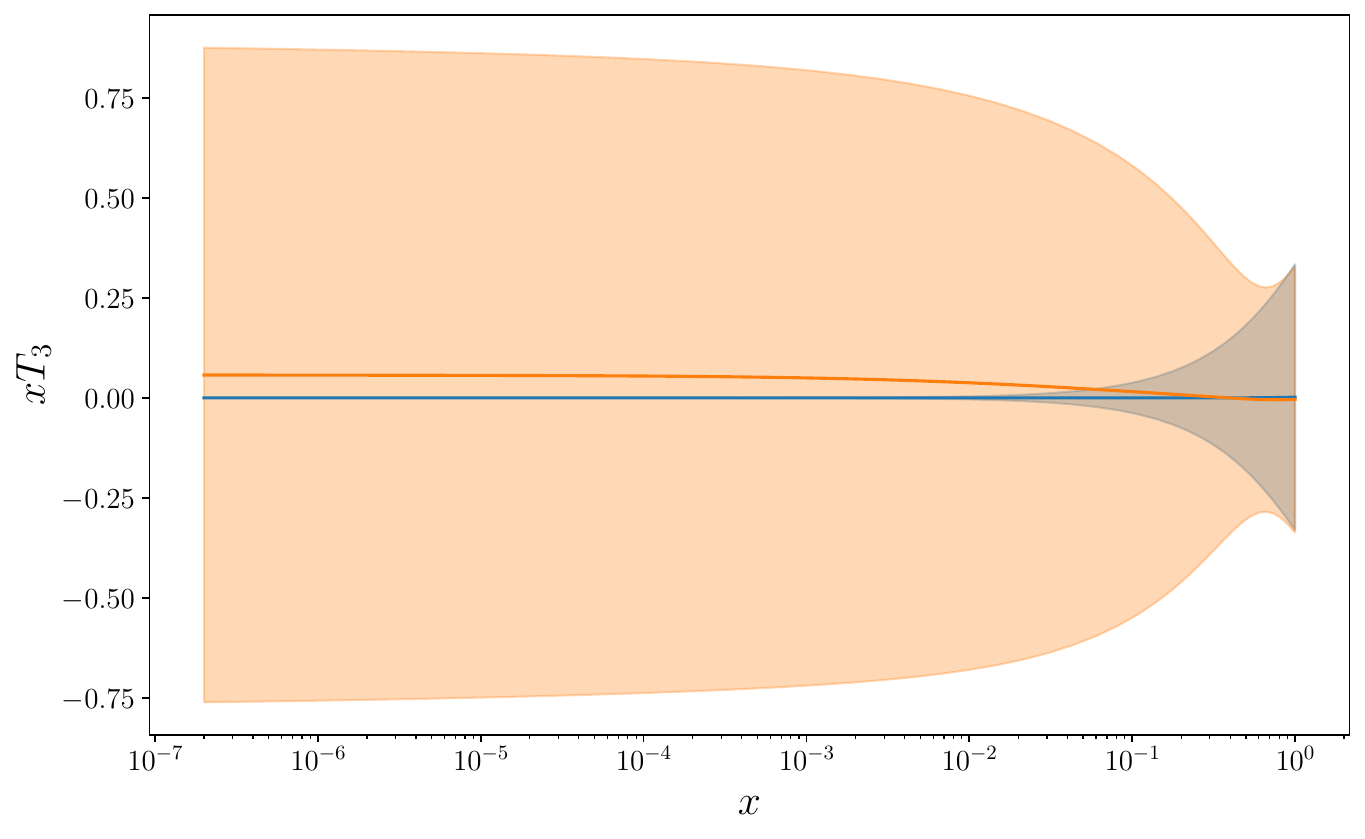}
    \caption{The output of the ensemble of neural networks at initialisation
    using the NNPDF architecture in linear (left) and logarithm (right) scale.
    We compare the case of linear input $f(x)$ (blue) and the case of
    scaled input $f(x, \log x)$ (orange). The solid lines represent
    the mean value computed over an ensemble of 100 replicas, while the shaded
    bands represent the one-sigma uncertainty computed as the variance over the
    same ensemble. In the figure, we show $xT_3$ as used in the following
    sections.}
    \label{fig:prior}
\end{figure}

As a consequence of the symmetry of the probability distribution, the mean value
of the fields at initialisation needs to vanish, while their variance at each
point $x_\alpha$ is given by the diagonal matrix elements of $K^{(\ell)}$. In
Fig.~\ref{fig:OutputDist}, the expected distribution is compared against the
empirical distribution of output fields for a selected value of $x$, using two
ensembles of replicas with $\nreps=100$ and $\nreps=1000$, respectively.
Inspecting the figures, we conclude that the recursion formula,
Eq.~\eqref{eq:RecursionForK}, accurately reproduces the output distribution of
the NNPDF networks at initialisation, provided that a sufficiently large
ensemble of replicas is used to sample the distribution. Finally,
Fig.~\ref{fig:prior} shows the mean and variance of the output at initialisation
across all values of $x$ for an ensemble of $\nreps=100$ neural networks
generated using the NNPDF architecture. We compare two cases: linear input
$f(x)$ and scaled input $f(x, \log x)$ as defined in
Eq.~\eqref{eq:InitLayerPhi}. The central value is computed according to
Eq.~\eqref{eq:ReplicaEnsemble},
\begin{align}
    \label{eq:MeanValAtInit}
    \bar{f}_{i\alpha} = \bar{f}_{i}(x_\alpha) = \frac{1}{\nreps} \sum_{k=1}^{\nreps} f^{(k)}_i(x_\alpha)\, ,
\end{align}
and the variance $\sigma^2_{i\alpha}$ is computed using the same formula with
\begin{align}
    \label{eq:VarAtInit}
    O(f) = \frac{\nreps}{\nreps-1} \left(f_i(x_\alpha) - \bar{f}_{i}(x_\alpha)\right)^2\, .
\end{align}
As is clear from the figure, the choice of input scaling has a significant impact
of the prior uncertainty, especially in the small-$x$ region. In the following,
we neglect this effect and focus on the case of linear input $f(x)$.

\FloatBarrier

\section{Training Dynamics and the Neural Tangent Kernel}
\label{sec:Training}

Having defined the physics goals and summarised some properties
of the neural network at initialisation, we now turn to the optimisation
process. In the context of machine learning, specifically when dealing with
neural networks, optimisation is an iterative algorithm that updates the
parameters of the network in order to minimise a figure of merit defined
appropriately. Due to the large number of parameters that characterise a neural
network, and the complex functional form induced by the recursive definition of the network, 
the figure of merit (also known as \textit{error function},
\textit{loss function}, or simply \textit{loss}) is a non-convex
high-dimensional function of the parameters, 
leading to numerical challenges in the minimisation task. In
addition, in order to avoid \textit{over-} and \textit{under-learning}, 
training algorithms are complemented by a so-called \textit{stopping criterion},
which specifies the optimal condition to end the training process.

In practice, minimization is performed using gradient methods where the
direction towards the minimum is defined by the gradient of the loss function.
These methods are usually improved by including, for instance, stochasticity and
information on previous iterations~\cite{amary1993sdg,Nadam2016,kingma2014adam}.
A detailed overview of these extended gradient methods is beyond the scope of
this work. In the context of PDF determinations, the NNPDF collaboration makes
intensive use of these tools and the reader is encouraged to refer to
Ref.~\cite{NNPDF:2021njg} for an extensive discussion.

Our main aim in this paper is understanding the dynamics driving the training
process. Indeed, while these algorithms have achieved remarkable empirical
success, a theoretical understanding of the optimization process remains
elusive. Therefore, we work with the
simplest gradient method, \ie, Gradient Descent (GD). Furthermore, we
consider a reduced dataset for which predictions can be computed using one flavor
combination of the PDFs, so that the integral in Eq.~\eqref{eq:TheoryPred}
reduces to
\begin{equation}
    \label{eq:TheoryPredSingleFlav}
    T_I[f] = \int dx\, C_{I}(x) f(x)\,.
\end{equation}
The details of the dataset and the definition of the flavour combination
considered in this work are provided in Appendix~\ref{app:dataset}. 
Extensions of this analysis to theoretical predictions that are
quadratic in the PDFs, multiple flavour combinations, alternative minimizers,
and cross-validation tools are left for future work.

Finally, we emphasise that the results in this section, while obtained having in mind
neural networks, apply to any fixed parametrization, including
fixed functional forms~\cite{Ablat:2024hbm,Bailey:2020ooq,Alekhin:2017kpj} or
kernels~\cite{Costantini:2025wxp}.

\subsection{Training in Functional Space}
\label{sec:GradFlow}

For analytical tractability, GD is described as a continuous flow of the
parameters $\theta_{\mu}$ in training time $t$ along the negative
gradient of the loss function $\mathcal{L}$. Here the index $\mu$ runs
over weights and biases in all layers of the network. For sufficiently small
learning rates $\eta$, this continuous flow approximates the discrete GD
trajectory in parameter space, as extensively discussed in
Ref.~\cite{barrett2022igr}. The continuous Gradient Flow (GF) is then given by
\begin{align}
    \label{eq:GradientFlowDef}
    \ddt &\theta_{t,\mu} = -\partial_\mu \mathcal{L}_t\, ,
\end{align}
where $\theta_{t,\mu}$ and $\mathcal{L}_t$ identify respectively the parameters
and the loss function at training time $t$. In Eq.~\eqref{eq:GradientFlowDef},
$\partial_\mu \equiv \partial/\partial \theta_\mu$ denotes the partial
derivative with respect to the single parameter $\theta_\mu$. The equation is
therefore a scalar relation that holds component-wise for each $\mu$. We
distinguish between the continuous training time $t$ and the discrete epochs of
GD, the latter denoted using the capital letter $T$. The two are related through
the learning rate, $t = \eta T$ with $\eta = 10^{-5}$, and in the following we
will use them interchangeably.

We focus here on quadratic loss functions that are obtained as the negative
logarithm of Gaussian data distributions around their theoretical predictions,
\begin{align}
    \label{eq:QuadLoss}
    \mathcal{L}_t = \frac12 \left(Y - T[f_t]\right)^T C_Y^{-1} \left(Y - T[f_t]\right)\, ,
\end{align}
where $f_t$ is the output of the network at training time $t$, obtained
from the time-dependence of the internal parameters. Here $C_Y$ is the
covariance of the data, which includes statistical and systematic errors given
by the experiments and also any theoretical error (\eg, missing higher
orders in the theoretical predictions). Indices that are summed over are
suppressed to improve the clarity of the equations. Note that the loss function
at training time $t$ is computed using the theoretical prediction $T[f_t]$, \ie,
the result of Eq.~\eqref{eq:TheoryPred} computed using the fields at training
time $t$. For a quadratic loss, the gradient is
\begin{align}
    \partial_\mu \mathcal{L}_t = - \left(\partial_\mu f_t\right)^T \left(\frac{\partial T}{\partial f}\right)_t
      C_Y^{-1} \epsilon_t\, ,
\end{align}
where, writing explicitly the data index,
\begin{align}
    \label{eq:EpsDef}
    \epsilon_{t,I} = Y_I - T_I[f_t]\, , \quad I=1, \ldots, \ndat\, .
\end{align}
For the specific case of a quadratic loss function, the gradient is proportional
to $\epsilon_t$, which is the difference between the theoretical prediction and
the data at training time $t$. If at some point during the training the
theoretical predictions reproduce all the data, the training process ends. 

A further simplification is obtained in the case of data that depend linearly on
the unknown function $f$. In the specific case of NNPDF fits, the integrals in
Eq.~\eqref{eq:TheoryPred} are approximated by a Riemann sum over the grid of $x$
points,
\begin{align}
    \label{eq:FKTabDef}
    T_I[f] \approx \sum_{i=1}^{\nflav}\sum_{\alpha=1}^{\ngrid} \FKtab_{Ii\alpha} f_{i\alpha}\, ,
\end{align}
and hence
\begin{align}
    \label{eq:dTbydf}
    \left(\frac{\partial T_I}{\partial f_{i\alpha}}\right)_t =
        \FKtab_{Ii\alpha}\, ,
\end{align}
which is independent of $t$. A few algebraic steps allow the flow of parameters
$\theta$ to be translated into a flow for the fields,
\begin{align}
    \label{eq:NTKFlow}
    \ddt &f_{t,i_1\alpha_1} = (\partial_\mu f_{t,i_1\alpha_1}) \ddt \theta_\mu =
      \Theta_{t,i_1\alpha_1i_2\alpha_2}
      \FKtabT_{i_2\alpha_2I} \left(C_Y^{-1}\right)_{IJ} \epsilon_{t,J}\, ,
\end{align}
where we have defined the Neural Tangent Kernel~\cite{jacot2018neural}
\begin{align}
    \label{eq:NTKDef}
    \Theta_{t,i_1\alpha_1i_2\alpha_2} = \sum_\mu
    \partial_\mu f_{t,i_1\alpha_1} \partial_\mu f_{t,i_2\alpha_2}\, .
\end{align}
As it will become clearer later, the NTK provides a powerful framework for
understanding neural network dynamics during training. Originally developed by
Jacot et al.~\cite{jacot2018neural} to analyse infinite-width feed-forward
networks, the NTK theory has since been extended to diverse architectures
including convolutional networks~\cite{arora2019exact} and recurrent
networks~\cite{alemohammad2021recurrent}. This theoretical framework has proven
invaluable for characterizing learning dynamics and generalization properties
across various network designs. We will see in the following and in
Sec.~\ref{sec:LazyTraining} how the NTK can also provide useful insights in the
context of PDF fitting.

In order to facilitate the discussion in Sec.~\ref{sec:AnlyticalLazySolution},
Eq.~\eqref{eq:NTKFlow} can be rewritten in a more compact form. We first omit
the indices and write, for instance,
\begin{align}
  \left(\frac{\partial T}{\partial f}\right)_t = \FKtab\, , \quad
  \Theta_t = \left(\partial_\mu f_t\right) \left(\partial_\mu f_t\right)^T\, .
  \label{eq:dTdfForLinearObs}
\end{align}
Then, using the definition of the error in Eq.~\eqref{eq:EpsDef}, we can rewrite
Eq.~\eqref{eq:NTKFlow} as 
\begin{align}
    \label{eq:FlowEquationNoIndices}
    \ddt f_t = -\Theta_t M f_t + b_t\, ,
\end{align}
where
\begin{align}
    M &= \FKtabT C_Y^{-1} \FKtab\, , \quad b_t = \Theta_t \FKtabT C_Y^{-1} Y\, .
    \label{eq:MandBDef}
\end{align}
Here $M$ is a positive-semidefinite matrix that depends only on the data
covariance and the FK tables that enter the theoretical predictions, while $b$
is a vector that depends (amongst other quantities) on the central value of the
data. We comment further on the role of $M$ in Sec.~\ref{sec:NTKRegularisation}.
Note that any vector $f$ that is in the kernel of $\FKtab$ is necessarily in the
kernel of $M$, $\ker M$. In turn, the vectors in $\ker M$ do not contribute to
the flow evolution, as seen explictly in Eq.~\eqref{eq:FlowEquationNoIndices}.

Before moving to the next subsection, a few comments are in order.
First, although derived in the context of neural networks, these equations do
not refer to a specific parametrization. Indeed, these remain valid even when
an explicit functional form is chosen to parametrize the PDFs, as in
Refs.~\cite{Ablat:2024hbm,Bailey:2020ooq,Costantini:2025wxp}. Second, it is
interesting to observe that the flow equation,
Eq.~\eqref{eq:FlowEquationNoIndices}, depends on two matrices, $\Theta$ and $M$.
The former encodes the model dependence, while the latter contains the physical
information. The interplay between these two matrices is crucial for
understanding the training dynamics, as discussed in
Sec.~\ref{sec:NTKAlign}. Finally, the NTK derived in Eq.~\eqref{eq:NTKDef} is
inherently time-dependent in a complex way, which precludes any attempt at
integrating Eq.~\eqref{eq:FlowEquationNoIndices} analytically. We come back
to this point in Sec.~\ref{sec:AnlyticalLazySolution}, after discussing the properties of the
NTK during training.

\FloatBarrier

\subsection{Inside the Training Dynamics: an NTK perspective}

From Eqs.~\eqref{eq:NTKDef} and~\eqref{eq:FlowEquationNoIndices}, we observe
that the NTK encodes the dependence on the architecture of the network and
governs its training dynamics. The analysis of the NTK properties is thus
crucial for understanding the behaviour of the network during training. We first
discuss the properties of the NTK at initialisation, before moving to the
training phase, where we provide a detailed study of the NTK in the context of
the NNPDF methodology.

\subsubsection{NTK at initialisation}
\label{sec:NTKAtInit}

Before training, the NTK is blind to data and depends on the $x$-grid of input
and on the architecture, as shown in Eq.~\eqref{eq:NTKDef}. The NTK
is a function of the fields $f$, which are stochastic variables described by
their joint probability distribution as discussed in Sect.~\ref{sec:Init}.
Therefore the NTK is also a stochastic variable, with its own probability
distribution, which we represent as usual as a set of replicas.

\begin{figure}[ht!]
  \centering
  \includegraphics[width=0.45\textwidth]{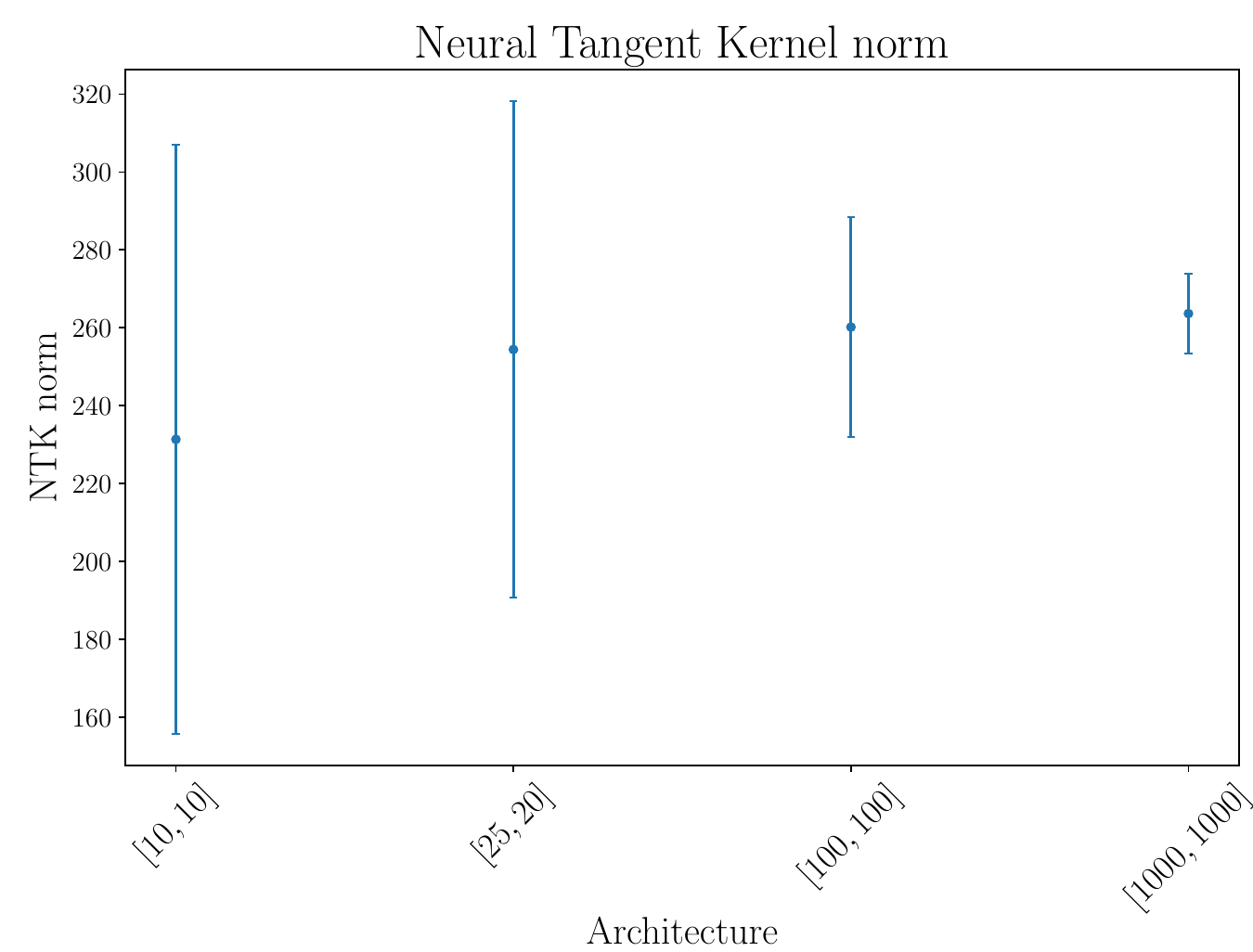}
  \includegraphics[width=0.45\textwidth]{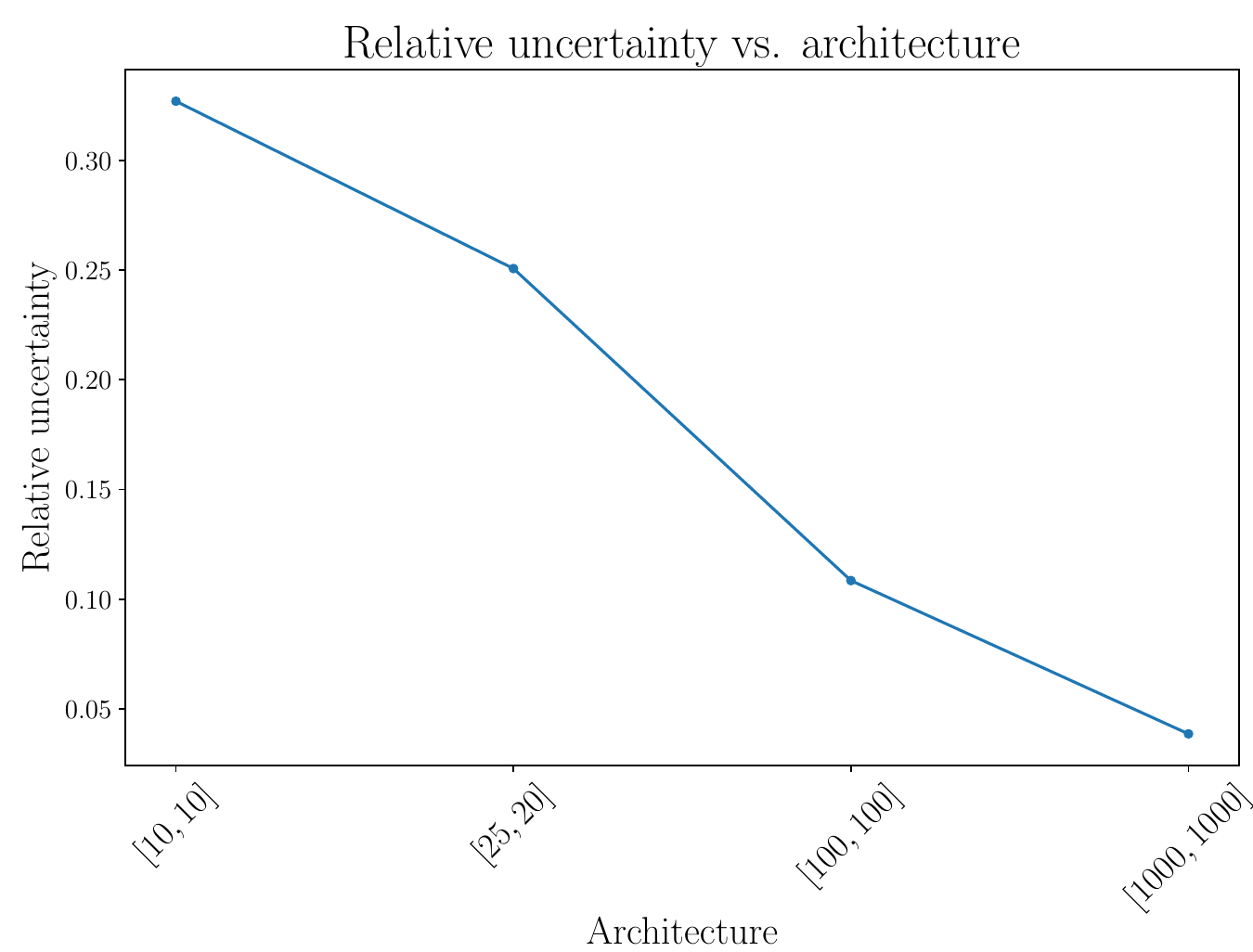}
  \caption{Frobenius norm of the NTK at initialisation, $\lVert \Theta_0
  \rVert$, as a function of the width of the network. On the left, the central
  values and uncertainty bands are obtained as the mean and one-sigma deviation
  of the ensemble of networks. The plot on the right shows the relative
  uncertainty. It is interesting to note the decrease of the relative uncertainty as the 
  architecture of the network is increased. For larger networks, the sensitivity to a change
  of the network parameters fluctuates less.}
  \label{fig:NTKInit}
\end{figure}

It is argued in the literature that, in the large-width limit, the variance of
the NTK over the set of replicas tends to zero with the width of the hidden
layers (see, \eg, \cite{Roberts:2021fes,Chiefa:2026TBA}). In order to quantify the
variation of the NTK, we start by computing the Frobenius norm of the NTK over
an ensemble of networks for different architectures. For each architecture, we
consider the standard deviation of the norm as a statistical
estimator of the variations of the NTK. The result is displayed in
Fig.~\ref{fig:NTKInit}. Even though the Frobenius norm is a coarse indicator of
the variations of the NTK, the figure shows clearly that the variance of the
norm becomes smaller with the size of the network, which is consistent with the
theoretical expectation that the NTK should not fluctuate for infinite-width
networks.\footnote{Note that, in addition to the scaling $\mathcal{O}(1/n)$
theoretically predicted for large networks, the uncertainty bands include
bootstrap errors due to the finite size of the ensemble. Using an ensemble of
100 replicas, the bootstrap error on the standard deviation is $\sim 10\%$.}

\begin{figure}[t]
  \centering
  \includegraphics[width=0.7\textwidth]{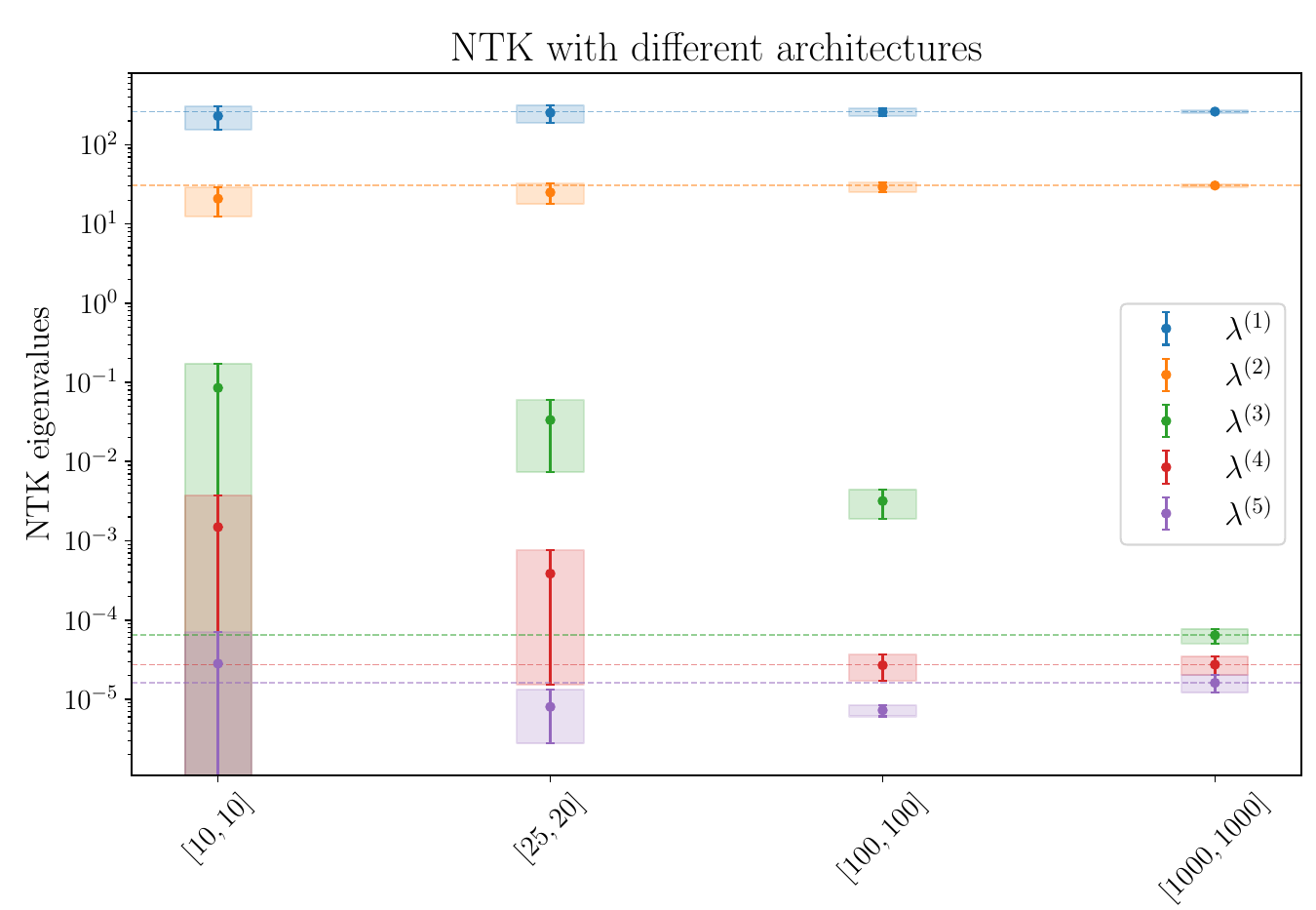}
  \caption{Spectrum of the NTK at initialisation for the architectures shown in
  Fig.~\ref{fig:NTKInit}. Error bands correspond to one-sigma uncertainties over
  the ensemble of networks. The hierarchy of the eigenvalues is independent of the size
  of the network. In agreement with the data in Fig.~\ref{fig:NTKInit}, the fluctuations 
  of the eigenvalues decrease as the width of the layers is increased. 
  }
  \label{fig:NTKSpectrum}
\end{figure}

A more quantitative description of the NTK at initialisation is provided by its
spectrum, which is shown in Fig.~\ref{fig:NTKSpectrum} for four different
architectures. Inspecting the figure, we see that the spectrum of the NTK is
heavily hierarchical, and only few eigenvalues are actually
non-zero.\footnote{Note that, due to the large difference in magnitude of the
eigenvalues, the finite precision used in our codes introduces noise in the
decomposition, so that small eigenvalues should be effectively considered zero.
We discuss the cut-off tolerance in Appendix~\ref{sec:cutoff}.} Such a hierarchy 
in the eigenvalues means that
only a small subset of active directions can inform the network during training,
as it will be discussed later. Note that, at least at initialisation, these
observations do not depend on the architecture: the eigenvalues in
Fig.~\ref{fig:NTKSpectrum} are mostly independent of the size of the network.
Even though the logarithmic scale on the vertical axis may hide some small variations, it 
is clear that most eigenvalues remain constant within the error bars. On the other hand, 
the logarithmic scale emphasises that there are several orders of magnitude between eigenvalues 
for a given architecture; that hierarchical structure does not depend on the architecture. 
There is a downward fluctuation of the third eigenvalue for the largest
architecture that we considered, but we do not have any evidence that this drop
is a physical feature of the system, rather than a fluctuation. Finally, the
variance of the set of eigenvalues over replicas decreases with increasing size,
as expected. 

\FloatBarrier

\subsubsection{NTK During Training}
\label{sec:NTKDuringTraining}

We now discuss the behaviour of the NTK during training. To this end, we are
going to adopt the so-called {\em closure tests} developed by the NNPDF
collaboration. A closure test uses synthetic data, generated using a known set
of PDFs, to train the neural network. The PDFs used for generating the data are
called here {\em input}\ PDFs. The results of the training are then compared to
the known input PDFs; the performance of the training algorithm and the NN
architecture are assessed by quantifying the comparison between trained PDFs and
input PDFs. Following the original presentation in Ref.~\cite{NNPDF:2014otw}, we
distinguish three levels of closure tests, which are defined by the complexity
of the data used to train the NNs. We use the standard NNPDF nomenclature and
refer to these three levels as level-0 (L0), level-1 (L1), and level-2 (L2)
closure tests, and we denote the input PDFs used to generate the data as $\fin$.
The definitions of these three levels of data are given in Appendix~\ref{app:dataset}.

For each of the closure-test data given above, we perform a fit of the
triplet combination $T_3$ using the simplified version of the NNPDF methodology that we 
discussed above. We initialise an ensemble of $\nreps = 100$ replicas with identical architecture,
training each replica independently using GD optimization. Throughout the
training process, we track the evolution of the NTK to understand how the
network's effective dynamics change as it learns the target function.

\paragraph{Onset of Lazy Training} 

\begin{figure}[t]
  \centering
  \includegraphics[width=0.60\textwidth]{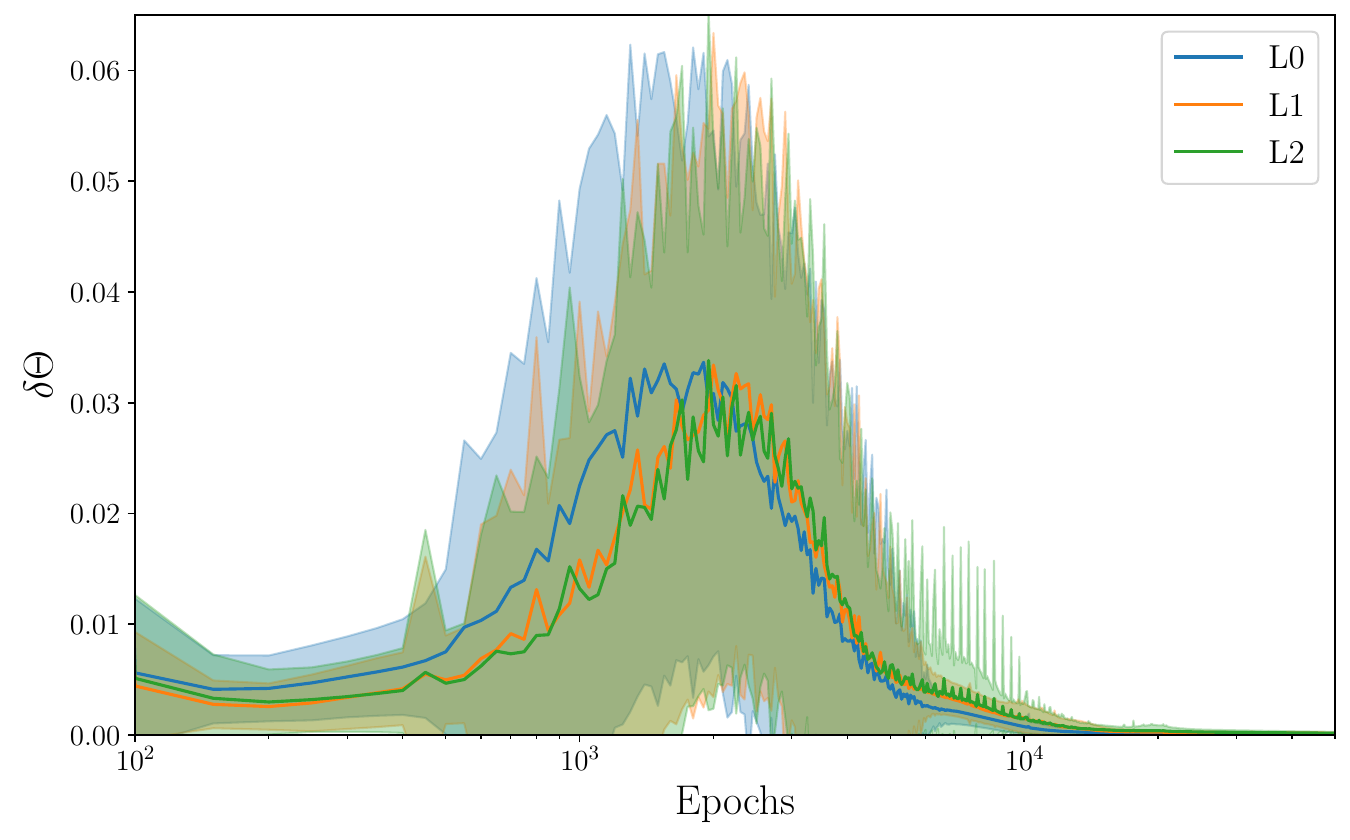}
  \caption{Relative variation of the NTK during training for L0, L1, and
  L2 data. Error bands correspond to one-sigma uncertainties over the ensemble
  of networks.}
  \label{fig:NTKTime}
\end{figure}

As a first estimator of the variation of the NTK, we show in
Fig.~\ref{fig:NTKTime} the Frobenius norm of the variation during training,
normalized by the Frobenius norm of the NTK itself, 
\begin{equation}
\delta \Theta_t = \frac{\lVert \Theta_{t+1} - \Theta_t \rVert}{\lVert \Theta_t \rVert} \;,
\label{eq:DeltaNTK}
\end{equation}
for the three different datasets, L0, L1, and L2. Inspecting the plot reveals that
the NTK undergoes significant changes during the initial phase of training, with
the relative variation $\delta \Theta_t$ reaching values as high as $6\%$. This
indicates that our settings differ from the standard picture of lazy training in
the context of very wide networks, as discussed, \eg, in
Refs.~\cite{jacot2018neural,Roberts:2021fes,lee2019wide}, where the NTK is expected to 
be independent of the flow time $t$. Remarkably, we do not
observe a dependence on how data have been generated, indicating that the NTK
dynamics is basically unaffected by the noise that affects the data. 

After this initial phase -- corresponding approximately to the first 20,000
epochs in our experiment -- the NTK tends to stabilize. These two regions will
be referred to as the \textit{rich} and \textit{lazy} training regimes,
respectively, in keeping with the standard terminology adopted in the literature
(see, \eg, Ref.~\cite{fort2020dlvk} where two similar regimes were also
identified). We do not comment any further on the implications of the lazy
regime, and postpone the discussion to Sec.~\ref{sec:AnlyticalLazySolution}.

\FloatBarrier

\paragraph{Eigenvalues During Training}

\begin{figure}[ht]
  \centering
  \includegraphics[width=0.3\textwidth]{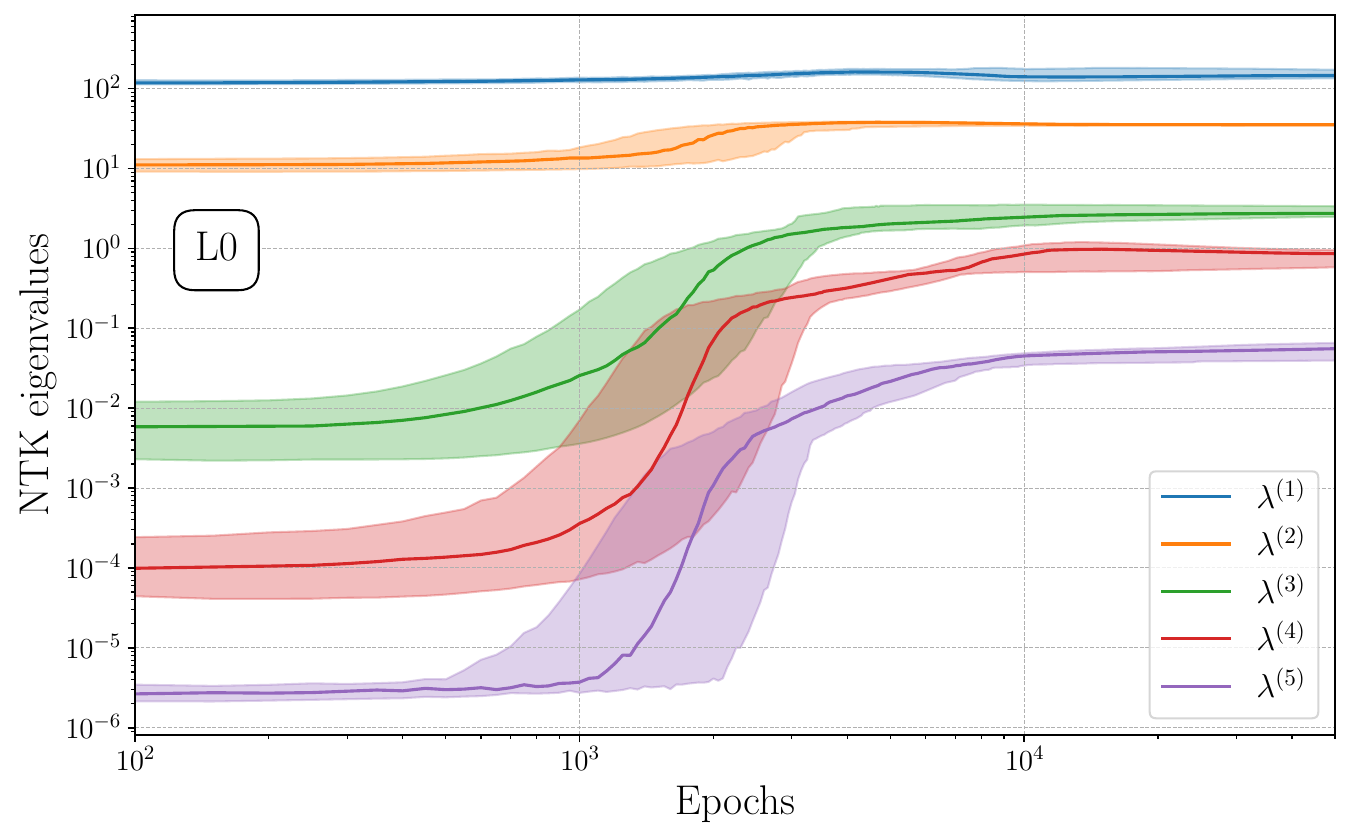}
  \includegraphics[width=0.3\textwidth]{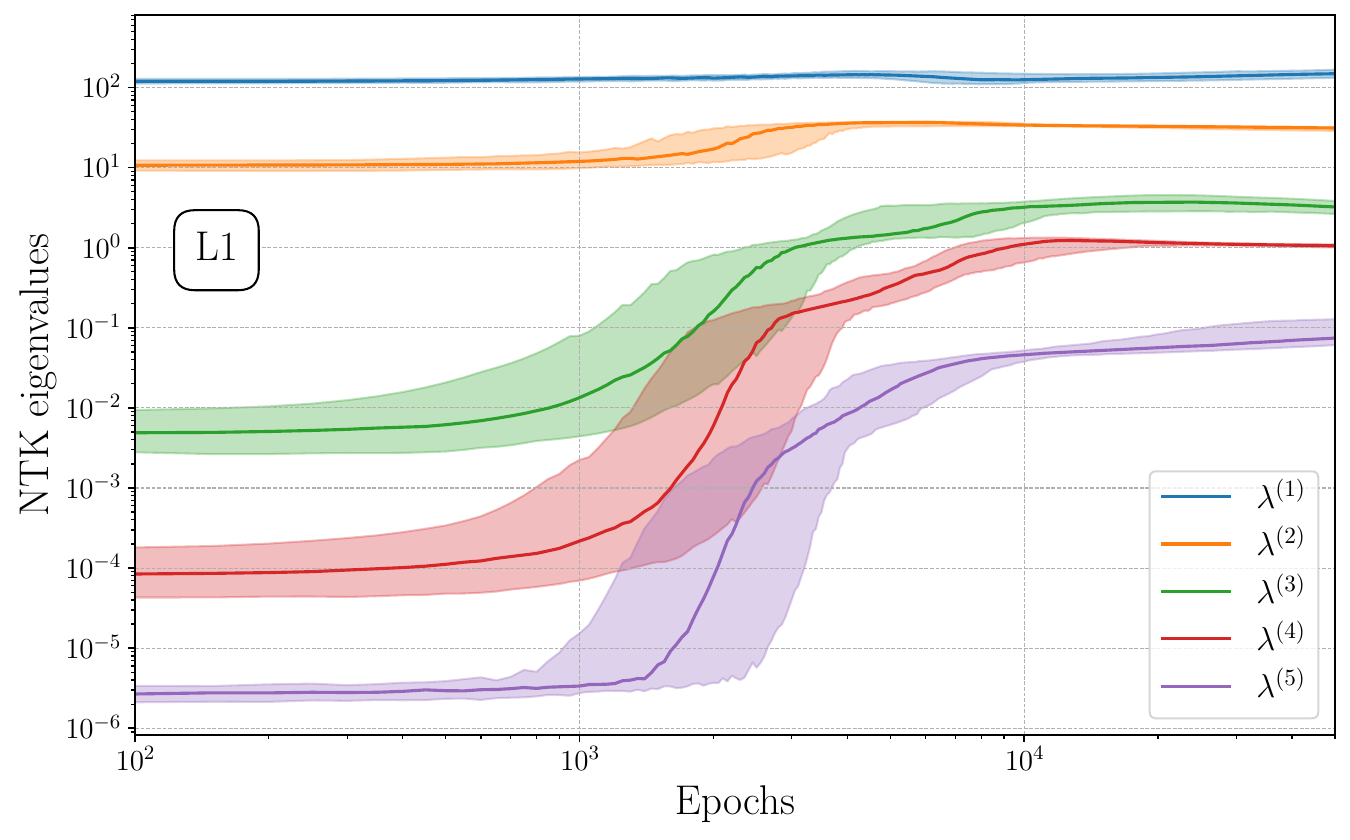}
  \includegraphics[width=0.3\textwidth]{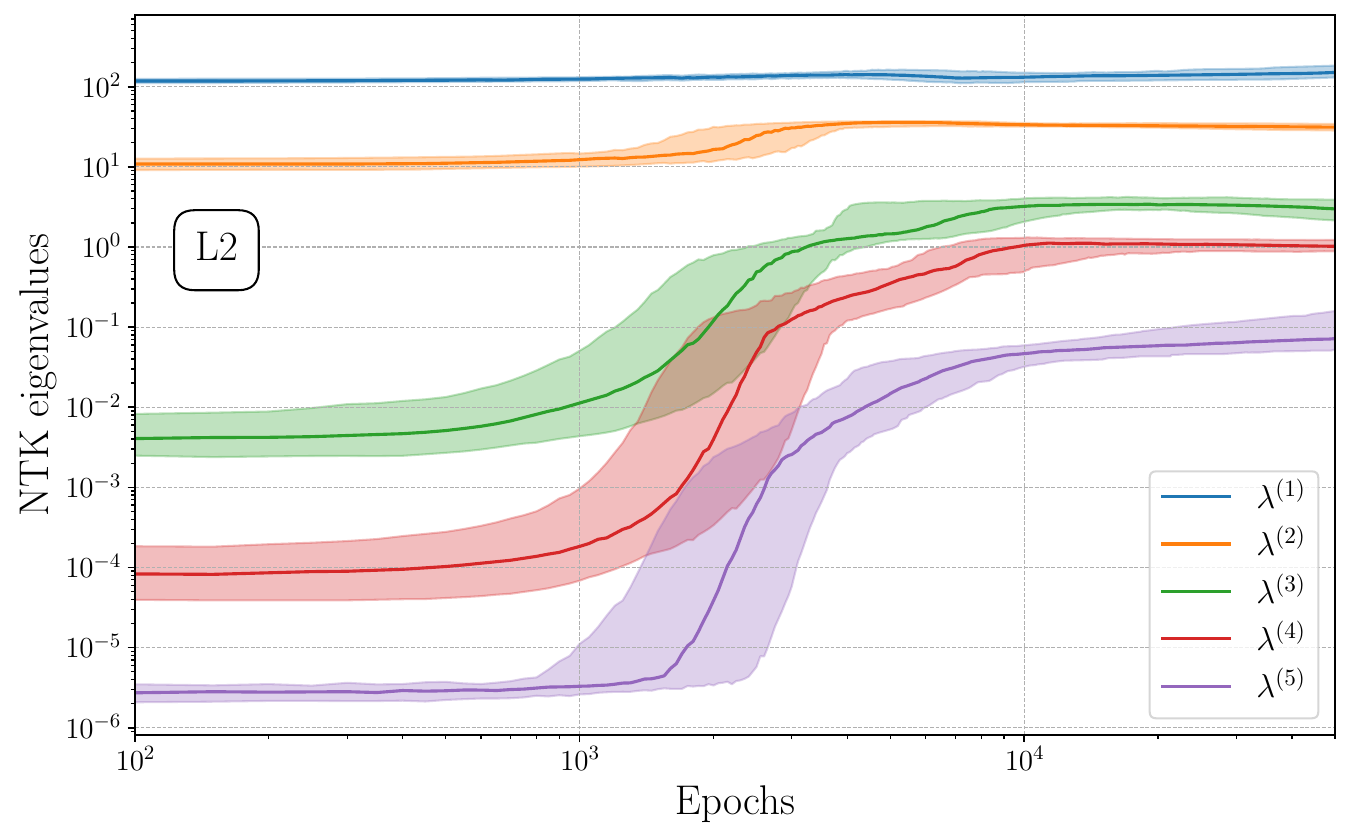} 
  \caption{Evolution during training of the first five eigenvalues of the NTK
  using L0 (left), L1 (center), and L2 (right) data. Solid lines represent the
  median over the ensemble of networks, while solid bands correspond to 68\%
  confidence level. Note that the subdominant eigenvalues $\lambda^{(3)}$, $\lambda^{(4)}$
  and $\lambda^{(5)}$ have increased by one or two orders of magnitude by the end of the rich 
  training phase.}
  \label{fig:NTKEigvalsTime}
\end{figure}

Further insight on the evolution of the NTK can be obtained by studying its
eigensystem as a function of the training time. In Fig.~\ref{fig:NTKEigvalsTime}
we report the variation of the first five eigenvalues of the NTK, using the
standard NNPDF architecture, for L0, L1, and L2 data. We see that the
hierarchical structure observed at initialisation is preserved, but the size of
the subdominant eigenvalues increases significantly in the early stages of
training -- by one or two orders of magnitude depending on the specific
eigenvalue. 

\FloatBarrier

\begin{figure}[ht]
  \centering
  \includegraphics[width=0.30\textwidth]{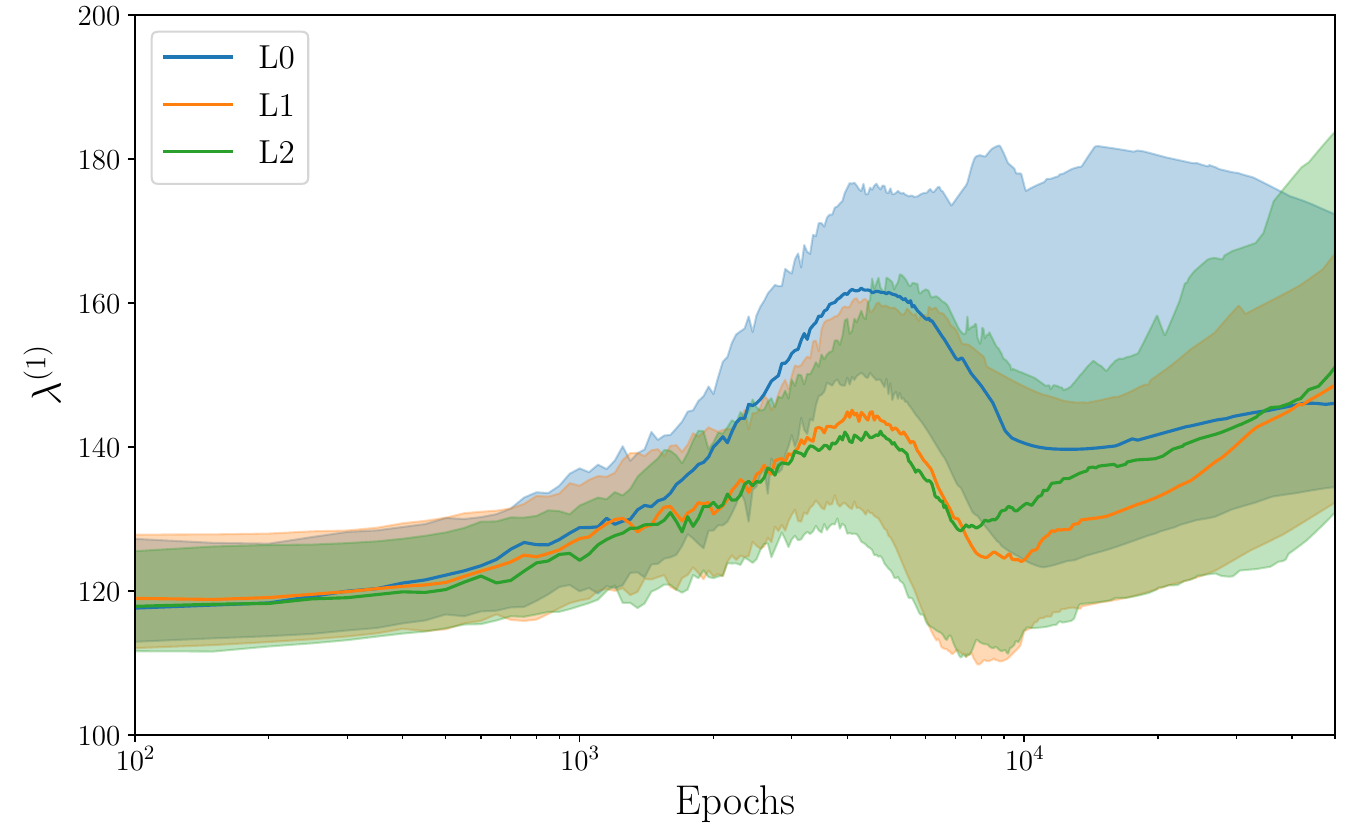}
  \includegraphics[width=0.30\textwidth]{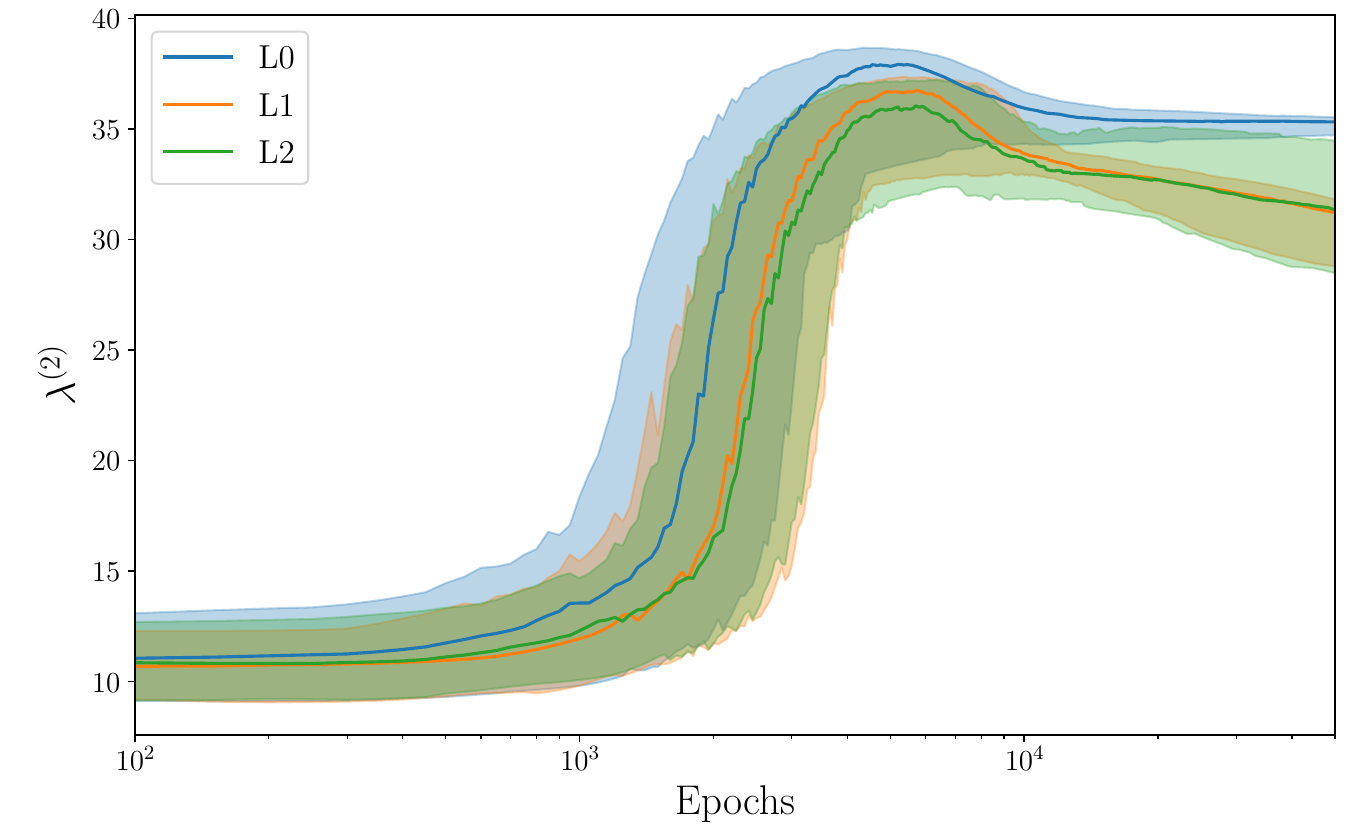}
  \includegraphics[width=0.30\textwidth]{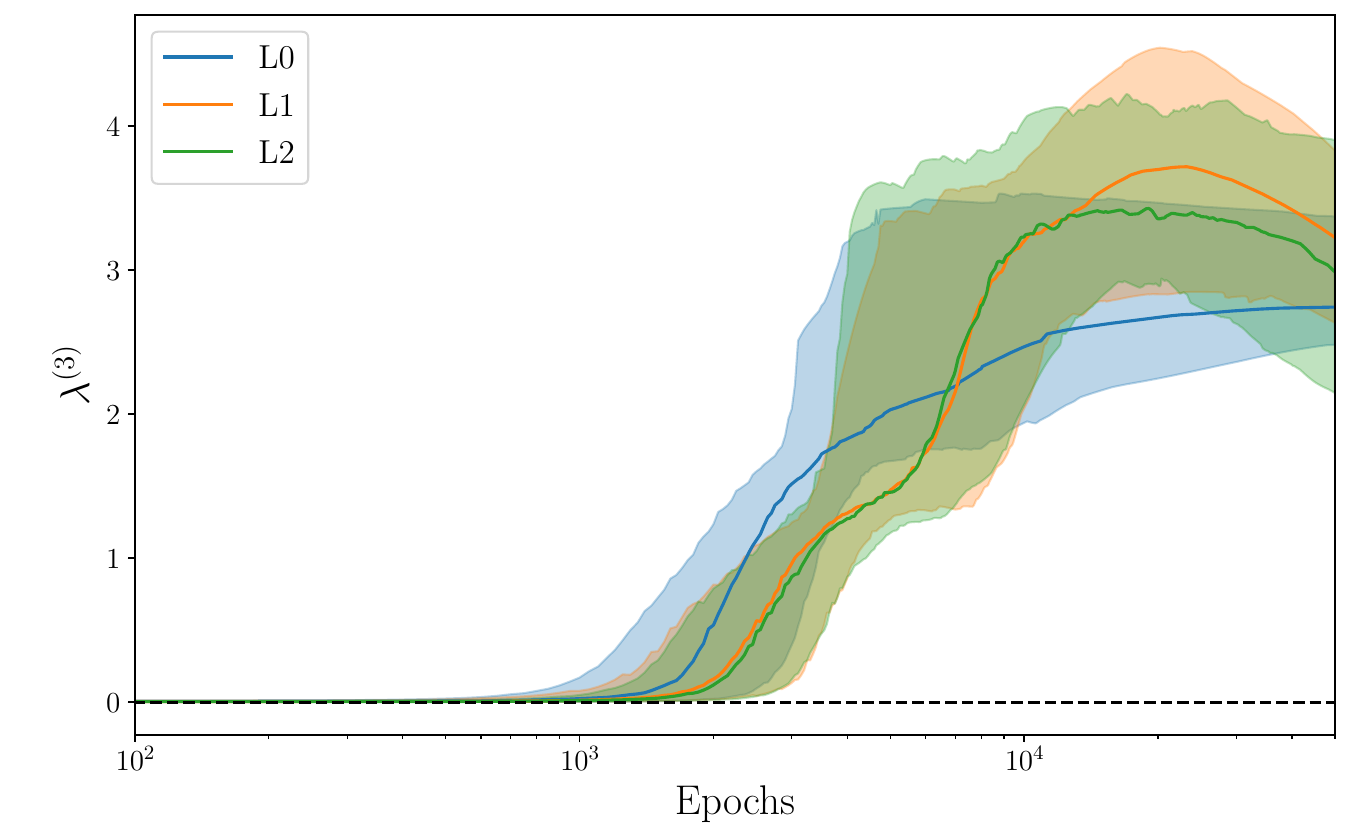}
  \includegraphics[width=0.30\textwidth]{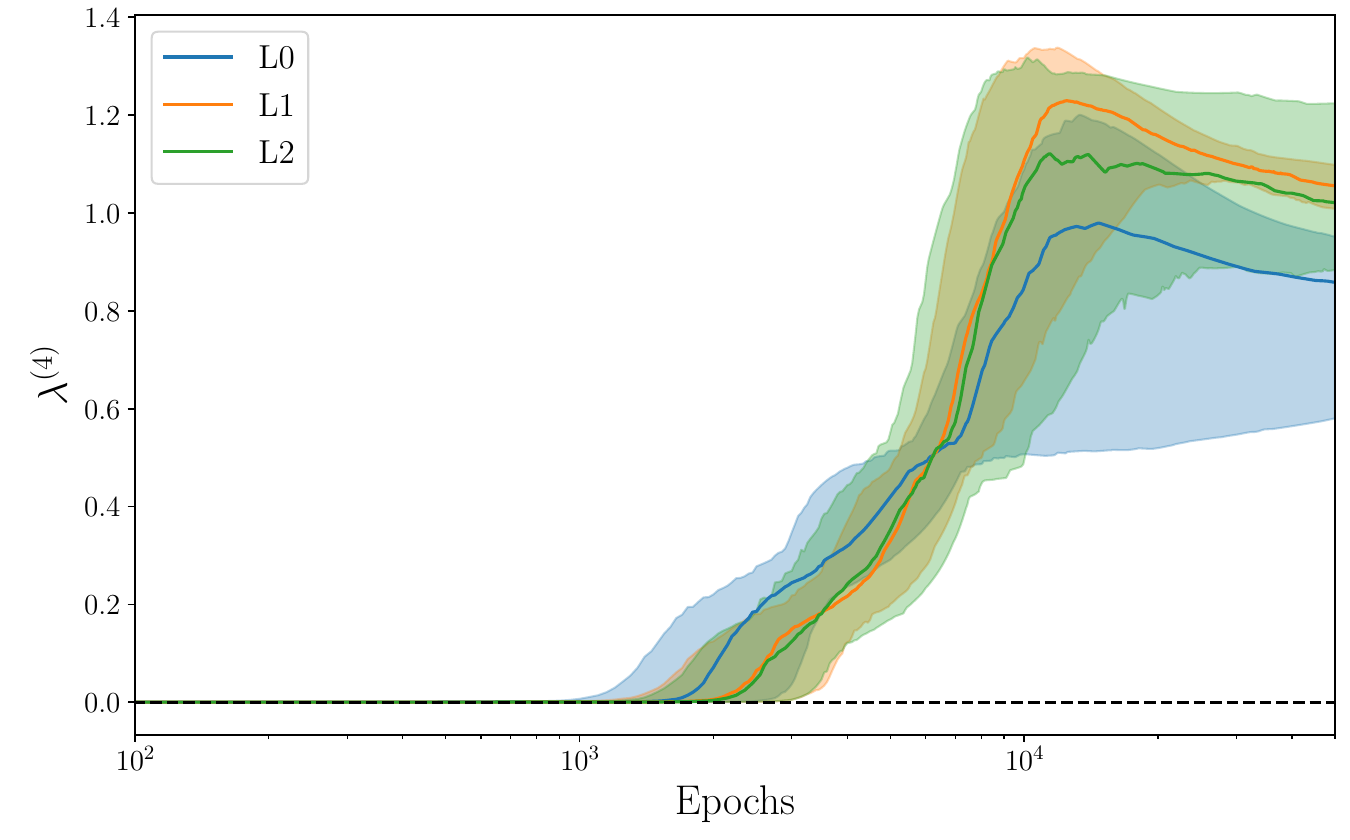}
  \includegraphics[width=0.30\textwidth]{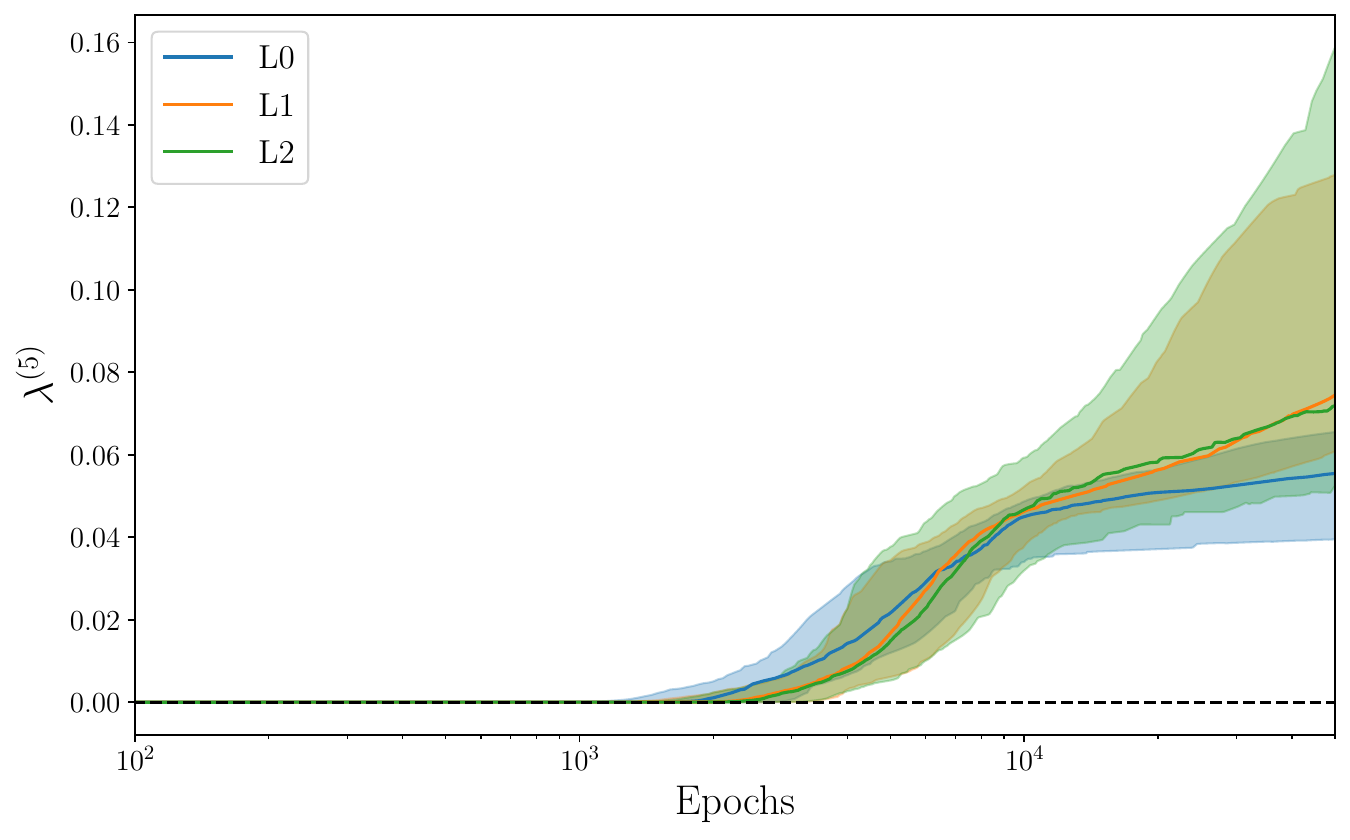}
  \caption{The first five eigenvalues of the NTK for L0, L1, and L2 data. Solid
  lines represent the median over the ensemble of networks, while solid bands
  correspond to 68\% confidence level. Each plot corresponds to a different eigenvalue, 
  as indicated by the label on the vertical axis. Note the different scales on the 
  vertical axes, which reflects the hierarchy of eigenvalues discussed above. Different
  colours correspond to different synthetic data, the agreement between these bands 
  confirms that the evolution of the eigensystem of the NTK does not depend on the 
  level of noise in the data.}
  \label{fig:EigvalsComparison}
\end{figure}

In Fig.~\ref{fig:EigvalsComparison}, the same first five eigenvalues of the NTK
are displayed for L0, L1, and L2 data. We observe a common pattern across all
data types, consistently with the observation made before in
Fig.~\ref{fig:NTKTime}. This indicates the NTK evolution is insensitive to
the noise included in the synthetic data. The increase of the subdominant
eigenvalues, combined with the analysis of Eqs.~\eqref{eq:FlowParallel}
and~\eqref{eq:FlowPerp} in Sect.~\ref{sec:LazyTraining}, suggests that 
more ``physical'' features become learnable before lazy training sets in.

\FloatBarrier

\begin{figure}[ht]
  \centering
  \includegraphics[width=0.30\textwidth]{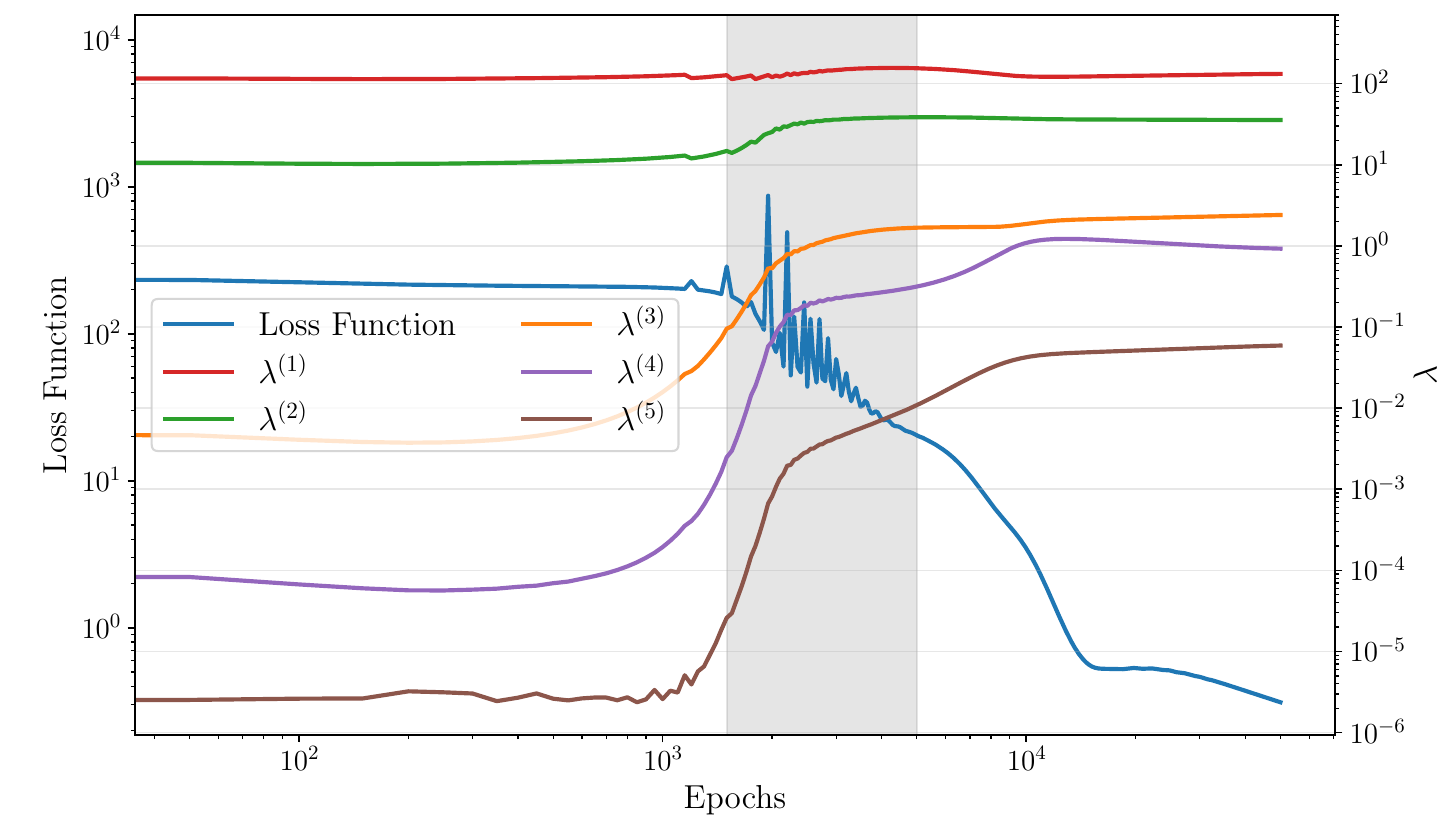}
  \includegraphics[width=0.30\textwidth]{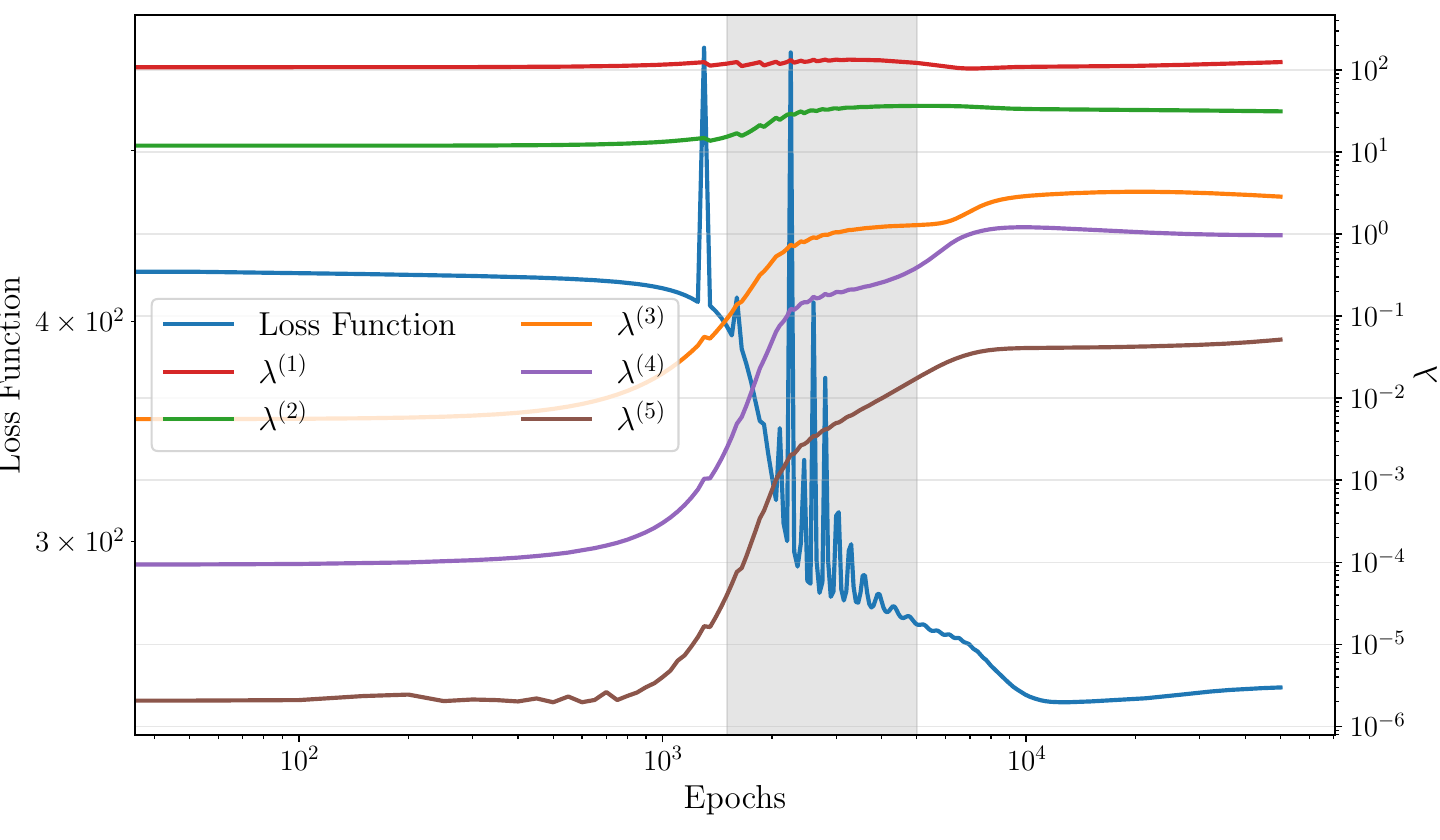}
  \includegraphics[width=0.30\textwidth]{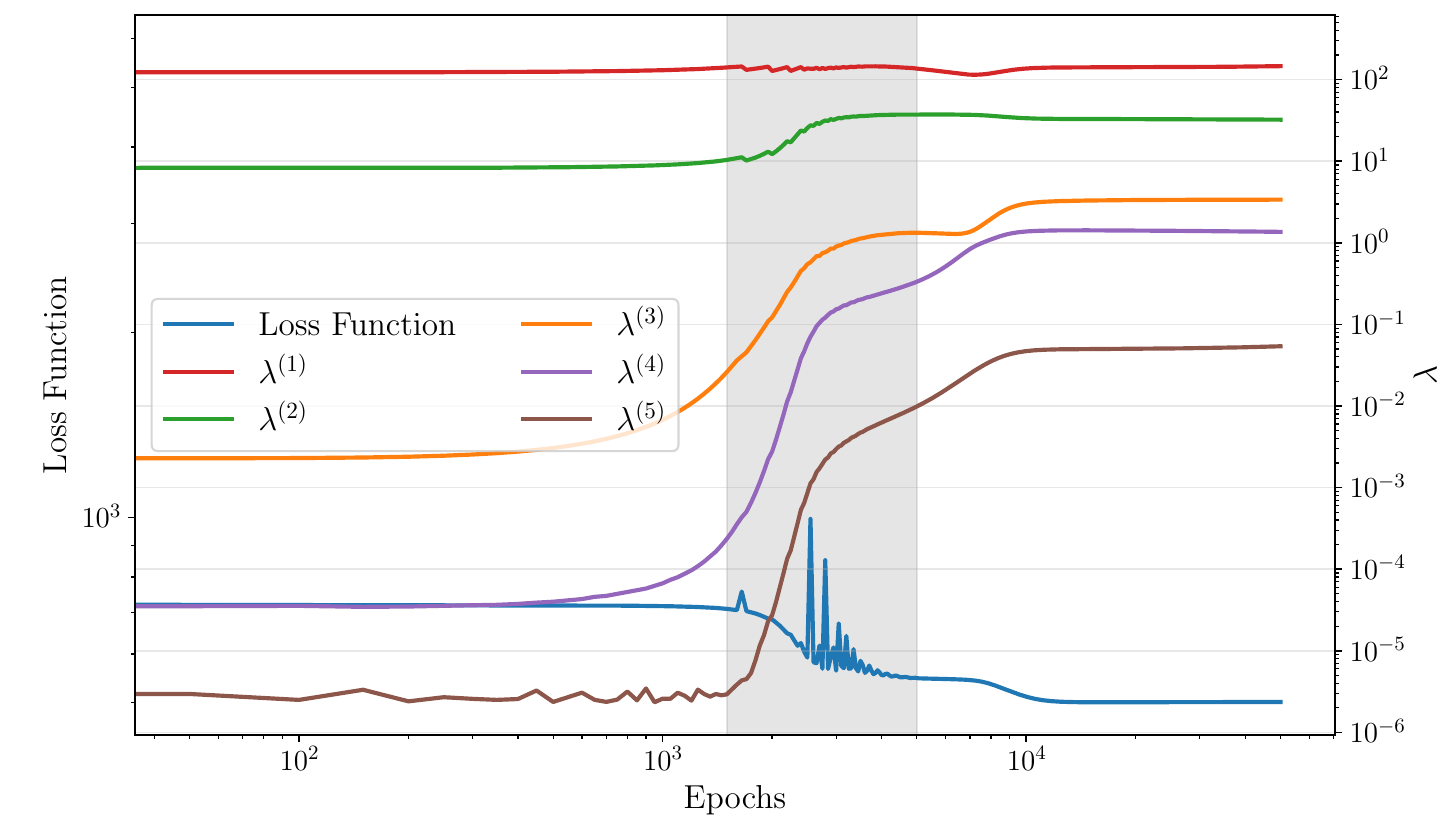}
  \caption{Variation of the loss function overlaid with the first five
  eigenvalues for a selected replica over the ensemble using L0 (left), L1
  (center), and L2 (right) data. Left scale refers to the loss, while the right
  scale refers to the eigenvalues.}
  \label{fig:Loss}
\end{figure}

\paragraph{Connection with the loss function} Finally, in Fig.~\ref{fig:Loss} we
show the variation of the loss function during training, overlaid with the first
five eigenvalues of the NTK, for a selected replica over the ensemble. It is
interesting to see that in correspondence with the sudden variation of the
subdominant eigenvalues, the loss function drops significantly, at the cost of
an instability localised in the descent. We interpret this as the network
learning new features, changing its internal representation to accommodate the
new information. After this initial phase, the eigenvalues stabilize and the
loss function decreases smoothly, as expected in the lazy training regime.

As it will be extensively discussed later in Sec.~\ref{sec:LazyTraining}, the
eigenvalues and eigenvectors of the NTK play a special role. Indeed, the output
$f$ can be decomposed into the basis of eigenvectors of the NTK. Hence the
eigenvectors corresponding to the larger eigenvalues can be interpreted as {\em
learnable}\ features, while the small (or zero) eigenvalues correspond to
directions in which the field $f$ never evolves during training.

\FloatBarrier

\subsubsection{Eigenvectors and Alignment of the NTK}
\label{sec:NTKAlign}

\begin{figure}[ht!]
  \centering
  \includegraphics[width=1\textwidth]{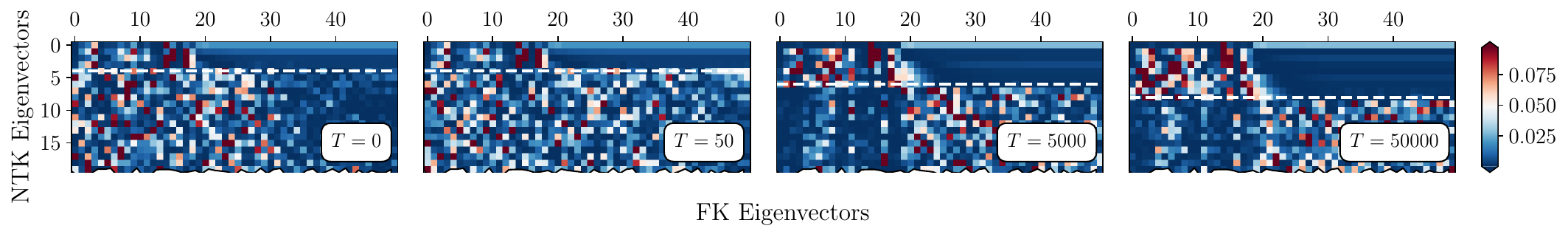}
  \caption{Matrix $A$ as defined in Eq.~\eqref{eq:MatrixA} for L2 data and for a
  single replica of the NTK. The matrix is shown at different epochs of the
  training process, indicated in the top of each panel. The white dashed line
  indicates the cut-off tolerance that we impose on the eigenvalues of the NTK
  (see Appendix~\ref{sec:cutoff}).}
  \label{fig:NtkMAlign}
\end{figure}

It has been argued above that there is a non-trivial interplay between the
eigenspace of the NTK and that of the matrix $M$. Indeed, the former encodes the
model dependence, while the latter yields physical information. Of course the
two matrices are independent at initialisation, and we do not expect any
alignment pattern between the two. However, this picture does change during
training, as the NTK evolves and the model learns the target function. To
quantify this alignment, we define the matrix $A$, 
\begin{equation}
  \label{eq:MatrixA}
  A_{kk'} = \left( z^{(k)}, v^{(k')} \right)^2 = \cos^2(\theta_{kk'}) \;,
\end{equation}
where $z^{(k)}$ and $v^{(k')}$ are the $k$-th and $k'$-th eigenvectors of the
NTK and $M$, respectively. The matrix $A$ is thus a measure of the alignment
between the eigenspaces of the two matrices. The rows of the matrix correspond
to the eigenvectors of the NTK, ordered by the value of the corresponding
eigenvalues, with the eigenvectors corresponding to the larger eigenvalues at
the top of the matrix. The columns correspond to eigenvectors of the matrix $M$,
also ordered by the values of the corresponding eigenvalues, with the largest
eigenvalues to the left in this case. In Fig.~\ref{fig:NtkMAlign}, we show the
matrix $A$ at different epochs of the training for L2 data and a single NTK
replica. 

The blue rectangle in the top right corner of the matrix shows that the
eigenvectors of the NTK corresponding to the largest eigenvalues are orthogonal
to the eigenvectors of $M$ that are in the kernel of $M$, \ie, the directions
that do not contribute to the observables. It is useful to remember that the
largest eigenvalues of the NTK correspond to the directions that are orthogonal
to $\ker\Theta$, \ie, the directions that are learnable during the training
process. In order to have a robust training process, we expect these learnable
directions to align with the directions that actually contribute to the loss
functions, which are the ones corresponding to the largest eigenvalues of $M$.
Consistently with this intuition, we see that the size of this blue rectangle
increases with training time. In particular, it is clear from our plot that it
becomes deeper by the onset of the lazy training regime: more of the learnable
directions -- the {\it features}\ that the network can learn -- are aligned with
the directions that contribute most to the observables.

\begin{figure}[ht!]
  \centering
  \includegraphics[width=0.60\textwidth]{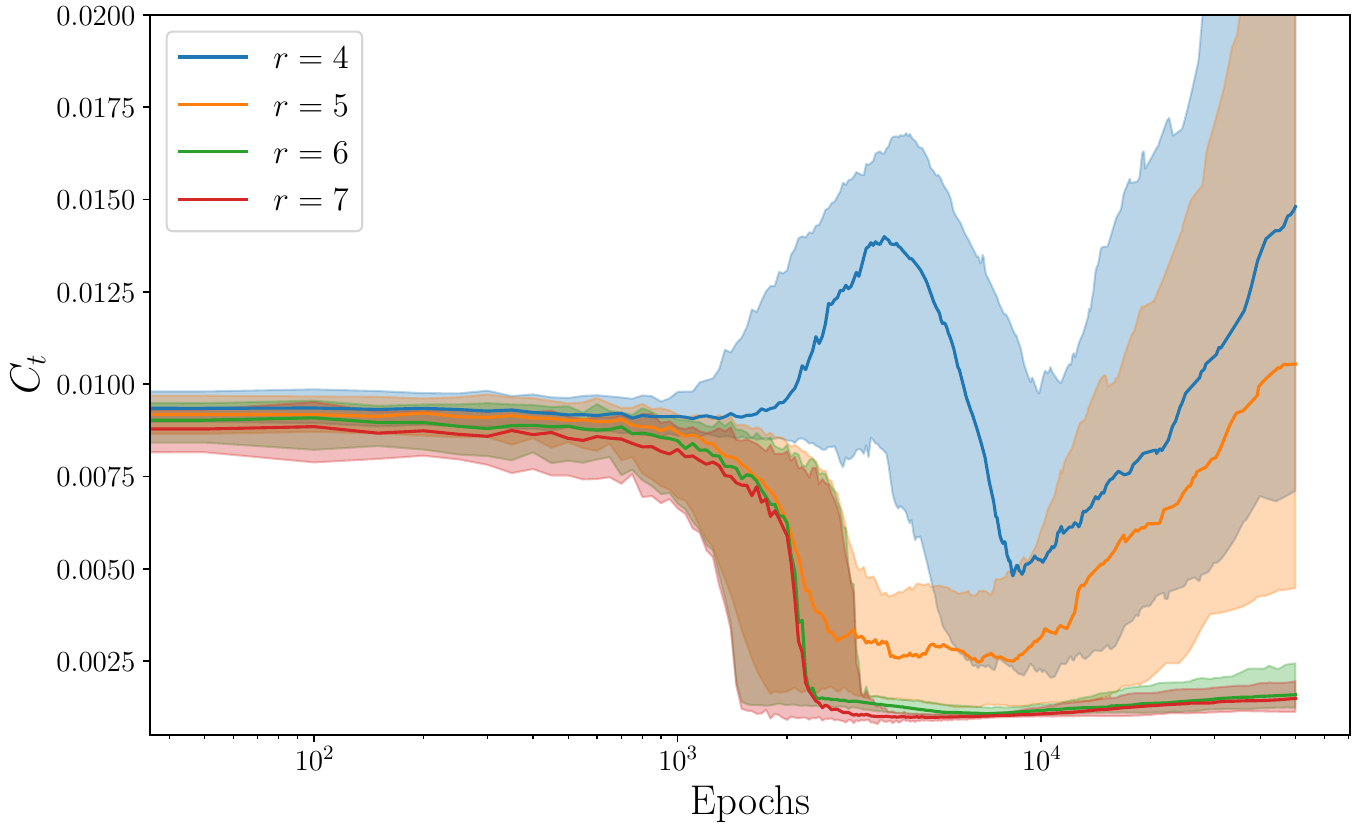}
  \caption{Reconstruction error $C_t(r)$ as defined in Eq.~\eqref{eq:NTKCoverage}
  as a function of the training time and for different numbers of eigenvectors
  $r$. Note that $\fin$ lies entirely in the subspace spanned by the first four
  eigenvectors of the NTK by the onset of the lazy training regime. We see that
  the NTK has aligned its features with the physically relevant directions of
  the problem.}
  \label{fig:NTKAlignFin}
\end{figure}
We remark that the eigenvectors of the NTK form an orthonormal basis in
$\mathbb{R}^{\ngrid}$ at any epoch of the training process. We have also shown
that only a subset of these eigenvectors contribute to the training dynamics. We
may thus wonder how expressive the vectors belonging to the subspace orthogonal
to the kernel are, \ie, how well the internal representation of the neural
network can reconstruct the input function $\fin$ used to generate the data. In
order to quantify this aspect, we define a new figure of merit
\begin{equation}
  C_t(r) = 1 - \sum_{k=1}^{r} \frac{(z^{(k)}, \fin)^2}{\Vert \fin \Vert^2} \,,
  \label{eq:NTKCoverage}
\end{equation}
which measures the reconstruction error of the input function $\fin$ when
projected onto the subspace spanned by the first $r$ eigenvectors of the NTK at
training time $t$. We show $C_t(r)$ as a function of the training time and for
different choices of $r$ in Fig.~\ref{fig:NTKAlignFin}. Inspecting the figure,
we see that in the early stages of training the reconstruction error does not
change significantly with time. This behaviour is shared across all values of
$r$. In fact, in this first phase the NTK has not yet found a suitable internal
representation and the inclusion of more eigenvectors corresponding to yet
undiscovered directions -- those associated with a small eigenvalue as in
Fig.~\ref{fig:NTKEigvalsTime} -- does not result in an improvement of the
reconstruction of $\fin$. Conversely, once the onset of lazy training is
approached, we identify two distinct behaviours depending on the number of used
eigenvectors. If we include the eigenvectors up to the modes discovered during
training (\eg, $r=6$ and $r=7$ in the figure), the reconstruction error drops
significantly and remains almost constant throughout the rest of the learning
process. On the other hand, if we consider fewer eigenvectors (\eg, $r=4$ and
$r=5$ in the figure), the reconstruction error becomes larger and grows
indefinitely with training time. This results show us that the subset of
eigenvectors orthogonal to $\ker \Theta$ is capable of providing an effective
lower dimensional basis, provided that the new modes discovered by the NTK
during training are included. Note that the reconstruction of the input function
improves as long as the included eigenvectors do not belong to $\ker \Theta$ --
being noise, they would not bring any physical information.

\begin{figure}[ht!]
  \centering
  \includegraphics[width=0.90\textwidth]{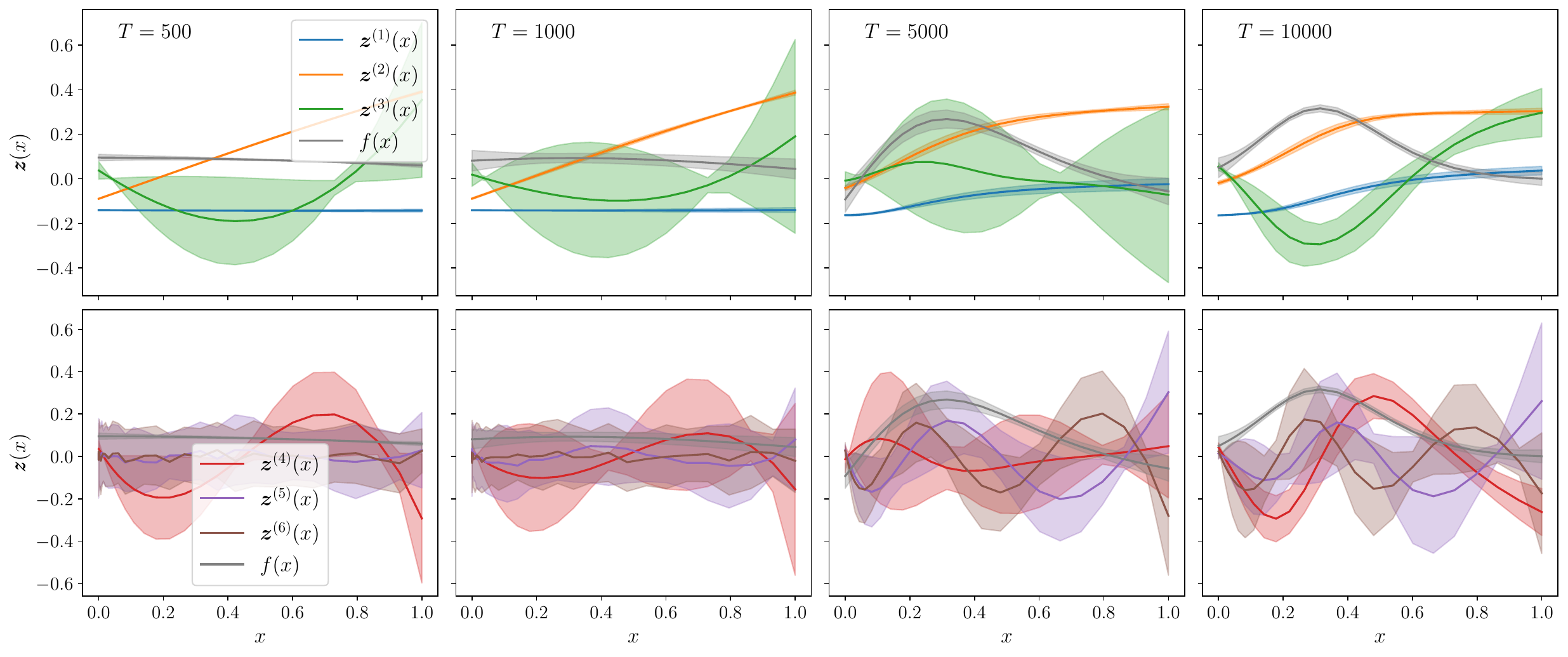}
  \caption{First five eigenvectors of the NTK at different training times and as
  function of the input $x$-grid. We also show the output of the network at the
  same training time, which is displayed in gray. L1 data is used.}
  \label{fig:NTKEigVecs}
\end{figure}
A complementary picture is displayed in Fig.~\ref{fig:NTKEigVecs}. Here, we
show the eigenvectors of the NTK at different training times as functions of the
$x$-grid, denoted by $z^{(i)}$. Together with the eigenvectors, we also show
the output of the trained neural network at the corresponding training time.
From these plots, we see that as the training progresses, the shape of the
eigenvectors becomes more structured in order to reproduce the output function.
Again, this conclusion supports the observations made previously on various
occasions, that during training the neural network is changing its internal
representation and the NTK encodes this information.
\FloatBarrier

\subsection{Regularisation of the inverse problem and the role of the NTK}
\label{sec:NTKRegularisation}
The discussion above characterised the training dynamics of the network in terms
of the NTK. It did not, however, address how the combination of a
parametrisation and an optimisation process can solve the ill-defined inverse
problem of PDF fitting. Before concluding this section, we now discuss two
complementary perspectives that expose the ill-defined nature of the inverse
problem and the role of the parametrisation in regularising it. The discussion
also introduces the eigenspace decomposition that is used extensively in
Sec.~\ref{sec:LazyTraining}.

\paragraph{The inverse problem in functional space.}
One could attempt to determine $f$ by directly minimising the loss in
Eq.~\eqref{eq:QuadLoss} in functional space. Taking the derivative with
respect to $f$ and setting it to zero yields the stationary condition
\begin{equation}
  \label{eq:chi2_inv_sol}
  M f = \FKtab^T C_Y^{-1} Y\, ,
\end{equation}
with $M$ defined as in Eq.~\eqref{eq:MandBDef}. Here the time dependence is
dropped, as no training process is involved. The transpose $\FKtab^T$ acts
as the \emph{adjoint} of the forward map $\FKtab$: it takes a vector in data
space, weighted by the data uncertainties, and returns the linear combination in
PDF space that the data probe. Crucially, $\FKtab^T$ is not the inverse of
$\FKtab$. Its range is the orthogonal complement of $\ker\FKtab$, so it carries
no information about PDF directions in $\ker\FKtab$ — precisely the directions
to which the data are insensitive. The composite operator $M=\FKtab^T C_Y^{-1}
\FKtab$ inherits this blind spot, with $\ker M=
\ker\FKtab$.\footnote{Geometrically, $M$ is the metric that measures how
distinguishable two models are in terms of their predictions. In fact, using L0
data the loss can be written as $\mathcal{L} = \tfrac{1}{2}(\fin - f)^T M (\fin
- f)$: two PDFs are \emph{indistinguishable} when they differ by a vector in
$\ker M$.} The solution of Eq.~\eqref{eq:chi2_inv_sol} is therefore not unique.
It decomposes as $f = \fker + \fimm$, where $\fker \in \ker M$ is undetermined
by the data and the particular solution in $\text{Im}\, M$ is
\begin{equation}
    \label{eq:chi2_inv_sol_moore_penrose}
    \fimm = M^+\FKtab^T C_{Y}^{-1} Y\,,
\end{equation}
with $M^+$ the Moore-Penrose pseudo-inverse of $M$. The pseudo-inverse acts as a
deconvolution operator on $\text{Im}\, M$, and acquires a statistical
interpretation: assuming $Y\sim\mathcal{N}(\bar{Y}, C_Y)$, we have
\begin{equation}
  \begin{split}
    \label{eq:chi2_inv_sol_moore_penrose_stats}
    &\mathbb{E}[\fimm] = M^+ \FKtab^T C_{Y}^{-1} \bar{Y}\,,\\
    &\textrm{Cov}[\fimm, \fimm^T] = M^+ M M^+ = M^+\,,
  \end{split}
\end{equation}
which shows that $M^+$ is itself the covariance of the reconstructed solution in
$\text{Im}\, M$.\@
\begin{figure}[t]
    \centering
    \begin{subfigure}[b]{0.45\textwidth}
      \centering
      \includegraphics[width=\textwidth]{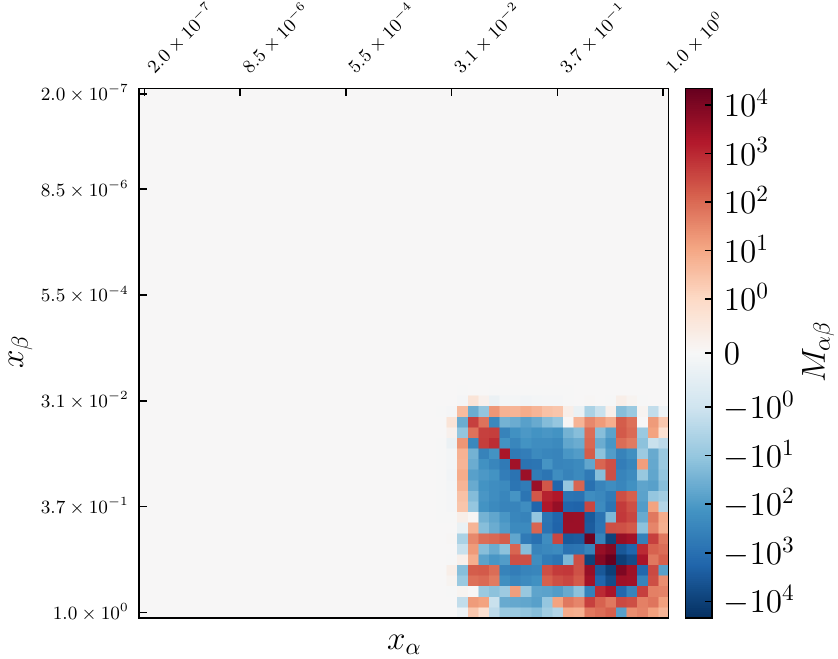}
      \subcaption{}
      \label{fig:M_matrix}
    \end{subfigure}
    \hspace{10pt}
    \begin{subfigure}[b]{0.45\textwidth}
      \centering
      \includegraphics[width=\textwidth]{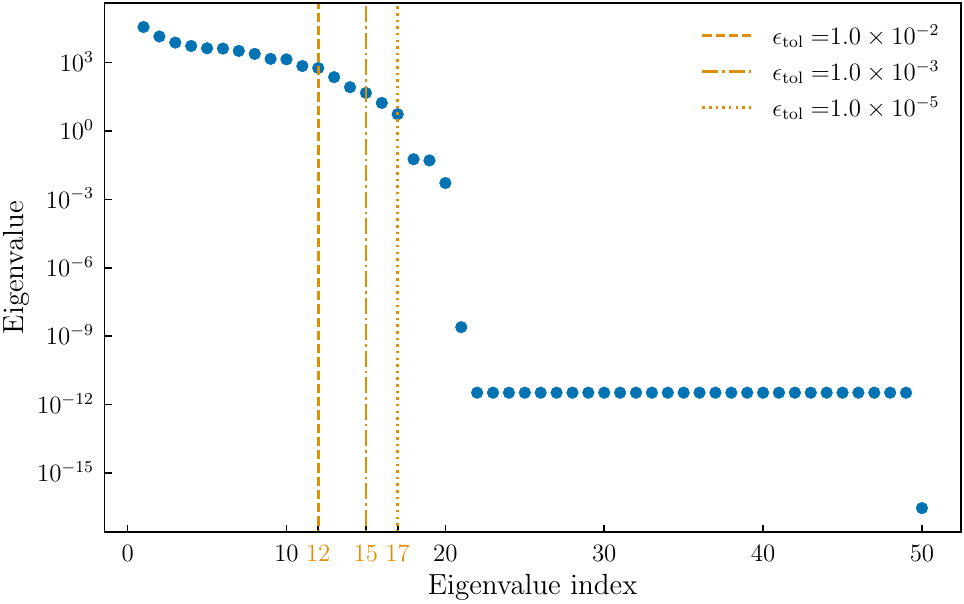}
      \subcaption{}
      \label{fig:M_spectrum}
    \end{subfigure}
    \caption{{\bf(a)} Entries of the operator $M$ for the BCDMS dataset.
    {\bf(b)} Singular values of the operator $M$, shown in logarithmic scale.
    The three vertical dashed lines correspond to the three choices of
    $\epsilon_{\rm tol}$ discussed in the main text. The points crossed by these
    lines are included in the pseudo-inverse.}
    \label{fig:M}
\end{figure}
\begin{figure}[t!]
    \centering
    \includegraphics[width=0.45\textwidth]{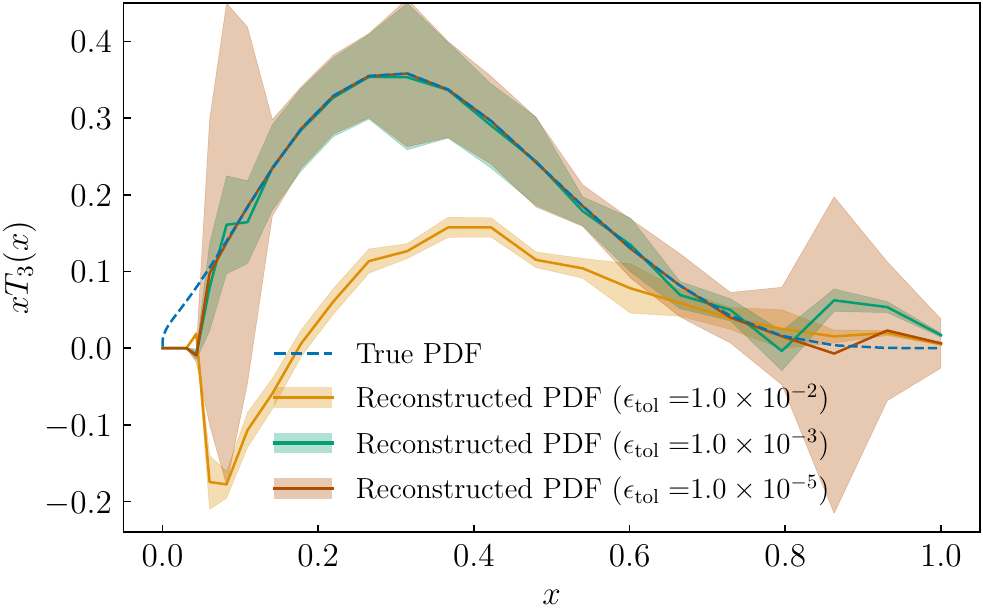}
    \includegraphics[width=0.45\textwidth]{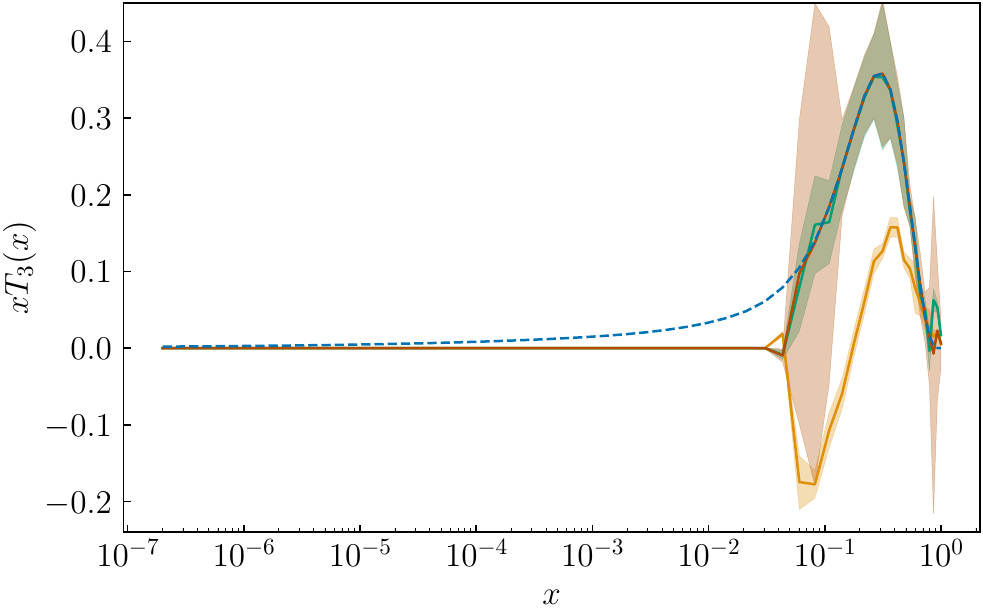}
    \caption{Solution of the inverse problem as in
    Eq.~\eqref{eq:chi2_inv_sol_moore_penrose} with the relative thresholds as in
    Fig.~\ref{fig:M}~\subref{fig:M_spectrum}. The dashed line indicates the underlying law used
    to generate the artificial data.}
    \label{fig:inversion_svd}
\end{figure}
In practice, the spectrum of $M$ for realistic datasets is highly hierarchical
and only a small subspace is numerically accessible.
Fig.~\ref{fig:M}~\subref{fig:M_matrix} displays $M$ for the BCDMS dataset used
in this work, while Fig.~\ref{fig:M}~\subref{fig:M_spectrum} shows its singular
values: many directions are essentially null and the non-zero eigenvalues span
several orders of magnitude. The effective rank of $M$ is therefore set by a
relative threshold $\epsilon_{\rm tol} \times \max(s)$, the choice of which is
arbitrary and propagates to both central values and uncertainties of the
reconstructed solution. As shown in Fig.~\ref{fig:inversion_svd}, a tight
threshold ($\epsilon_{\rm tol}=10^{-2}$) excludes too many modes and underfits
the underlying law; a loose one ($\epsilon_{\rm tol}=10^{-5}$) recovers it but
with unstable uncertainties at the boundary; intermediate values are a
compromise but still leave uncertainties unstable across the $x$ range. On top
of the arbitrariness of $\epsilon_{\rm tol}$, the solution provides no robust
estimate in the extrapolation region (where any $\fker \in \ker \FKtab$ can be
added) and is not constrained to be smooth across the boundary between the
constrained and unconstrained regions. These pitfalls motivate the use of more
robust methods, such as the one studied in this work.

\paragraph{Regularisation through a linear parametrisation}
A simple example can help illustrate the role of the parametrisation in
regularising the inverse problem. Consider a function linear in the parameters,
\begin{equation}
    \label{eq:LinearModel}
    f(x;\theta) = \sum_{i=1}^{n} \theta_i \phi_i(x),
\end{equation}
with fixed basis functions $\phi_i(x)$ (\eg, polynomials).\footnote{In the ML
literature, the basis functions $\phi_i(x)$ are often referred to as features.}
Evaluated on the $x$-grid, the function reads
\begin{equation}
    \label{eq:LinearModelCompact}
    f = \Phi \theta,
\end{equation}
where $\Phi_{\alpha i} \equiv \phi_i(x_{\alpha})$ is the Jacobian of $f$ with
respect to $\theta$, evaluated on the $x$-grid with $\alpha = 1, \dots, \ngrid$. Here we are using the same
compact and index-less notation introduced in Sec.~\ref{sec:GradFlow}. The
minimum of the loss in Eq.~\eqref{eq:QuadLoss} with respect to the parameters
$\theta$ satisfies
\begin{equation}
    \label{eq:LinearSystem}
    \Phi^T \FKtab^T C_Y^{-1}
    \left(Y - \FKtab f(\tilde{\theta})\right) = 0\, ,
\end{equation}
where $\tilde{\theta}$ denotes the vector of optimal parameters. By observing
that, for the linear model in Eq.~\eqref{eq:LinearModelCompact}, the NTK reduces
to the time-independent matrix $\Theta = \Phi \Phi^T$, we can multiply both sides
of the previous equation by $\Phi$,
\begin{equation}
    \label{eq:LinearSystemNTK}
    \Theta M f(\tilde{\theta}) = \Theta \FKtab^T C_Y^{-1} Y,
\end{equation}
which recasts the stationary condition in terms of the NTK, similar to the flow
equation in Eq.~\eqref{eq:FlowEquationNoIndices}. In general, $\Theta$ is only
positive \emph{semi}-definite, with a non-trivial kernel whenever $n < \ngrid$
-- as illustrated by the spectrum at initialisation in
Fig.~\ref{fig:NTKSpectrum}. The eigenvalue equation,
\begin{equation}
    \label{eq:ThetaEigensystem}
    \Theta z^{(k)} = \lambda^{(k)} z^{(k)},
\end{equation}
provides a basis of $\mathbb{R}^{\ngrid}$ separating the kernel of the NTK from
its image. The kernel collects the directions in functional space that the
parametrisation cannot reach, irrespective of the data. We introduce the
notation
\begin{align}
    \label{eq:ParallelComponents}
    &f^\parallel_{k} = \left(z^{(k)}, f\right)\, , \quad \text{if}\ \lambda^{(k)} = 0\, , \\
    \label{eq:OrthogonalComponents}
    &f^\perp_{k} = \frac{1}{\sqrt{\lambda^{(k)}}} \left(z^{(k)}, f\right)\, , \quad
        \text{if}\ \lambda^{(k)} \neq 0\, ,
\end{align}
with the scalar product defined as $\left(f'_{t'}, f_t\right) = \sum_{i,\alpha}
f'_{t',i\alpha} f_{t,i\alpha}$. In this linear case, $f = \Phi \theta$ lies by
construction in $\rm{im}\,\Phi \equiv \rm{im}\, \Theta$ for \emph{any} choice of
the parameters $\theta$. Hence, the component $f^{\parallel} = 0$
identically: the parametrisation forces the solution to be orthogonal to the
kernel of the NTK. Projecting Eq.~\eqref{eq:LinearSystemNTK} onto $z^{(k)}$ for
$\lambda^{(k)} \neq 0$, and denoting by $d_\perp$ the dimension of the subspace
orthogonal to $\text{ker}\ \Theta$, yields
\begin{equation}
  \label{eq:LinearSystemProjected}
  \sum_{k'}^{d_\perp} H_{kk'}^{\perp} f_{k'}^{\perp} =  \sqrt{\lambda^{(k)}} \left(
    z^{(k)},
    \FKtab^T C_{Y}^{-1} Y
  \right),
\end{equation}
where we have introduced the $d_\perp\times d_\perp$ symmetric matrix
\begin{equation}
  \label{eq:HPerpDef}
  H^\perp_{kk'} = \sqrt{\lambda^{(k)}} \left(z^{(k)}, M z^{(k')}\right) \sqrt{\lambda^{(k')}}\,.
\end{equation}
We refer to $H^\perp$ as the \emph{flow} (or \emph{training}) \emph{Hamiltonian}
for reasons that become clear in the next section: training takes place in the
subspace orthogonal to $\ker\Theta$. Its dynamics is set by an interplay of
the architecture (encoded in the NTK) and the data (encoded in $M$), thus
dictating the dynamics of the training process. The crucial observation is that
$H^\perp$ is invertible whenever $\ker M \cap \text{Im}\,\Theta = \{0\}$.
Outside pathological choices of basis $\{\phi_i\}$ -- where the parametrisation
is characterised by non-zero directions $z^{(k)}$ such that $M z^{(k)} = 0$ for
some $\lambda^{(k)} \neq 0$ -- this condition is satisfied. The projected system
is then solved unambiguously,
\begin{equation}
  \label{eq:LinearSystemProjectedInverted}
  f_k^{\perp} = \sum_{k'}^{d_\perp} \left(H^\perp\right)^{-1}_{kk'}
    \sqrt{\lambda^{(k')}}
    \left(z^{(k')},\, \FKtab^T C_{Y}^{-1} Y\right),
\end{equation}
and the full solution is reconstructed as
\begin{equation}
  f = \sum_{k}^{d_\perp} \sqrt{\lambda^{(k)}}\, f_k^{\perp}\, z^{(k)}\, .
\end{equation}
Comparing with Eq.~\eqref{eq:chi2_inv_sol_moore_penrose}, the role of the
parametrisation is now made explicit: the singular operator $M$ has been
replaced by its projection $H^\perp$ onto the image of the NTK, which is
non-singular under mild assumptions. The parametrisation thus regularises
the inverse problem by restricting the solution to the subspace of
functions it can represent \emph{and} where the data-induced metric is
non-degenerate.

\paragraph{From the linear analogue to the NN case.}
The linear model captures the essential mechanism by which a parametrisation
regularises the inverse problem, but it differs from the NN case in two
important respects. First, in a NN the NTK $\Theta_t$ depends on the training
time $t$, so the decomposition in Eqs.~\eqref{eq:ParallelComponents} and
\eqref{eq:OrthogonalComponents} is itself time-dependent, and modes can transfer
between $\ker \Theta_t$ and $\text{Im}\, \Theta_t$ during the rich phase.
Second, when the parametrisation populates $\ker\Theta_t$, the parallel
component $f_t^{\parallel}$ no longer vanishes: in the lazy regime, where the
NTK becomes approximately constant, gradient descent has no gradient along
$\ker\Theta$, and $f_t^{\parallel}$ is frozen at the value $f_0^{\parallel}$
inherited from the onset of the lazy training. These unconstrained directions
are not pinned down by the loss but by the initial condition, and the randomness
propagates from the initialisation into the trained solution. We make this
picture quantitative in Sec.~\ref{sec:LazyTraining}, deriving a closed-form
expression for $f_t$ in the lazy regime using the same eigenspace decomposition
introduced here.

\section{Lazy Training in NNPDF}
\label{sec:LazyTraining}

In the previous section we observed that the NTK is able to capture the main
features of the training process, and that its time evolution is characterised
by a rapid initial transient, followed by a slower evolution during the rest of
the training. We also showed, by means of a simplified analogy, the role of the
parametrisation in regularising the ill-defined inverse problem. We now turn our
attention to this last stage of the training, where the NTK has stabilised and
becomes approximately constant. In doing so, we build upon the results presented
in Refs.~\cite{jacot2018neural,lee2019wide} and extend them to the case of
NNPDF. In the following, we derive the analytical solution of the flow equation,
which allows us to write an explicit expression for the trained field as a
function of the field at initialisation and the data.

\subsection{Analytical Results}
\label{sec:AnlyticalLazySolution}
The lazy training regime is characterised by a slow-evolving NTK. We denote as
$t_{\rm ref}$ the time at which the onset of this regime occurs. The NTK is then
\textit{frozen} to its value at $t_{\rm ref}$, and from this time onward the NTK
is taken to be constant
\begin{equation}
  \Theta_t = \Theta_{t_{\rm ref}} \equiv \Theta, \quad \textrm{for } t \geq t_{\rm ref}\,,
\end{equation} 
and we use the same convention as in Sec.~\ref{sec:Training} to distinguish
between the continuous time $t_{\rm ref}$ and the discrete epoch $T_{\rm ref}$.
The flow equation can then be written as
\begin{align}
  \ddt f_t = -\Theta M f_t + b\, ,
  \label{eq:FlowEqTwo}
\end{align}
where $M$ and $b$ are defined as in Eq.~\eqref{eq:MandBDef}. Note that now
neither $\Theta$ nor $b$ depend on the training time $t$. We solve this
first-order linear differential equation by projecting $f_t$ onto the basis
spanned by the eigenvectors of the NTK, defined in
Eq.~\eqref{eq:ThetaEigensystem}. Specifically, we distinguish the two components
$f^\parallel_{t,k}$ and $f^\perp_{t,k}$ introduced in
Eqs.~\eqref{eq:ParallelComponents} and~\eqref{eq:OrthogonalComponents}, now
supplemented with explicit time dependence.

\begin{figure}[t]
  \centering
  \includegraphics[width=0.3\textwidth]{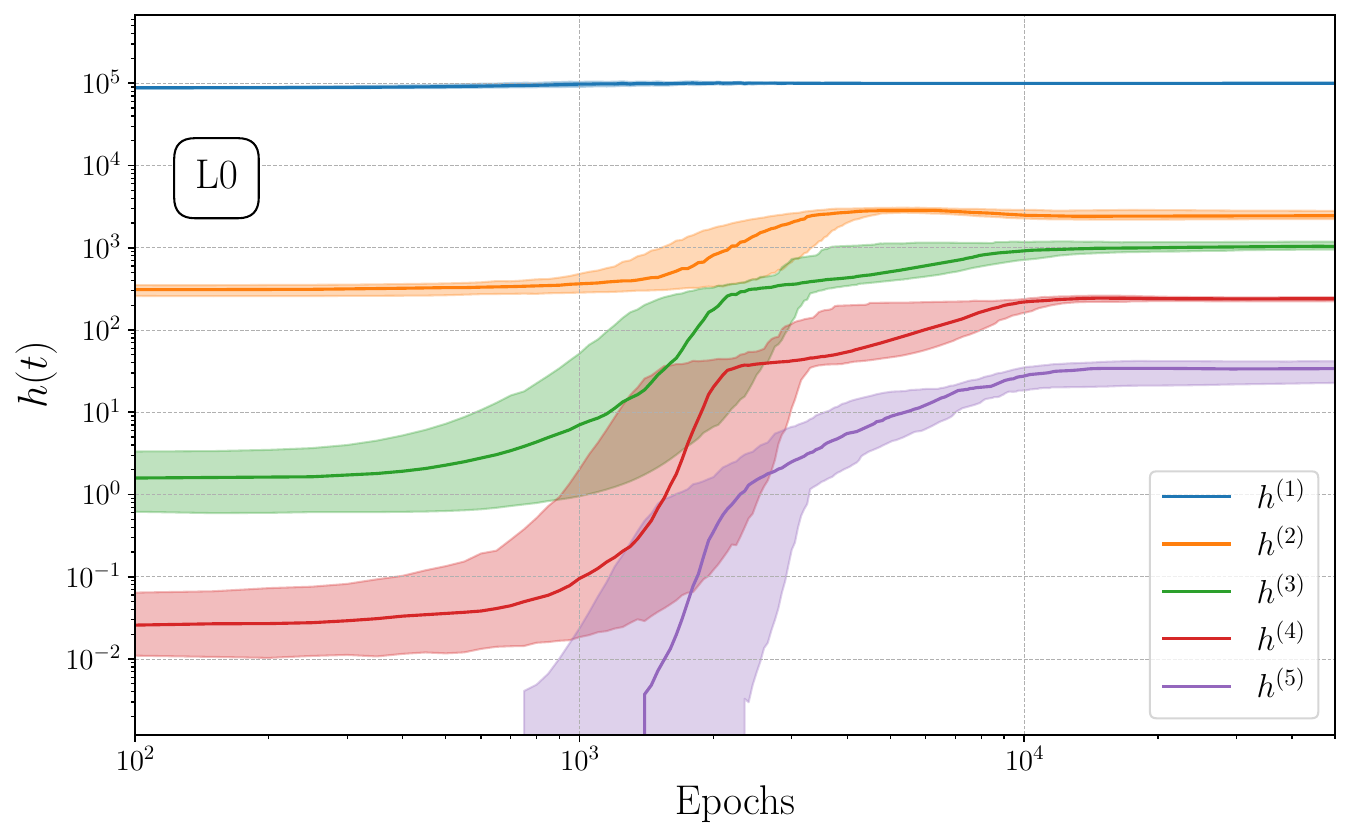}
  \includegraphics[width=0.3\textwidth]{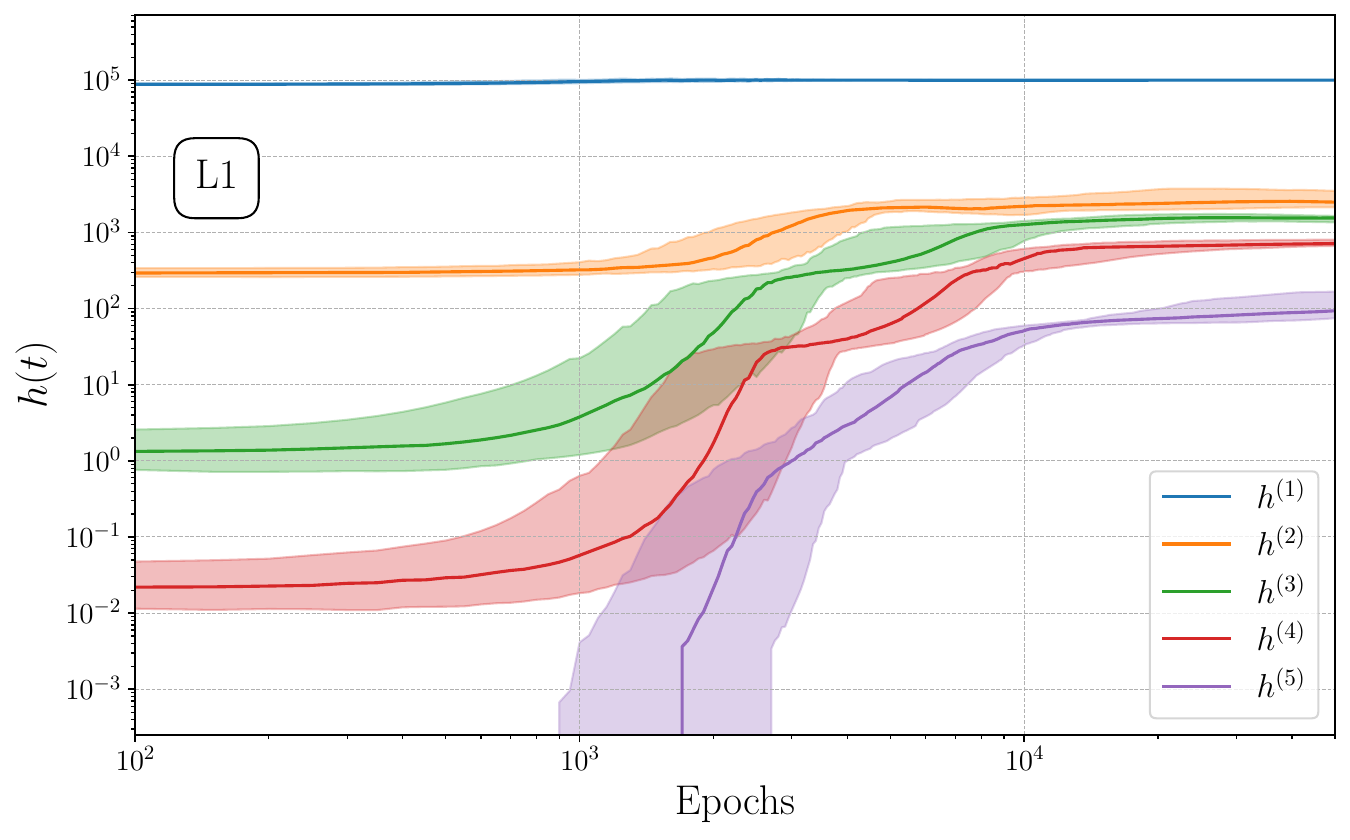}
  \includegraphics[width=0.3\textwidth]{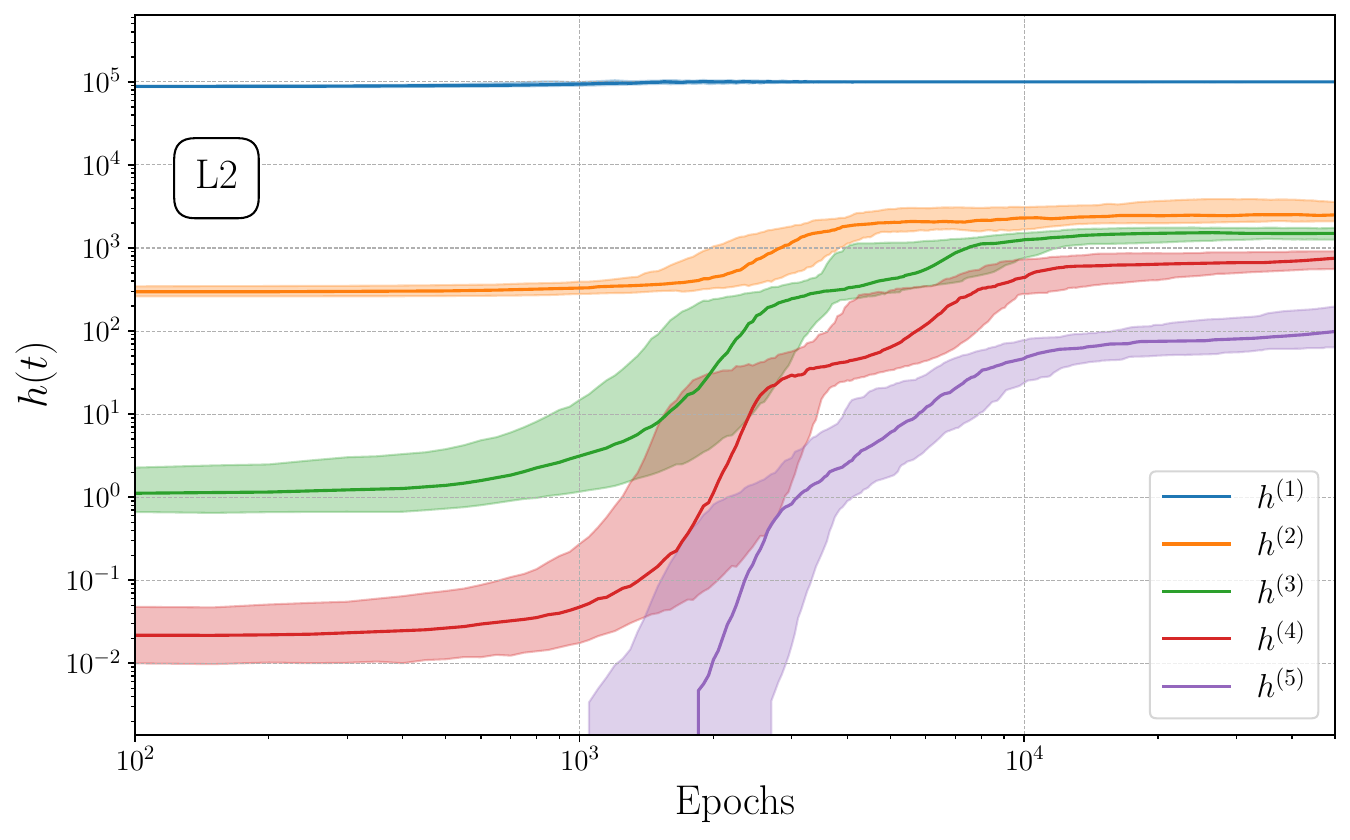}
  \caption{Evolution during training of the first five eigenvalues of
  $H^{\perp}$ using L0 (left), L1 (center), and L2 (right) data. Solid lines
  represent the median over the ensemble of networks, while solid bands
  correspond to 68\% confidence level. Note that the subdominant eigenvalues
  $\lambda^{(3)}$, $\lambda^{(4)}$ and $\lambda^{(5)}$ have increased by one or
  two orders of magnitude by the end of the rich training phase.}
  \label{fig:HPerpEigvalsTime}
\end{figure}
\begin{figure}[t!]
  \centering
  \includegraphics[width=0.90\textwidth]{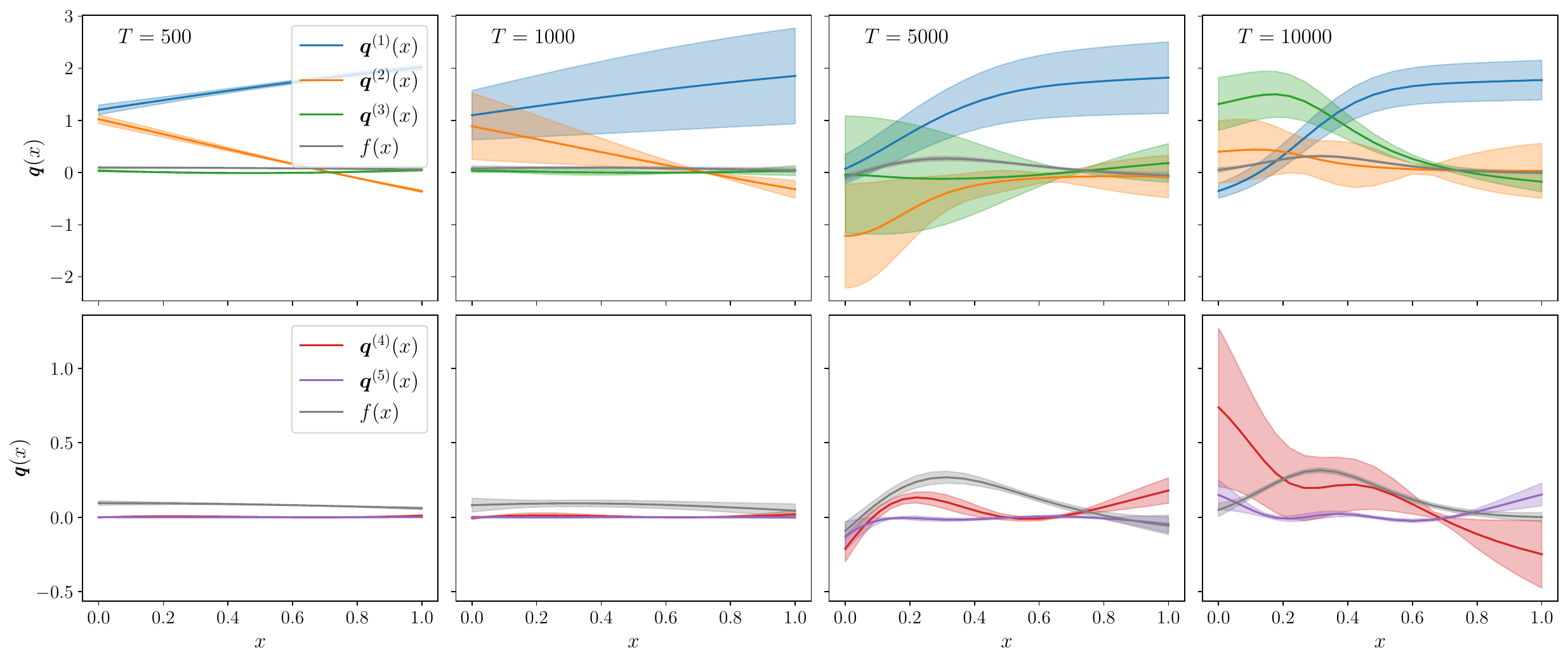}
  \caption{First five eigenvectors of $H^{\perp}$ projected onto PDF space,
  denoted as $q^{(i)}$, for different training times and as a function of the
  input $x$-grid. We also show the output of the network at the same training
  time, which is displayed in grey. L1 data is used.}
  \label{fig:HEigVecs}
\end{figure}
One can readily see that the components in the kernel of $\Theta$, $\text{ker}\
\Theta$, do not evolve during training,\footnote{Despite this result having been
obtained using the frozen NTK, it is worth mentioning that at any time during
training the kernel of the NTK is always defined and in general non-empty.
Hence, even in the initial stages of the training process, there is a component 
that is completely determined by the initial condition, \ie, by the prior distribution in
functional space.}
\begin{align}
    \label{eq:FlowParallel}
    \ddt f^\parallel_{t,k} = 0
        \quad \Longrightarrow \quad f^\parallel_{t,k} = f^\parallel_{0,k}\, .
\end{align}
This means that the final solution is affected by an irreducible noise that is
purely dictated by the initial condition. The flow equation for the orthogonal
components can be written as
\begin{align}
    \label{eq:FlowPerp}
    \ddt f^\perp_{t,k} = - H^\perp_{kk'} f^\perp_{t,k'}
        + B^\perp_{k}\, ,
\end{align}
where $H^{\perp}$ is the \emph{training Hamiltonian} defined in
Eq.~\eqref{eq:HPerpDef} and
\begin{equation}
    B^\perp_k = -\sqrt{\lambda^{(k)}} \left[\sum_{k'}^{d_{\parallel}}\left(z^{(k)}, M z^{(k')}\right) f^\parallel_{0,k'}
        - \left(z^{(k)}, \FKtabT C_Y^{-1} Y\right)\right],
\end{equation}
where $d_\parallel$ denotes the dimension of the subspace $\text{ker}\ \Theta$.
The indices on quantities that have a $\perp$ suffix only span the space
orthogonal to the kernel of $\Theta$, while the indices on quantities that have
a $\parallel$ suffix span the kernel. As discussed in
Sec.~\ref{sec:NTKRegularisation}, $H^{\perp}$ is positive definite under the
mild assumption $\ker M \cap \text{im}\,\Theta = \{0\}$. Therefore, its
eigenvalues and eigenvectors,
\begin{equation}
    H^\perp_{kk'} w^{(i)}_{k'} = h^{(i)} w^{(i)}_{k}\,,
\end{equation}
can be used to solve the flow equation similar to the linear-in-parameters
analogue of Sec.~\ref{sec:NTKRegularisation}. In Fig.~\ref{fig:HPerpEigvalsTime}
we show the evolution during training of the first five eigenvalues of
$H^{\perp}$ for the three different closure datasets. In Fig.~\ref{fig:HEigVecs}
we show, for different training times, the first five eigenvectors of
$H^{\perp}$ projected onto the PDF space according to
\begin{equation}
  q_{\alpha}^{(i)} = \sum_{k=1}^{d_\perp} \sqrt{\lambda^{(k)}} z^{(k)}_\alpha w^{(i)}_k,
\end{equation}
which follows from the notation introduced in
Eq.~\eqref{eq:OrthogonalComponents}. It should not come as too much of a
surprise that the eigenvalues and eigenvectors of $H^{\perp}$ have a similar
behaviour to those of the NTK (see Figs.~\ref{fig:NTKEigvalsTime} and
\ref{fig:NTKEigVecs}), from which they are constructed. However, we see that the
eigenvalues $h^{(i)}$ are are roughly three orders of magnitude larger than
those of the NTK.

\begin{figure}[ht!]
  \centering
  \includegraphics[width=0.9\textwidth]{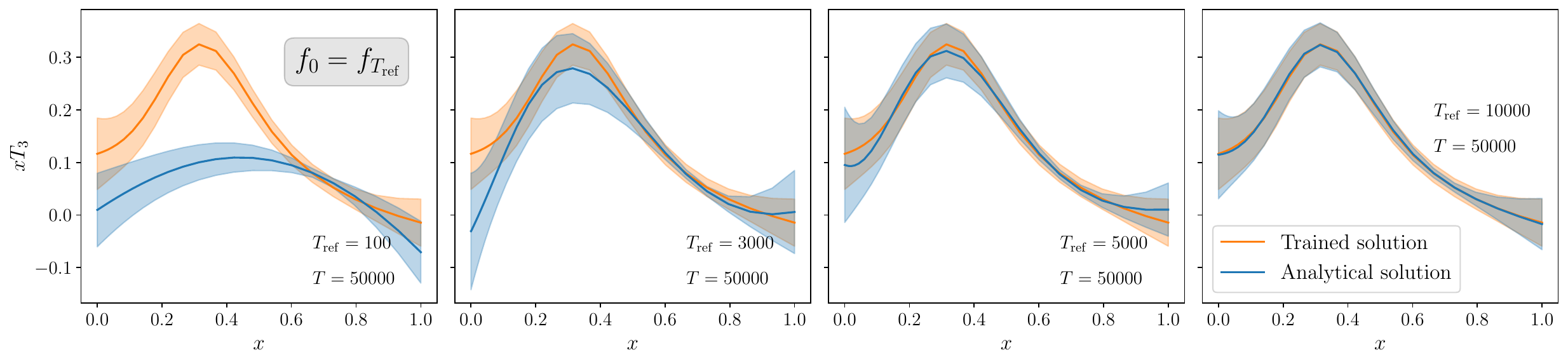} 
  \caption{Comparison of the trained and analytical evolution at the end of
  training. Each panel corresponds to a different frozen NTK, whereby the
  analytical solution is computed starting from $f_{T_{\rm ref}}$. The orange
  curve represents the final trained function after 50000 iterations of GD, and
  is the same across panels. Error bands represent one-sigma uncertainties
  across replicas. L2 data is used.}
  \label{fig:OnsetLazyL2}
\end{figure}

The solution to the flow equation, whose derivation is detailed in
Appendix~\ref{app:derivation}, can be written as
\begin{align}
    \label{eq:AnalyticSol}
    f_{t,\alpha}
        = U(t)_{\alpha\alpha'} f_{0,\alpha'} + V(t)_{\alpha I} Y_{I}\,,
\end{align}
where the evolution operators $U(t)$ and $V(t)$ have lengthy, yet explicit,
expressions that we also report in Appendix~\ref{app:derivation}.
Eq.~\eqref{eq:AnalyticSol} is the main result of this section. It shows that the
training process can be described as the sum of a linear transformation of the
initial fields $f_{0,\alpha}$, and a linear transformation of the data $Y_I$.
The two transformations depend on the flow time $t$ and are given by the
evolution operators $U(t)$ and $V(t)$. Fig.~\ref{fig:OnsetLazyL2} compares the
analytical solution with the trained function at the end of training, for
different choices of the frozen NTK. The NN is trained using the numerical GD
until $t_{\rm ref}$, at which point the NTK is frozen. The evolution time $t$
used in the analytical solution is the difference between the total training
time and $t_{\rm ref}$; the initial condition for the analytical solution is the
trained solution at $t_{\rm ref}$. Central value and uncertainty bands are
obtained by computing the analytical solution for each replica of the initial
condition and frozen NTK.\footnote{Unless stated otherwise, in this section
central values and uncertainties are always computed as ensemble averages across
replicas.} As expected, the closer $t_{\rm ref}$ is to the onset of the lazy
regime, the better the agreement between the analytical solution and the trained
function.

\FloatBarrier

\begin{figure}[ht]
    \centering
    \includegraphics[width=0.9\textwidth]{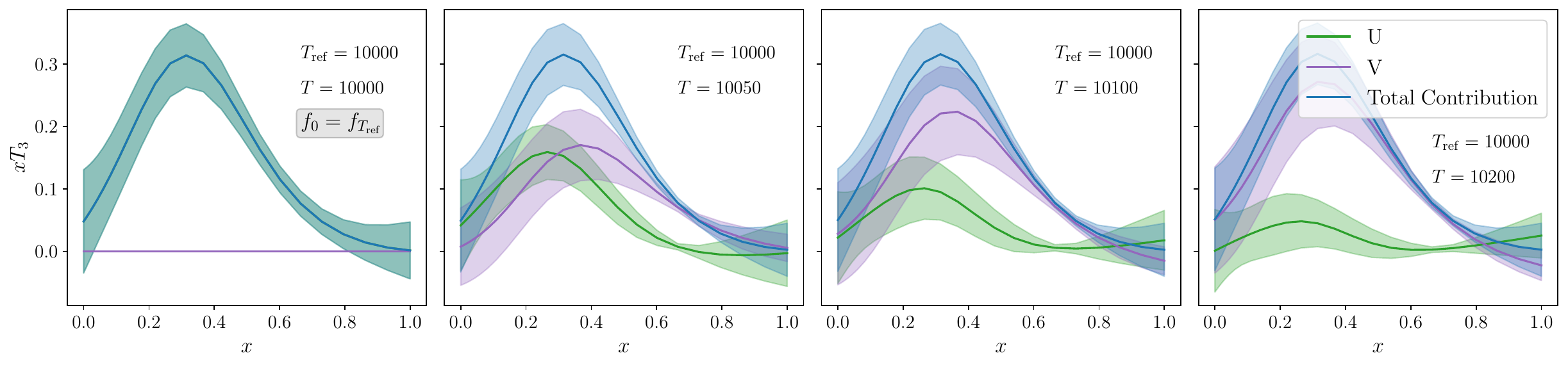} 
    \caption{Decomposition of the analytical solution into the two contributions
    from $U$ and $V$ at different training times. The frozen NTK is fixed across
    panels, and corresponds to the one at $T_{\rm ref} = 10000$. The initial
    condition for the analytical solution is always $f_{T_{\rm ref}}$. 
    As in Fig.~\ref{fig:OnsetLazyL2}, L2 data is used.}
    \label{fig:FrefDecompositionL2}
  \end{figure}

A complementary perspective is provided in Fig.~\ref{fig:FrefDecompositionL2},
where the analytical solution is decomposed into the two contributions from $U$
and $V$. In each panel, the initial condition $f_{t_{\rm ref}}$ is evolved
analytically for different training times by keeping the frozen NTK fixed. We
see that as training proceeds, the contribution from $U$ is progressively
suppressed, in accordance with the observation made above. On the other hand,
the contribution from $V$ grows and becomes dominant at later epochs, indicating
that the trained function is mostly determined by the data, rather than the
initial condition of the network. We also observe that such behaviour happens
quite rapidly -- in a training time interval $\Delta T \approx 200$ -- since the
time scales in the analytical solution are determined by the inverse of the
eigenvalues of $H^\perp$, which are typically large.

The analytical solution in Eq.~\eqref{eq:AnalyticSol} sheds new light on the
numerical training of a neural network. Given these results, it is natural to
ask whether the information encoded in the NTK alone can drive training,
independently of the initial condition, \ie, whether the analytical solution can
be used to perform kernel learning. We address this question in the following
section. \FloatBarrier

\subsection{Numerical Results}
\label{sec:NTKKernelLearning}

The results shown in Sec.~\ref{sec:AnlyticalLazySolution} and Sec.~\ref{sec:NTKDuringTraining}
support the idea that the NTK is capable to encode in its eigenvectors the
physical features learned during training. We now probe this idea further by
employing the analytical solution in Eq.~\eqref{eq:AnalyticSol} \textit{\`a la}
kernel learning, \ie, by applying it to an initial condition drawn from the prior distribution. 
In the following, we choose the initial condition to be an ensemble of
networks at initialisation as in Fig.~\ref{fig:prior}, whose architecture is the
same as the one used in the training. This represents our prior assumption on
the space of functions, which is then updated using the data and the NTK frozen
at $t_{\rm ref}$.

\subsubsection{Central Value and Covariance of the Trained Fields}
\label{sec:CentralAndCovariance}

The analytical solution in Eq.~\eqref{eq:AnalyticSol} is inherently stochastic,
since the frozen NTK at $t_{\rm ref}$ is actually obtained from an ensemble of
networks. As a consequence, the operators $U(t)$ and $V(t)$ are both random
variables. We can then characterize the distribution of the analytical solution
using the mean and variance across the ensemble, as shown in
Eq.~\eqref{eq:ReplicaEnsemble}. The central value of the analytical solution is thus
defined as
\begin{align}
    \label{eq:MeanValAtT}
    \bar{f}_{t,\alpha} = \mathbb{E}\left[f_{t,\alpha}\right]
        = \mathbb{E}\left[U(t)_{\alpha\alpha'} f_{0,\alpha'}\right]
            + \mathbb{E}\left[V(t)_{\alpha I} Y_I\right] \, .
\end{align}
More interestingly, we can also compute their covariance matrix at any time $t$,
\begin{align}
    \cov[f_t,f_t^T]
        &= \mathbb{E}\left[U(t) f_0 f_0^T U(t)^T\right] 
            - \mathbb{E}\left[U(t) f_0\right] \mathbb{E}\left[f_0^T U(t)^T\right]  \nonumber \\
        &\quad + \mathbb{E}\left[U(t) f_0 Y^T V(t)^T\right] 
            - \mathbb{E}\left[U(t) f_0\right] \mathbb{E}\left[Y^T V(t)^T\right] \nonumber \\
        &\quad + \mathbb{E}\left[V(t) Y f_0^T U(t)^T\right]
            - \mathbb{E}\left[V(t) Y\right] \mathbb{E}\left[f_0^T U(t)^T\right] \nonumber \\
    \label{eq:CovAtT}
        &\quad + \mathbb{E}\left[V(t) Y Y^T V(t)^T\right]
            - \mathbb{E}\left[V(t) Y\right] \mathbb{E}\left[Y^T V(t)^T\right] \, .
\end{align}
Note that the first and the fourth lines above yield symmetric matrices, while
the third line is just the transpose of the second, thereby ensuring that the
whole covariance matrix is the sum of three symmetric matrices and therefore is
symmetric, 
\begin{align}
    \label{eq:SumOfCovariances}
    \cov[f_t,f_t^T] = C_t^{(00)} + C_t^{(0Y)} + C_t^{(YY)}\, ,
\end{align}
where
\begin{align}
    \label{eq:C00term}
    C_t^{(00)} 
        &= \mathbb{E}\left[U(t) f_0 f_0^T U(t)^T\right] 
        - \mathbb{E}\left[U(t) f_0\right] \mathbb{E}\left[f_0^T U(t)^T\right]\, ,\\
    C_t^{(0Y)}
        &= \mathbb{E}\left[U(t) f_0 Y^T V(t)^T\right] 
        - \mathbb{E}\left[U(t) f_0\right] \mathbb{E}\left[Y^T V(t)^T\right] \nonumber \\
        \label{eq:C0Yterm}
        &\quad + \mathbb{E}\left[V(t) Y f_0^T U(t)^T\right]
            - \mathbb{E}\left[V(t) Y\right] \mathbb{E}\left[f_0^T U(t)^T\right] \, ,\\
    C_t^{(YY)}
        &= \mathbb{E}\left[V(t) Y Y^T V(t)^T\right]
        - \mathbb{E}\left[V(t) Y\right] \mathbb{E}\left[Y^T V(t)^T\right]\, .
        \label{eq:CYYterm}
\end{align}
Eq.~\eqref{eq:SumOfCovariances} shows explicitly the various contributions to 
the covariance matrix. Indeed, $C_t^{(00)}$ quantifies the
contribution to the covariance matrix that is purely due to the fluctuations of
the initial condition, while $C_t^{(YY)}$ quantifies the contribution that is
purely due to the statistical fluctuations of the data. The mixed term
$C_t^{(0Y)}$ accounts for the correlations between the two sources of
uncertainty.

\subsubsection{Convergence of the Analytical Solution}
\begin{figure}[t]
  \centering
  \includegraphics[width=0.9\textwidth]{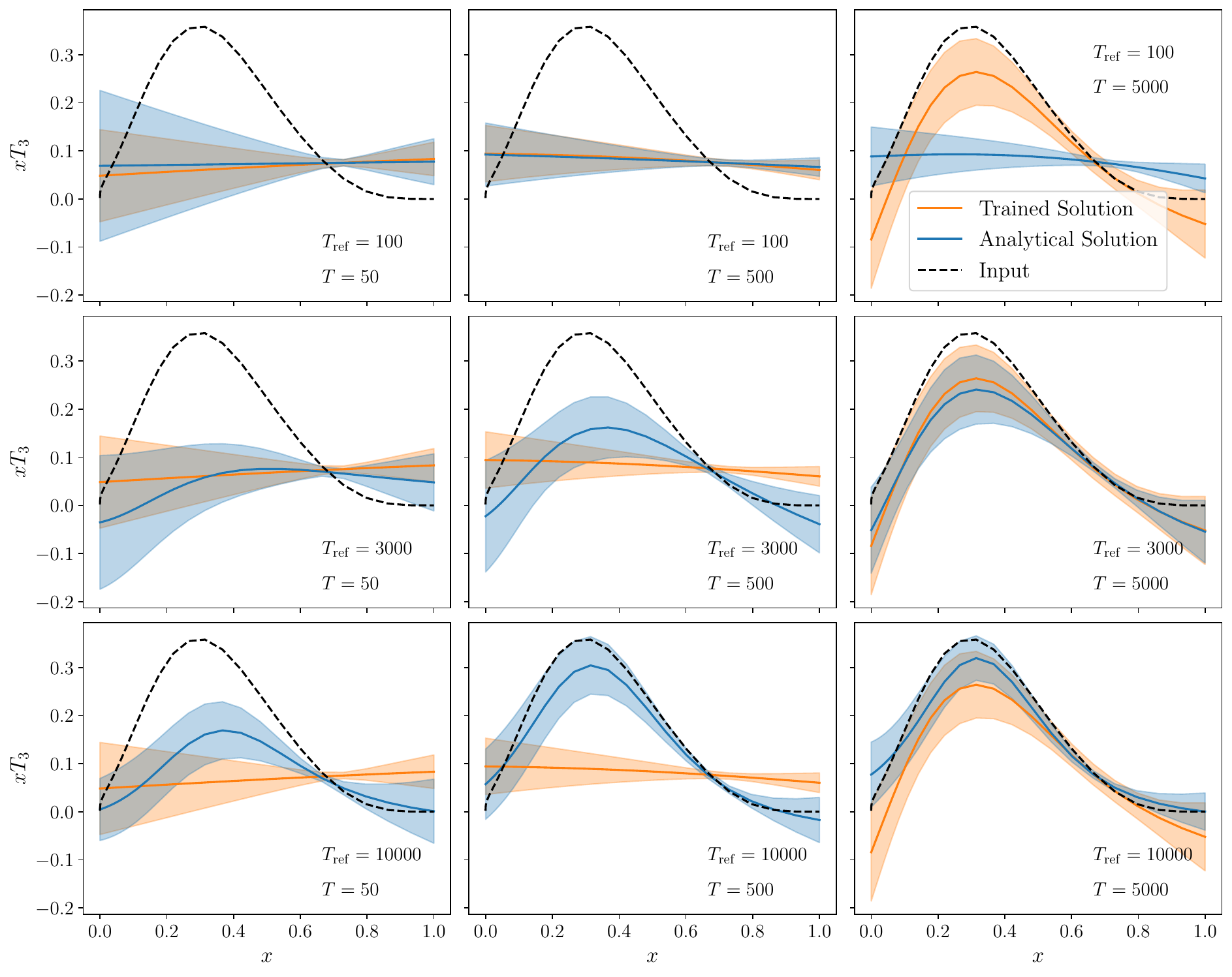} 
  \caption{Comparison of the trained (orange) and analytical (blue) evolution
  starting from an ensemble of networks at initialisation as the initial
  condition. Each row corresponds to a different frozen NTK, while the columns
  represent different training times. The dashed line represents the input 
  function used to generate the synthetic data, \ie, the {\em true}\ result. 
  L2 data is used.}
  \label{fig:EvolutionGridF0L2}
\end{figure}

We start by comparing the analytical solution (AS), obtained using an ensemble
of networks at initialisation as the initial condition, with the trained
solution (TS), obtained by training another ensemble of networks drawn
from the same prior distribution using GD. This comparison is shown in
Fig.~\ref{fig:EvolutionGridF0L2} for L2 data, where the rows in the grid
correspond to different frozen NTKs, while the columns represent numerical and
analytical evolution after $T=50, 500$ and 5000 epochs. These results deserve a
few comments.

The first observation is that the NTK at early stages of training is not able to
drive the prior towards the true function, as shown in the first and, though
less dramatically, in the second row of Fig.~\ref{fig:EvolutionGridF0L2}. This
is expected, as we extensively discussed in Sec.~\ref{sec:NTKDuringTraining},
since at early stages of training by GD the NTK has not yet aligned its 
internal representation with the data.

More significantly, we observe a significant discrepancy between the AS and the
TS even at $T=5000$. This can be explained as follows. At the beginning of
training, there is clearly no difference between AS and TS since both represent
the neural network output at initialisation (Sec.~\ref{sec:Init}), with
variations due only to different initialisation seeds. During early training
stages, AS and TS differ as expected. Indeed, the analytical solution is
computed using the frozen NTK at $t_{\rm ref}$, while the trained solution
evolves with an NTK that is still changing as shown in Sect.~\ref{sec:Training}.
Crucially, if the NTK at $t_{\rm ref}$ has already learned from the data and
aligned with the solution, the AS converges faster to the target, while the TS
requires additional epochs before evolving in the correct direction.

\subsubsection{Connection with Linear Methods}
\label{sec:ConnectWithLinear}
We can consider a simplifying limit of Eq.~\eqref{eq:MeanValAtT}, where the
initial condition $f_0$ and the data $Y$ are statistically independent from the
respective evolution operators $U(t)$ and $V(t)$. Note that the first term on
the right-hand side of Eq.~\eqref{eq:MeanValAtT} can only be non-zero because of
the correlations between $U(t)$ and $f_0$. In the absence of such correlations,
the first term would be given by the product of the expectation values and
therefore would vanish if $f_0$ is an ensemble of networks at initialisation.
Under these assumptions, we have
\begin{align}
    \label{eq:MeanUt}
    \bar{U}(t)
        &= \mathbb{E}\left[U(t)\right]\, , \\
    \label{eq:MeanVt}
    \bar{V}(t)
        &= \mathbb{E}\left[V(t)\right]\, ,
\end{align}
and
\begin{equation}
    \label{eq:MeanValAtTNoCorr}
    \bar{f}_{t,\alpha} = \bar{U}(t)_{\alpha\alpha'} \bar{f}_{0,\alpha'}
        + \bar{V}(t)_{\alpha I} Y_I = \bar{V}(t)_{\alpha I} Y_I \, .
\end{equation}
The second term in Eq.~\eqref{eq:MeanValAtT}, or equivalently
Eq.~\eqref{eq:MeanValAtTNoCorr}, explicitly shows the contribution of each data
point to the central value of the trained fields at each value of $x_{\alpha}$.
It is worthwhile remarking that in this limit, the central value from the set of
trained networks is a linear combination of the data points, with coefficients
given by the evolution operator $V(t)_{\alpha I}$.

\begin{figure}[ht!]
  \centering
  \includegraphics[width=0.95\textwidth]{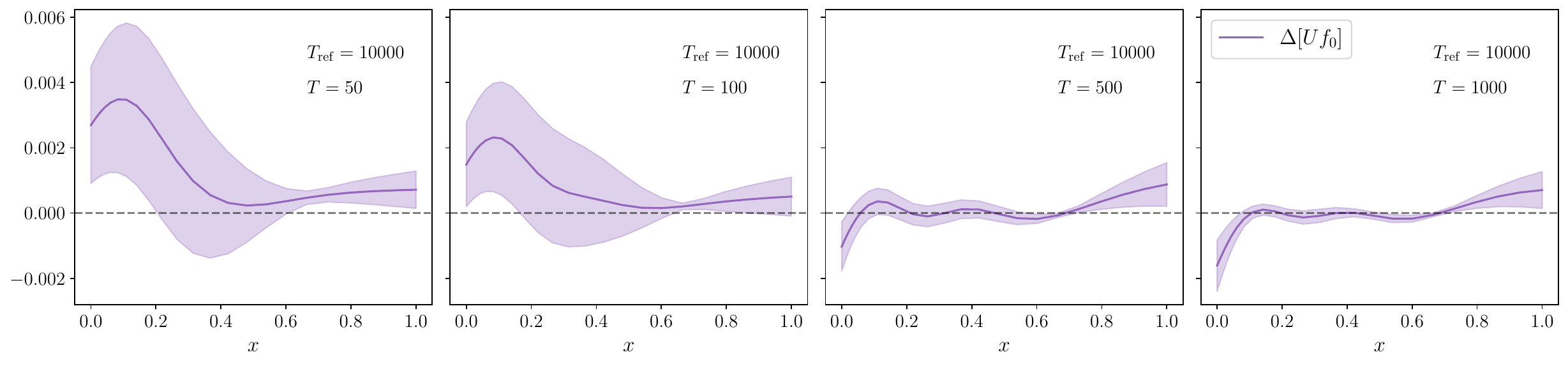}
  \includegraphics[width=0.95\textwidth]{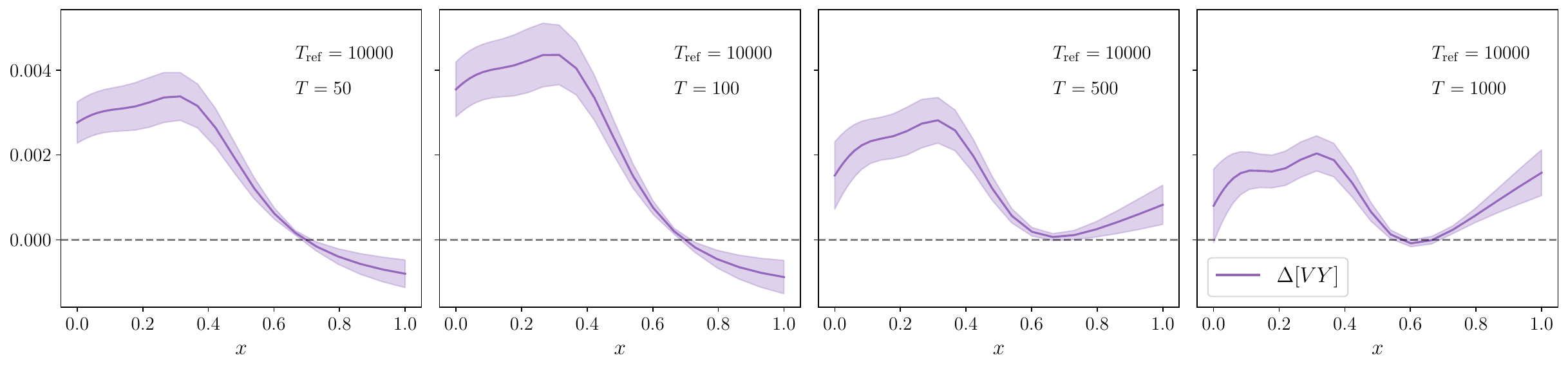}
  \caption{Behaviour of $\Delta [U(t)f_0]$ and $\Delta [V(t)Y]$, as defined in
  Eqs.~\eqref{eq:DeltaExpValUtF0} and~\eqref{eq:DeltaExpValVtY}, as functions of
  the training time. The operators $U(T)$ and $V(T)$ are constructed by taking
  the NTK at $T_{\rm ref} = 10000$, which is fixed across panels. The
  uncertainties are extracted from the bootstrap ensemble as discussed in the
  text.}
    \label{fig:xT3_exp_val}
\end{figure}

In the absence of general theorems, we verify this assumption empirically. From
the ensemble of replicas, we generate bootstrap samples and compute the
following two estimators,
\begin{align}
    \label{eq:DeltaExpValUtF0}
    \Delta[U(t)f_0] &= \mathbb{E}\left[U(t) f_{0}\right] 
      - \mathbb{E}\left[U(t) \right] \mathbb{E}\left[f_{0}\right]\, , \\
    \label{eq:DeltaExpValVtY}
    \Delta[V(t)Y] &= \mathbb{E}\left[V(t) Y\right] 
      - \mathbb{E}\left[V(t) \right] \mathbb{E}\left[Y\right]\, .
\end{align}
for different training times, using the same frozen NTK and L2 data. The results
are shown in Fig.~\ref{fig:xT3_exp_val} for the $U$ (upper panel) and $V$ (lower
panel) contributions. The error bands are computed using bootstrap error. By
inspecting the figures, we see two distinct patterns emerging. For the operator
$U$, $\Delta[U(t) f_0]$ is different from zero for small training times, and thus
the correlations between $U(t)$ and $f_0$ are non-negligible. However, as
training proceeds, $\Delta[U(t) f_0]$ becomes compatible with zero within the error
bars, suggesting that the correlations are progressively suppressed. The case of
the $V$ operator is even more striking, as $\Delta[V(t) Y]$ is clearly
non-negligible across all training times, although it also shows a decreasing
trend as training proceeds. This suggests that the correlations between $V(t)$
and $Y$ cannot be neglected. 

Let us conclude this brief discussion by noting that Eq.~\eqref{eq:MeanValAtTNoCorr} 
resembles the structure of a linear method, like Backus-Gilbert or 
Gaussian Processes. We
believe that this observation, as well as the results shown in
Fig.~\ref{fig:xT3_exp_val}, deserves further investigations, which we leave for
future work. Understanding the relation between different methods is important 
in order to assess the robustness of the solution. 

\subsubsection{Error decomposition}
\label{sec:ErrorDec}
\begin{figure}[ht!]
  \centering
  \includegraphics[width=0.80\textwidth]{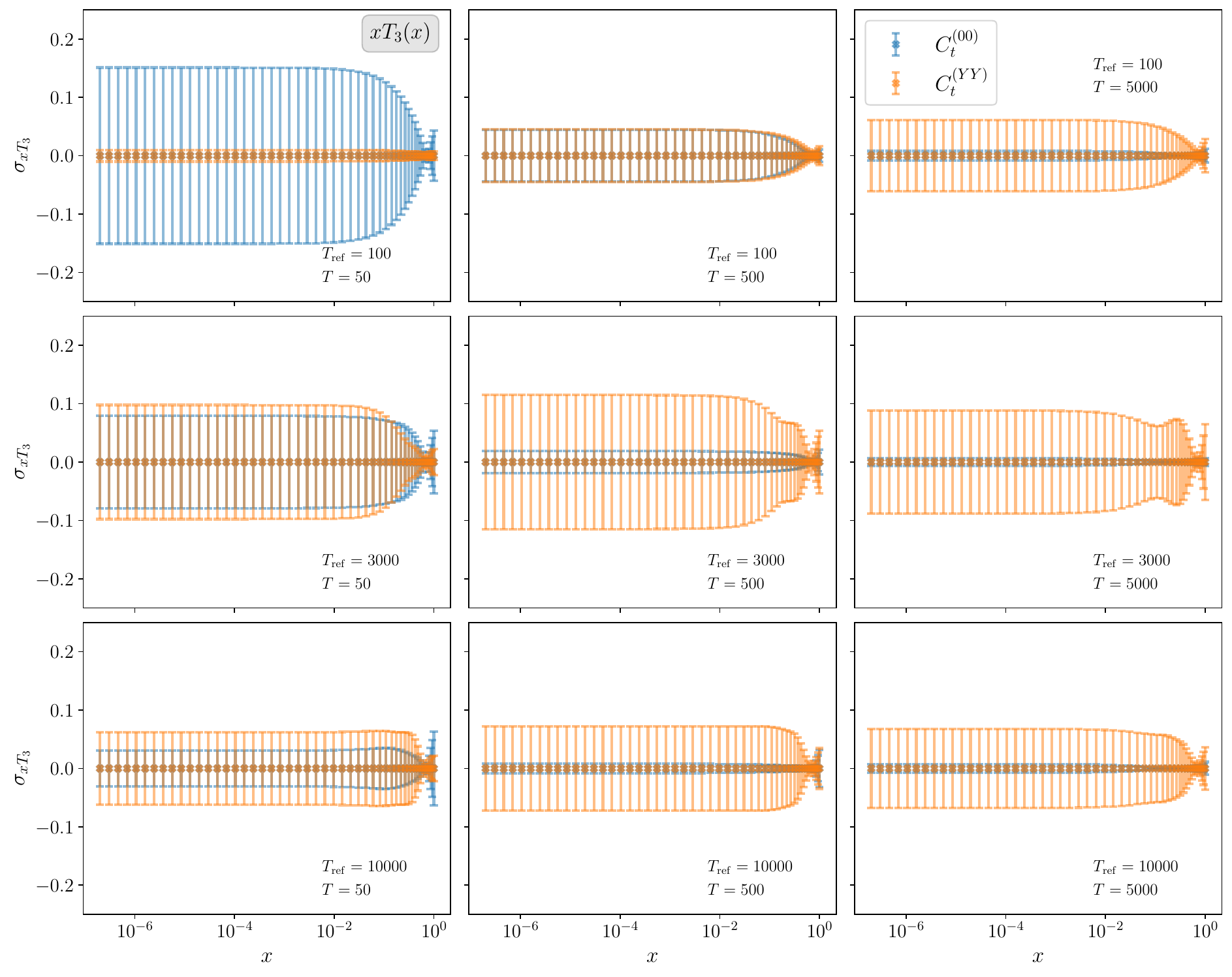}
  \caption{Decomposition of the error budget of the trained fields into the two
  components from the initial condition (blue) and from the data (orange), as
  defined in Eqs.~\eqref{eq:C00term} and~\eqref{eq:CYYterm}. Each row
  corresponds to a different frozen NTK, while the columns represent different
  training epochs. L2 data is used. We see that if the NTK is taken at later
  stages of training, the contribution from the initial condition is severely
  suppressed towards the end of training. }
  \label{fig:ErrorBudgetL2}
\end{figure}

The analytical expression for the covariance matrix,
Eq.~\eqref{eq:SumOfCovariances}, allows us to monitor the relative size of the
three contributions as training proceeds. For a properly trained ensemble of
networks, the covariance of the trained fields should be dominated by the
statistical error on the data. We show the diagonal entries of the two
contributions $C_t^{(00)}$ (blue band) and $C_t^{(YY)}$ (orange band) to the
error budget in Fig.~\ref{fig:ErrorBudgetL2}, for different frozen NTKs (rows)
and different training epochs (columns), using L2 data as before. We do not show
the mixed term $C_t^{(0Y)}$ since it is negligible with respect to the other two
contributions -- the two sources of uncertainty are largely uncorrelated. In
general, we observe that towards the end of the training process the
contribution from the data $C_t^{(YY)}$ becomes dominant with respect to the
contribution coming from the initial condition $C_t^{(00)}$. We also see that
the suppression of the initial condition is more severe and happens earlier when
the frozen NTK is taken at later stages of training.

\begin{figure}[ht!]
  \centering
  \includegraphics[width=0.80\textwidth]{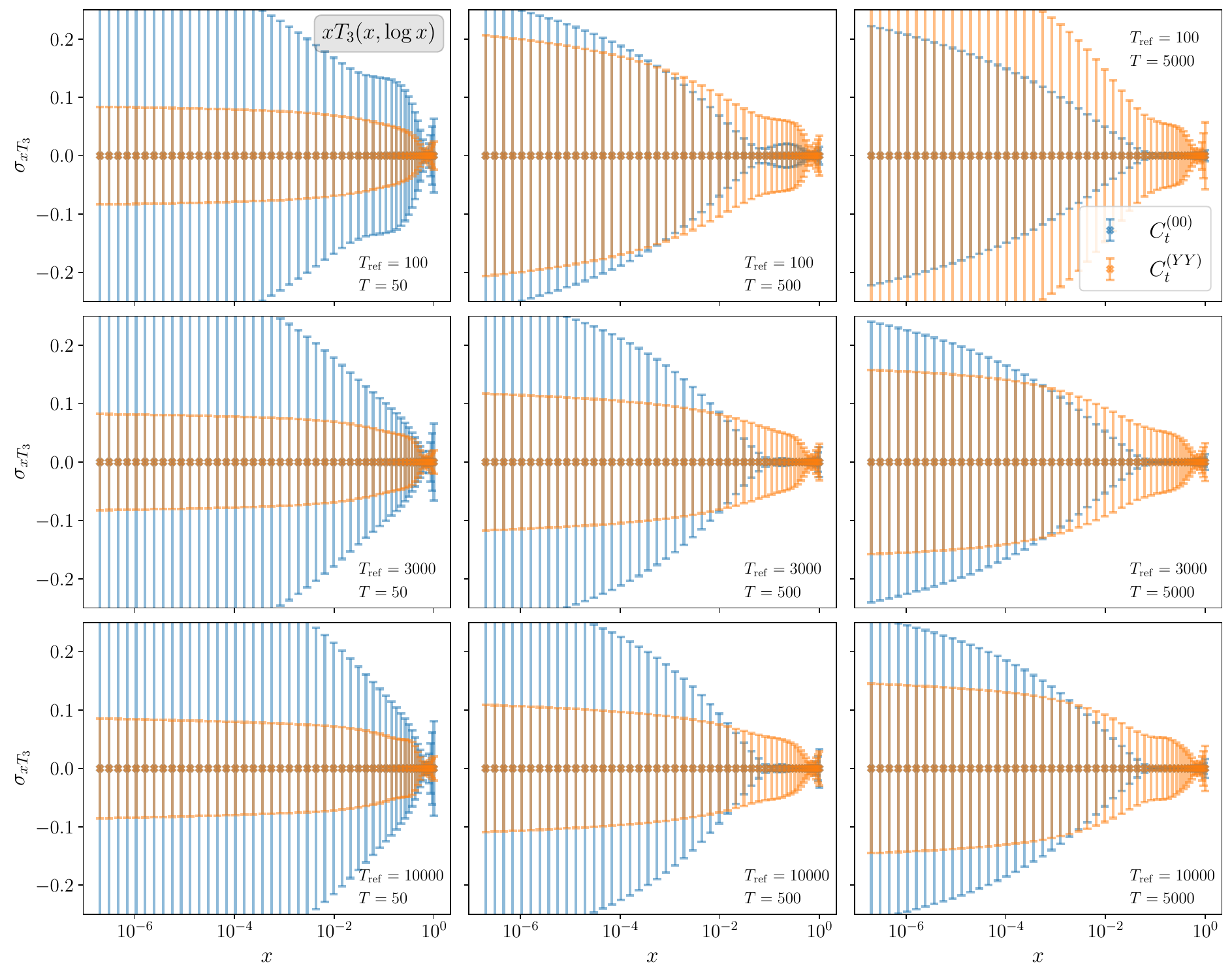}
  \caption{Similar to Fig.~\ref{fig:ErrorBudgetL2}, but now for the case of scaled
  input $f(x, \log x)$.}
  \label{fig:ErrorBudgetL2Logx}
\end{figure}

In order to study the dependence of the error decomposition on the initial
condition, we repeat the same analysis for the case of scaled input $f(x, \log
x)$, as described in Sec.~\ref{sec:Init}. Hence, the frozen NTK is taken from an
ensemble of trained networks with input layer as in Eq.~\eqref{eq:InitLayerPhi}.
The initial condition $f_0$ of the analytical solution is drawn from the same
prior distribution as the trained solution, and corresponds to the orange curve
in Fig.~\ref{fig:prior}. The resulting error decomposition is shown in
Fig.~\ref{fig:ErrorBudgetL2Logx} for L2 data, where panels are ordered as in
Fig.~\ref{fig:ErrorBudgetL2}. Inspecting the figures, we observe that now the
contribution of the initial condition becomes dominant in the small-$x$ region,
even for large training times and irrespective of the epoch at which the NTK is
frozen. This result reflects the behaviour of the prior distribution at
small-$x$, where indeed error bands are significantly enlarged with respect to
the case of linear input. Interestingly, even the contribution from the data,
$C^{(YY)}_t$, increases at small-$x$ towards larger training times. This can be
explained by observing that, despite not being explicitly dependent on the
initial condition $f_0$, the evolution operator $V(t)$ is constructed from a
frozen NTK that has encoded the dependence on the architecture through the
training process (see Eq.~\eqref{eq:VDef}). That the difference between
Figs.~\ref{fig:ErrorBudgetL2} and \ref{fig:ErrorBudgetL2Logx} is primarily
localised at small-$x$ showcases that, for the region left uncovered by the
data, the methodology is not able to suppress the dependence on the initial
condition. In fact, where the information from the data is available
(corresponding roughly to $x \gtrsim 0.01$), the dependence on the initial
condition lessens as the analytical solution evolves (left to right). This
reduction occurs more rapidly for frozen NTKs taken at later stages of training
(top to bottom), showing that there is a non-trivial interplay between the
information provided by the data and that acquired by the NTK.

These studies reveal the intricate connection between the prior distribution and
the uncertainties of PDFs. In the region constrained by data, the error is
dominated by the statistical error on the data, rather than by the fluctuations
of the initial fields. This is an important step in our study of the error
estimates. It guarantees that the error bars computed from the ensemble of
trained PDFs are not biased by the choice of prior, which depends on the selected
architecture, activation function, and probability distributions for the biases
and weights at initialisation.


\FloatBarrier

\subsection{Bias-Variance Decomposition}
\label{sec:BiasVarNumerical}
Finally, we can use the analytical solution to compute the bias and variance of
the trained ensemble of networks. Following the definition of
Ref.~\cite{NNPDF:2021njg}, we define the bias as
\begin{align}
  \mathcal{B}_t 
  &= \frac{1}{N_{\rm dat}} \left( \bar{T}[f_t] - Y_{\rm L0} \right)^T C_Y^{-1} \left( \bar{T}[f_t] - Y_{\rm L0} \right) \\
  &= \frac{1}{N_{\rm dat}} \left( \bar{f_t} - \fin \right)^T M \left( \bar{f_t} - \fin \right)
  \label{eq:BiasDef}
\end{align}
where $\bar{T}[f_t] = \FKtab \bar{f}_t$ is the theoretical prediction computed
using the central value of the trained fields at time $t$,
Eq.~\eqref{eq:MeanValAtT}, and $Y_{\rm L0}$ are the data points generated using
the input function $\fin$, $Y_{\rm L0}= \FKtab \fin$ (see
Appendix~\ref{app:dataset}). In going from the first to the second line, we use
the definition of the matrix $M$ given in Eq.~\eqref{eq:MandBDef}, which includes
the physical information encoded in the FK tables. The bias thus assumes a
clear interpretation -- it measures the ability of the training process in
inverting the forward map used to generate the data. In fact, we recall that the
inversion is not exact even with level-0 data; the minimum of the loss is achieved for
\begin{equation}
  \fin_{M^{\perp}} = M^{+} (FK)^T C_Y^{-1} Y_{L0} = M^{+} M \fin \neq \fin \,,
  \label{eq:NaiveInversion}
\end{equation}
where $M^{+}$ is the pseudo-inverse of $M$. As in the case of the NTK, there is
a residual component of the input vector in the null eigenspace of $M$,
$\fin_{M^{\parallel}}$, that is not constrained by the forward map, similarly to
the residual noise introduced by the NTK when training the network,
Sec.~\ref{sec:AnlyticalLazySolution}. Given that this component is not constrained by
the data, the parametrization, together with the fitting methodology, has
complete freedom to fix it. In our analytical solution,
Eq.~\eqref{eq:AnalyticSol}, this freedom is brought in by the linear
transformation of the fields $f_0$. On the other hand, $\fin_{M^{\perp}}$
receives its contribution from the data and from the theoretical predictions,
resembling the linear transformation of the input data in
Eq.~\eqref{eq:AnalyticSol}. This is even more evident if one compares
Eq.~\eqref{eq:NaiveInversion} with Eq.~\eqref{eq:VDef}
\begin{equation}
  \mathbb{E} \left[ V(t) Y \right] - \fin_{M^{\perp}}
    = \mathbb{E} \left[ \mathcal{M}(t) \FKtabT C_Y^{-1} Y \right] 
    - M^{+} \FKtabT C_Y^{-1} Y_{L0} \,,
\end{equation}
where we recall that $Y$ is different from $Y_{L0}$, in general. From this
expression, it follows that $\mathcal{M}(t)$ can be seen as a regularised
variant of $M^{+}$, with the training time $t$ being the regularisation
parameter.

These arguments motivate us to reshape Eq.~\eqref{eq:BiasDef} as follows
\begin{equation}
  \mathcal{B}_t = \mathcal{B}_{t,U} + \mathcal{B}_{t,V} + \mathcal{B}_{t,UV} \,,
  \label{eq:BiasDecomposition}
\end{equation}
where
\begin{align}
  \mathcal{B}_{t,U} &= \frac{1}{N_{\rm dat}} 
  \bigl( \mathbb{E}\left[U(t)f_0\right] - \fin_{M^{\parallel}} \bigr)^T
  M 
  \bigl( \mathbb{E}\left[U(t)f_0\right] - \fin_{M^{\parallel}} \bigr) \,, \\
  \mathcal{B}_{t,V} &= \frac{1}{N_{\rm dat}} 
  \bigl( \mathbb{E}\left[V(t)Y\right] - \fin_{M^{\perp}} \bigr)^T
  M
  \bigl( \mathbb{E}\left[V(t)Y\right] - \fin_{M^{\perp}} \bigr) \,, \\
  \mathcal{B}_{t,UV} &= \frac{2}{N_{\rm dat}} 
  \bigl( \mathbb{E}\left[U(t)f_0\right] - \fin_{M^{\parallel}} \bigr)^T
  M
  \bigl( \mathbb{E}\left[V(t)Y\right] - \fin_{M^{\perp}} \bigr) \,.
\end{align}

Continuing in this vein and following Ref.~\cite{NNPDF:2021njg}, we define the
variance as
\begin{align}
  \mathcal{V}_t 
  &= \frac{1}{N_{\rm dat}} \mathbb{E} \biggl[\bigl( \bar{T}[f_t] - T[f_t] \bigr)^T C_Y^{-1}\bigl( \bar{T}[f_t] - T[f_t] \bigr)\biggr] \\
  &= \frac{1}{N_{\rm dat}} \mathbb{E} \biggl[\bigl( \bar{f_t} - f_t \bigr)^T M \bigl( \bar{f_t} - f_t \bigr)\biggr] \,,
\end{align}
where in the second line we use the same reasoning as in Eq.~\eqref{eq:BiasDef}.
This expression therefore serves as a measure of the spread of the trained
fields in the space of data. Similarly to the bias, we can decompose the
variance into three components
\begin{equation}
  \mathcal{V}_t = \mathcal{V}_{t,U} + \mathcal{V}_{t,V} + \mathcal{V}_{t,UV} \,,
  \label{eq:VarianceDecomposition}
\end{equation}
where now the contributions are
\begin{align}
  \mathcal{V}_{t,U}
    &= \frac{1}{\ndat} \mathbb{E} \biggl[
      \bigl( \mathbb{E} [U(t)f_0] - U(t)f_0 \bigr)^T M \bigl( \mathbb{E} [U(t)f_0] - U(t)f_0 \bigr)
    \biggr]\,, \\
  \mathcal{V}_{t,V}
    &= \frac{1}{\ndat} \mathbb{E} \biggl[
      \bigl( \mathbb{E} [V(t)Y] - V(t)Y \bigr)^T M \bigl( \mathbb{E} [V(t)Y] - V(t)Y \bigr)
  \biggr]\,, \\
  \mathcal{V}_{t,UV}
    &= \frac{2}{\ndat} \mathbb{E} \biggl[
      \bigl( \mathbb{E} [U(t)f_0] - U(t)f_0 \bigr)^T M \bigl( \mathbb{E} [V(t)Y] - V(t)Y \bigr)
  \biggr]\,.
\end{align}

\begin{figure}[!h]
  \centering
  \includegraphics[width=0.90\textwidth]{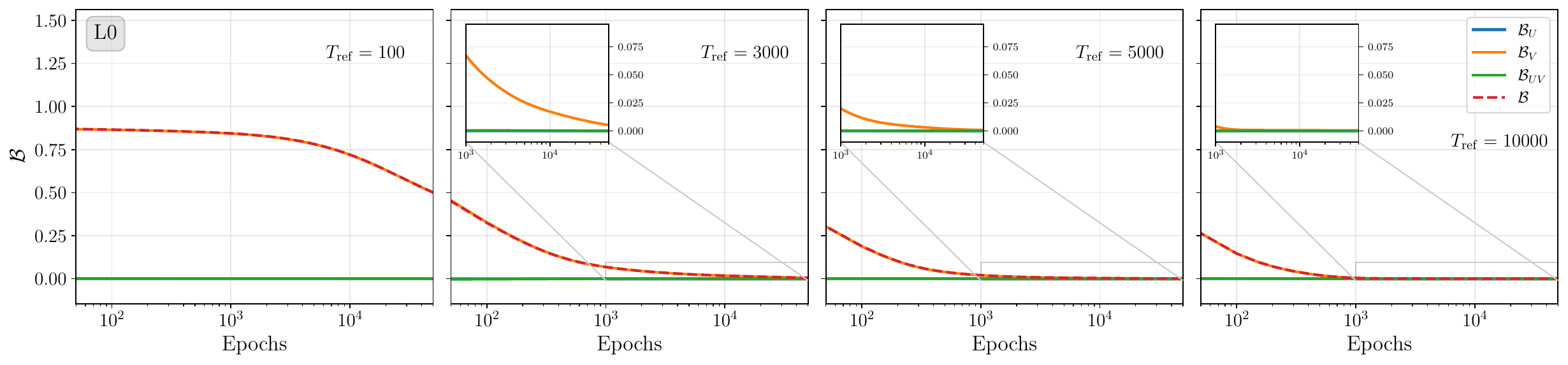}
  \includegraphics[width=0.90\textwidth]{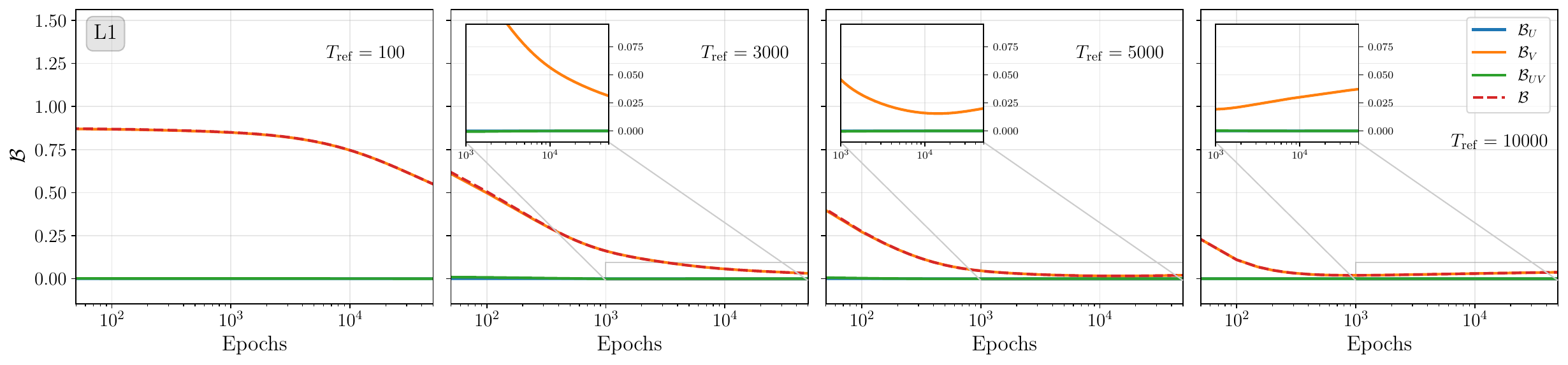}
  \includegraphics[width=0.90\textwidth]{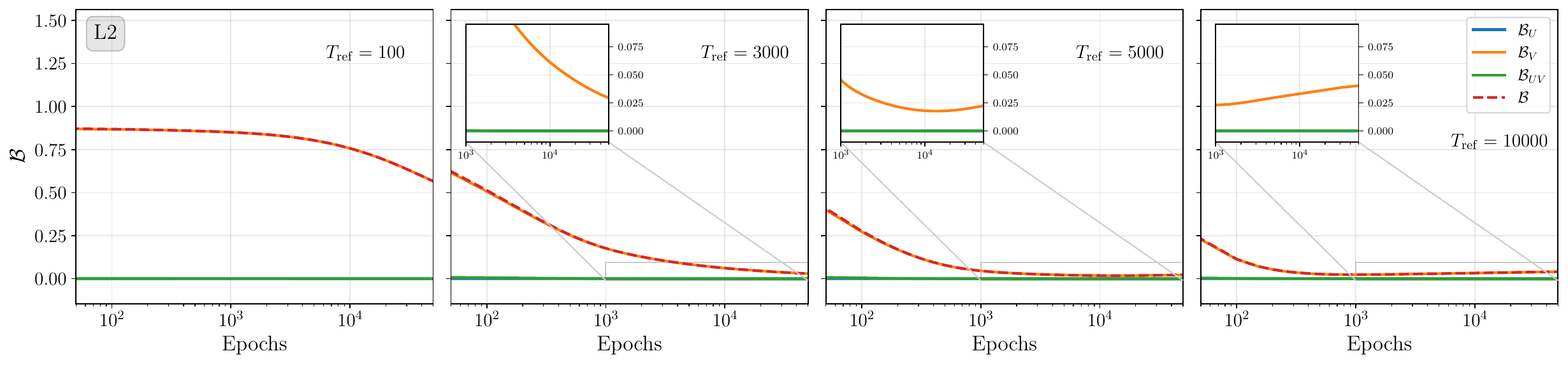}
  \caption{Bias decomposition as in Eq.~\eqref{eq:BiasDecomposition} as a function
  of the training time for L0 (top), L1 (middle) and L2 (bottom) data. The curves
  are computed using the analytical result in Eq.~\eqref{eq:AnalyticSol}. Each
  column corresponds to a different frozen NTK.
  \label{fig:BiasDecomposition}}
\end{figure}

\begin{figure}[!h]
  \centering
  \includegraphics[width=0.90\textwidth]{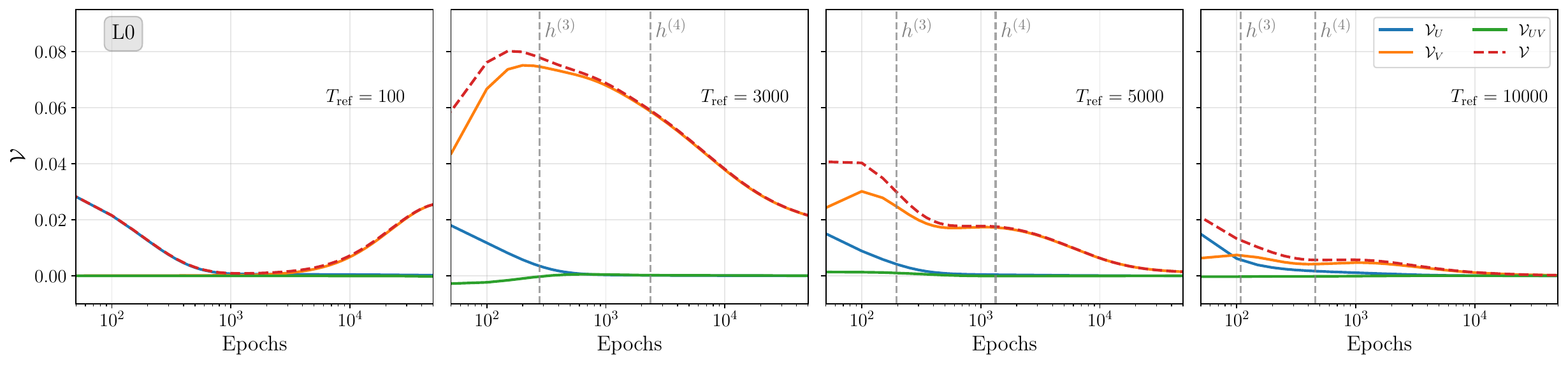}
  \includegraphics[width=0.90\textwidth]{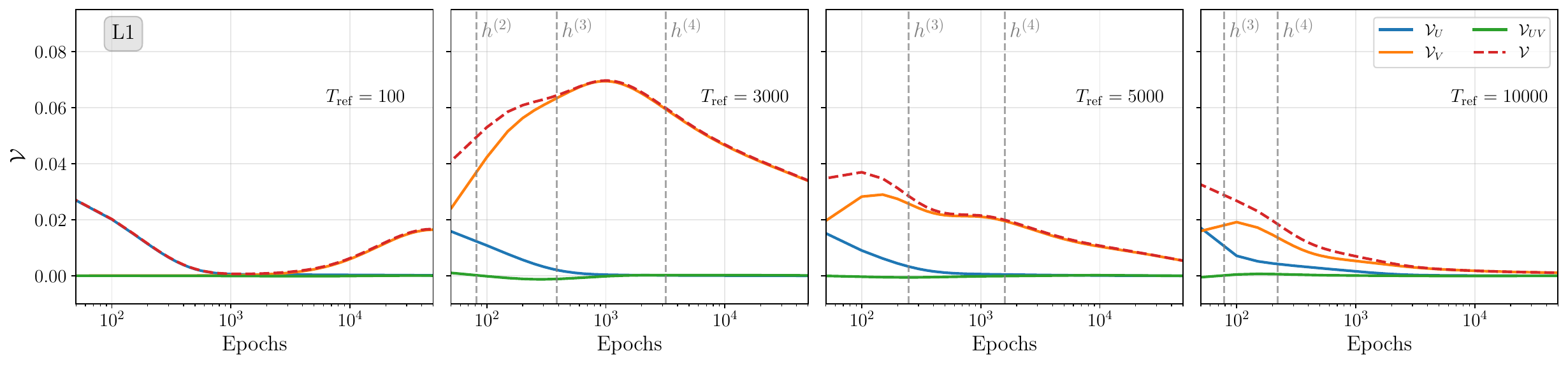}
  \includegraphics[width=0.90\textwidth]{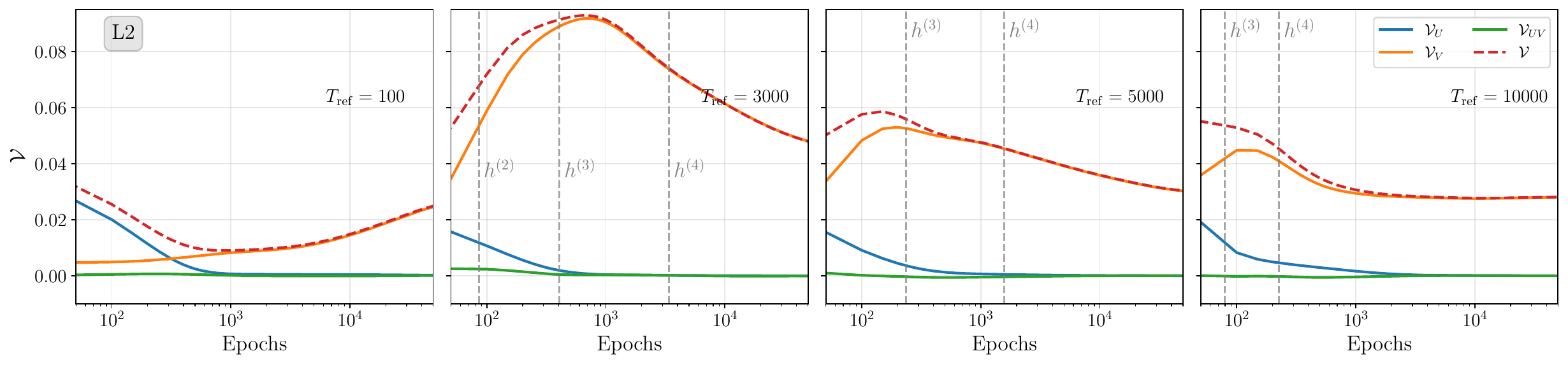}
  \caption{Variance decomposition as in Eq.~\eqref{eq:VarianceDecomposition} as a function
  of the training time for L0 (top), L1 (middle) and L2 (bottom) data. The curves
  are computed using the analytical result in Eq.~\eqref{eq:AnalyticSol}. Each
  column corresponds to a different frozen NTK. The vertical dashed lines in grey
  indicate the inverse of the eigenvalues of $H^\perp$, $1/h^{(i)}$.
  \label{fig:VarianceDecomposition}}
\end{figure}

Bias and variance, together with their decompositions introduced above, are
shown in Figs.~\ref{fig:BiasDecomposition} and~\ref{fig:VarianceDecomposition},
respectively. We show these curves as functions of the training time for
different frozen NTKs and for L0, L1 and L2 data. We now discuss these figures
in turn.

Starting from the bias, we see that $\mathcal{B}_{t,V}$ contributes the most to
the total budget of the bias, in accordance with the observations made in
Fig.~\ref{fig:FrefDecompositionL2}. Furthermore, the bias decreases more rapidly
when the NTK is frozen at later stages of training, as the NTK has had more time
to align with the data and therefore is more effective in reducing the bias. We
also observe that the behaviour of the bias at larger epochs is different
depending on the closure level. For L0 data, the bias decreases monotonically
with training time, approaching zero asymptotically. This is expected, since L0
data do not contain any noise and the bias can be reduced indefinitely by
training longer. On the other hand, the bias for L1 and L2 data presents a
minimum at large epochs, after which the bias increases again (see the
inset in each figure). Here the dependence on the frozen NTK determines the
location of the minimum -- the later the NTK is frozen, the earlier the bias
reaches its minimum. That both L1 and L2 have a similar behaviour tells us that
the central value of the analytical solution is mostly insensitive to the
experimental uncertainty propagated through the Monte Carlo fluctuations of the
data.
 
Turning now to the variance, we see again that the dominant contribution comes
from $\mathcal{V}_{t,V}$, unless the NTK is frozen at very early stages of
training. This agrees with Fig.~\ref{fig:ErrorBudgetL2}, where we observed that
the $U$ contribution is significant only at early epochs. In
general, the variance exhibits a peak at intermediate epochs, after which it
decreases. This stands out particularly for $T_{\rm ref} = 3000$ (second column
in Fig.~\ref{fig:VarianceDecomposition}), where the height of the peak has its
maximum for the case of L2 data. When the NTK is frozen at later stages of
training, the peak is less pronounced and located at earlier epochs. Note that,
contrary to the bias, the behaviour at large epochs of the variance differs
between L1 and L2 data. In fact, while the variance for L0 and L1 data
approaches zero asymptotically, for L2 data the variance converges to a residual
value shifted from zero. Finally, a closer inspection of the figures reveals a
strong correlation between the eigenvalues of $H^\perp$, introduced in
Sec.~\ref{sec:AnlyticalLazySolution}, and the transitions in the slope of the variance. Indeed,
the components of the analytical solution contribute with different timescales,
characterised by the inverse of the eigenvalues of $H^\perp$. This is
particularly evident for the case at $T_{\rm ref}=5000$, where in correspondence
of the timescale $1/h^{(i)}$, shown as vertical dashed lines in grey, the variance
shows a change in slope. This remains true across all closure-test data,
although slightly mitigated for L2 data.

\begin{figure}[!h]
  \centering
  \includegraphics[width=0.90\textwidth]{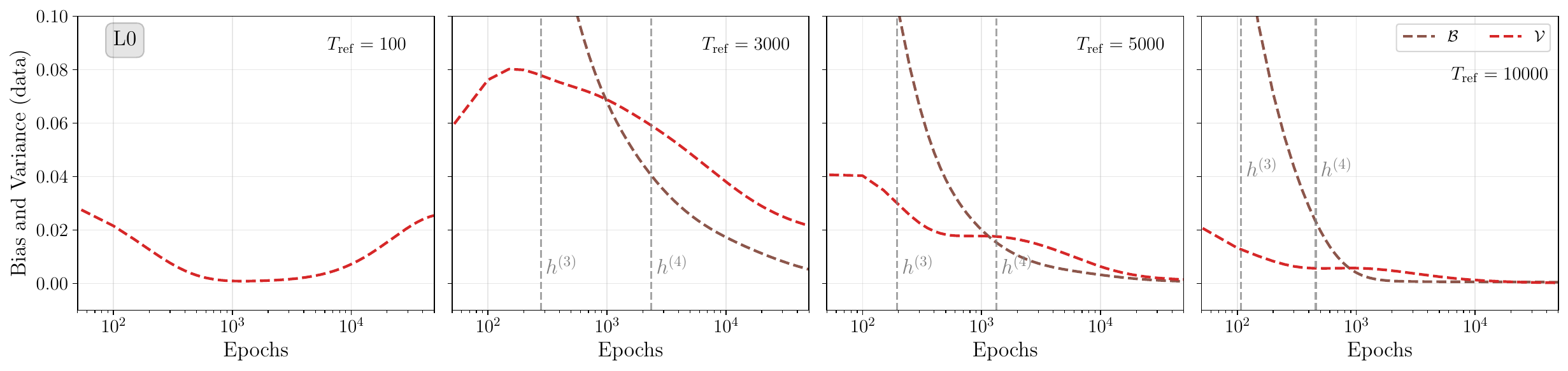}
  \includegraphics[width=0.90\textwidth]{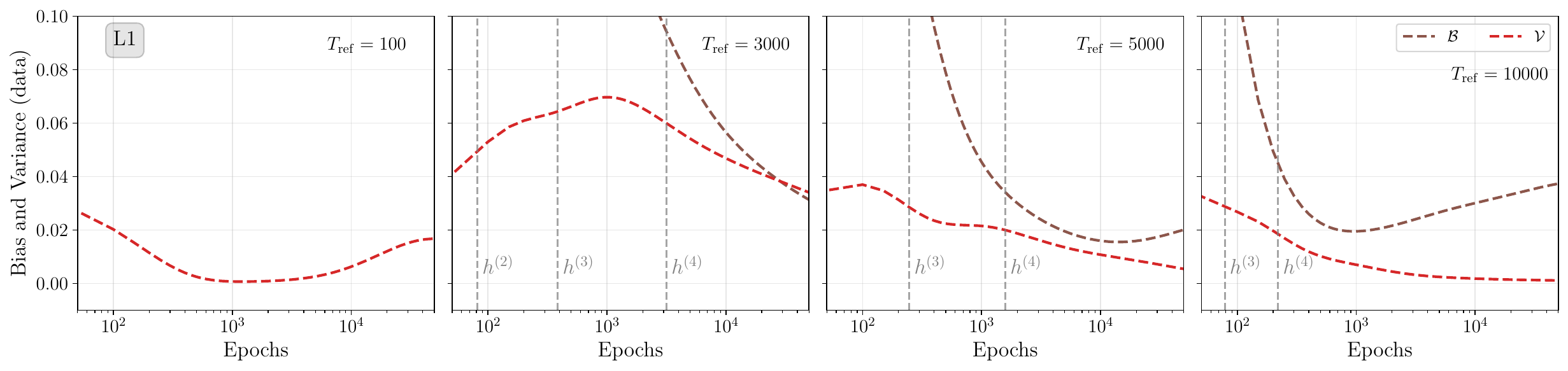}
  \includegraphics[width=0.90\textwidth]{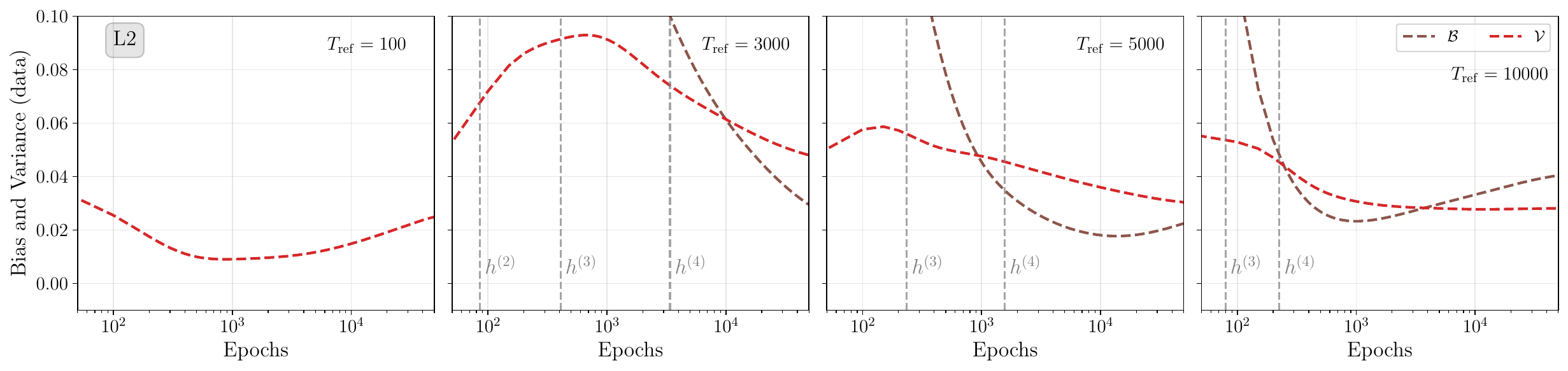}
  \caption{Comparison of the total contribution of bias and variance in the
  space of the data as a function of the training time for L0 (top), L1 (middle)
  and L2 (bottom) data. The curves are the same as in
  Figs.~\ref{fig:BiasDecomposition} and~\ref{fig:VarianceDecomposition},
  displayed here together. Each column corresponds to a different frozen NTK. In
  first column, the bias is off-scale. As in
  Fig.~\ref{fig:VarianceDecomposition}, the vertical dashed lines in grey
  indicate the inverse of the eigenvalues of $H^\perp$, $1/h^{(i)}$.
  \label{fig:BiasVarianceComparison}}
\end{figure}

The interplay between bias and variance, as well as the presence of extrema in
both curves, has practical implications in the choice of the optimal
stopping time. In Fig.~\ref{fig:BiasVarianceComparison} we compare the total
contributions of bias and variance as functions of the training time, for
different frozen NTKs, and for L0, L1 and L2 data. One major observation that
emerges from these figures is that the intersection point between bias and
variance -- which defines a bias-variance ratio equal to one -- is not
sufficient to determine whether the fit has converged optimally. For instance,
for L0 data with the NTK frozen at $T_{\rm ref}=3000$ (first row, third column),
the two curves intersect at around $T \sim 1000$ and $T> 10000$. Clearly, only
the latter represents a good fit. This becomes even more evident for L2 data
(third row), in particular when the NTK is frozen at $T_{\rm ref}=10000$ (last
column). Interestingly, there are no intersection points between bias and
variance for L0 data and when the NTKs are frozen at $T_{\rm ref} > 3000$. This
is a direct consequence of the observations made above, namely that the bias is
insensitive to the noise in the data, while the variance is. Indeed, as is clear
by comparing L1 and L2 data for $T_{\rm ref} = 10000$, adding the Monte Carlo
fluctuations in the data leaves the bias almost unchanged but increases the
variance, resulting in the intersection of the two curves.

We conclude by remarking that bias and variance are computed using the
analytical solution in Eq.~\eqref{eq:AnalyticSol}, hence from an ensemble of
frozen NTKs. The results discussed above provide evidence that the NTK is capable
of encoding not only the physical features of the data, but also the statistical
features of the methodology. Such a finding deserves further attention in
future assessments of PDF determinations.


\FloatBarrier

\section{Conclusions}
\label{sec:conclusions}

Our present age is marked by unprecedented advancements in machine learning
techniques, whose applications span many scientific domains -- PDF determination
being one of them. It is of paramount importance to understand how these new
techniques behave when applied to complex problems such as PDF determination.

In this work, we have taken a step forward in this direction. We have
investigated a novel treatment of the learning process in the context of PDF
fitting by exploring the training dynamics in the functional space of the neural
network. In fact, the NTK can be used to unravel complex dynamics obfuscated by
the training algorithms commonly employed in PDF fits. We have shown that the
properties of the NTK are highly entangled with the fitting results, and that a
proper understanding of its structure and time dependence can provide precious 
insights on the learning process. The identification of separate rich and lazy training 
phases is potentially a general feature of MLP training, which is worthy of highlighting  
beyond the specific application to PDF fits.
In the context of PDF fits, we have developed, under certain assumptions, an
analytical description of how the neural network evolves during training,
enabling us to better understand the NNPDF methodology and its dependence on the
underlying model architecture. To the best of our knowledge, such description
has not yet been attempted in the context of ill-defined inverse problems.

Yet, this work is far from being conclusive. As a pioneering study, we believe
that the most important contribution is the identification and initial
discussion of the role of the NTK in PDF fits, leading to a set of relevant
diagnostic metrics to be explored in future works. Indeed, many aspects merit
further investigations. First of all, we need to study in detail the outcome of
a similar analysis when applied to a more complex and realistic framework,
including multiple fitted flavors and real data. For data that depend linearly
on the PDFs, the extension of the formalism is straightforward. However, the
agreement with the analytical predictions need to be tested by performing the
numerical simulations. Further insights could be gained by exploring different
parametrizations beyond neural networks. These systematic studies will be
carried out in forthcoming works using the recent flexible and extensible
open-source framework \texttt{Colibri}~\cite{Costantini:2025agd}.

In spite of the simplified framework adopted in the present study, our findings
highlight the complexity and richness of the learning process; a quantitative
description of the learning process is needed in order to pin down the origin of PDF uncertainties. 
This poses a significant challenge in light of the improved precision of forthcoming
measurements. We believe that the tools presented here can help address this
gap. 

\section*{Acknowledgements}
We have enjoyed discussing these topics as we were developing this work, clarifying 
many aspects of our analysis. We are grateful to M Ubiali, T Giani and A Ramos for 
their critical comments, ideas and suggestions. LDD is pleased to acknowledge support
from the IFIC Valencia, for multiple enjoyable stays, where many of these ideas were 
scrutinised carefully. 
LDD is supported by an STFC Consolidated Grant (ST/T000600/1, ST/X000494/1). 

\newpage

\appendix
\appendix
\section{The BCDMS dataset for $T_3$}
\label{app:dataset}

The analysis presented in this work employs a set of synthetic data points
generated using a known underlying law $\fin$ that we seek to reconstruct.
Analogous to Ref.~\cite{Candido:2024hjt}, the pseudo-data are constructed by
convolving $\fin$ with FK tables whose kinematic dependence is specified by
measurements of the structure function $F_2$ on proton and deuterium targets
from the BCDMS collaboration~\cite{Benvenuti:1989fm}. By combining these
measurements, one can construct the observable $F_2^p - F_2^d$ that, at
next-to-next-to-leading order (NNLO) in QCD and under the assumption of
isoscalarity for the deuterium nucleus, provides a clean probe of the
non-singlet triplet PDF combination $T_3 = u^+ - d^+$, where $u^+ = u + \bar{u}$
and $d^+ = d + \bar{d}$. The factorisation formula in Eq.~\eqref{eq:TheoryPred}
then simplifies to 
\begin{equation}
F_2^p - F_2^d = C_{T_3} \otimes T_3,
\end{equation}
where $C_{T_3}$ is the corresponding Wilson coefficient computed in perturbative
QCD, and $\otimes$ is a short-hand notation that denotes the convolution as in
Eq.~\eqref{eq:TheoryPred}. The construction of the FK tables needed to compute
the predictions, as well as the construction of the covariance matrix, are
identical to Ref.~\cite{Candido:2024hjt}, to which the reader is referred for
further details. This yields a total of 333 data points, which then reduce to
248 points after applying the kinematic cuts.

Following the closure test framework developed by the NNPDF
collaboration~\cite{DelDebbio:2021whr,NNPDF:2021njg}, we generate data with
three different levels of noise, labelled as Level 0 (L0), Level 1 (L1) and
Level 2 (L2) data. Furthermore, we use the non-singlet triplet $xT_3$ from the
NNPDF4.0 parton set~\cite{NNPDF:2021njg} as the input law $\fin$. In the
following, we summarise the definition of the different levels of pseudo-data.

\paragraph{Level 0}
The pseudo-data are generated without any experimental noise, \ie, by using the
input function and the FK tables as follows
\begin{equation}
Y_{L0} =  \FKtab \fin.
\end{equation}
In this ideal scenario, the analysis should reproduce the input $\fin$, though
some residual reconstruction error may remain in the kinematic region not
covered by the FK tables. Level 0 assesses the intrinsic bias of the
methodology, as any neural network replica will be trained on the same data
points $Y_{L0}$.

\paragraph{Level 1}
In this case, the experimental noise is added on top of the L0 data, by sampling
from the multivariate normal distribution with the full experimental covariance
matrix $C_Y$ provided by the BCDMS collaboration
\begin{equation}
Y_{L1} =  Y_{L0} + \eta, \quad \textrm{where} \quad \eta \sim \mathcal{N}(0, C_Y).
\end{equation}
This case is closer to actual experimental data, where the ``true'' value is
blurred by the presence of noise. Note however that we are not yet propagating
the experimental uncertainties into the uncertainties of the fitted PDF, as the
added noise is fixed over all replicas.

\paragraph{Level 2}
Finally, we generate L2 pseudo-data by adding a different noise realisation to
each replica, sampled from the same multivariate normal distribution
\begin{equation}
Y_{L2}^{(k)} =  Y_{L1} + \xi^{(k)}, \quad \textrm{where} \quad \xi^{(k)} \sim \mathcal{N}(0, C_Y).
\end{equation}
This represents the most realistic scenario where both model and data
uncertainties are present. In this case, each neural network replica will be
trained on a different set of data points $Y_{L2}^{(k)}$.
\section{Dependence on the Architecture of the NTK}
\label{sec:NTKArchDep}

\begin{figure}[t]
  \centering
  \includegraphics[width=0.45\textwidth]{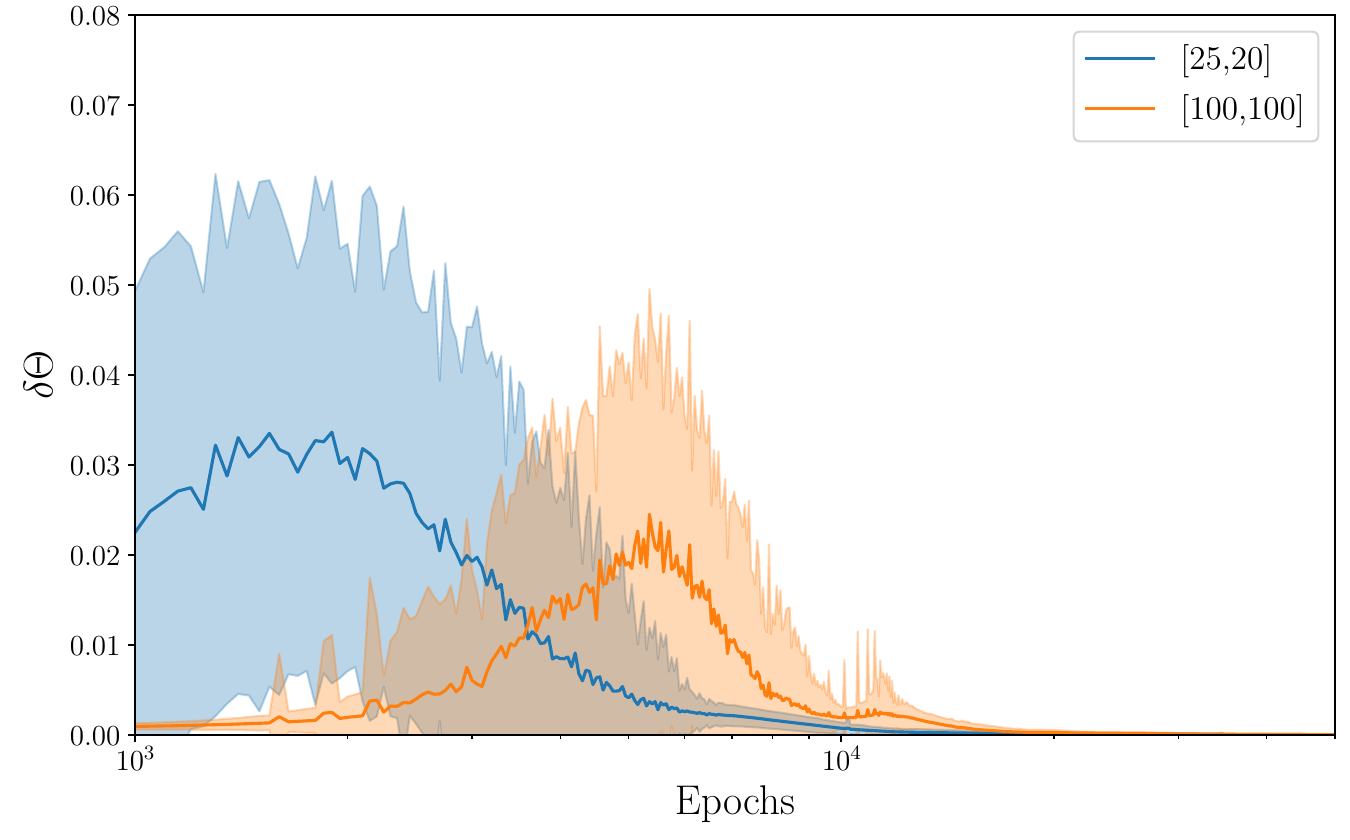}
  \includegraphics[width=0.45\textwidth]{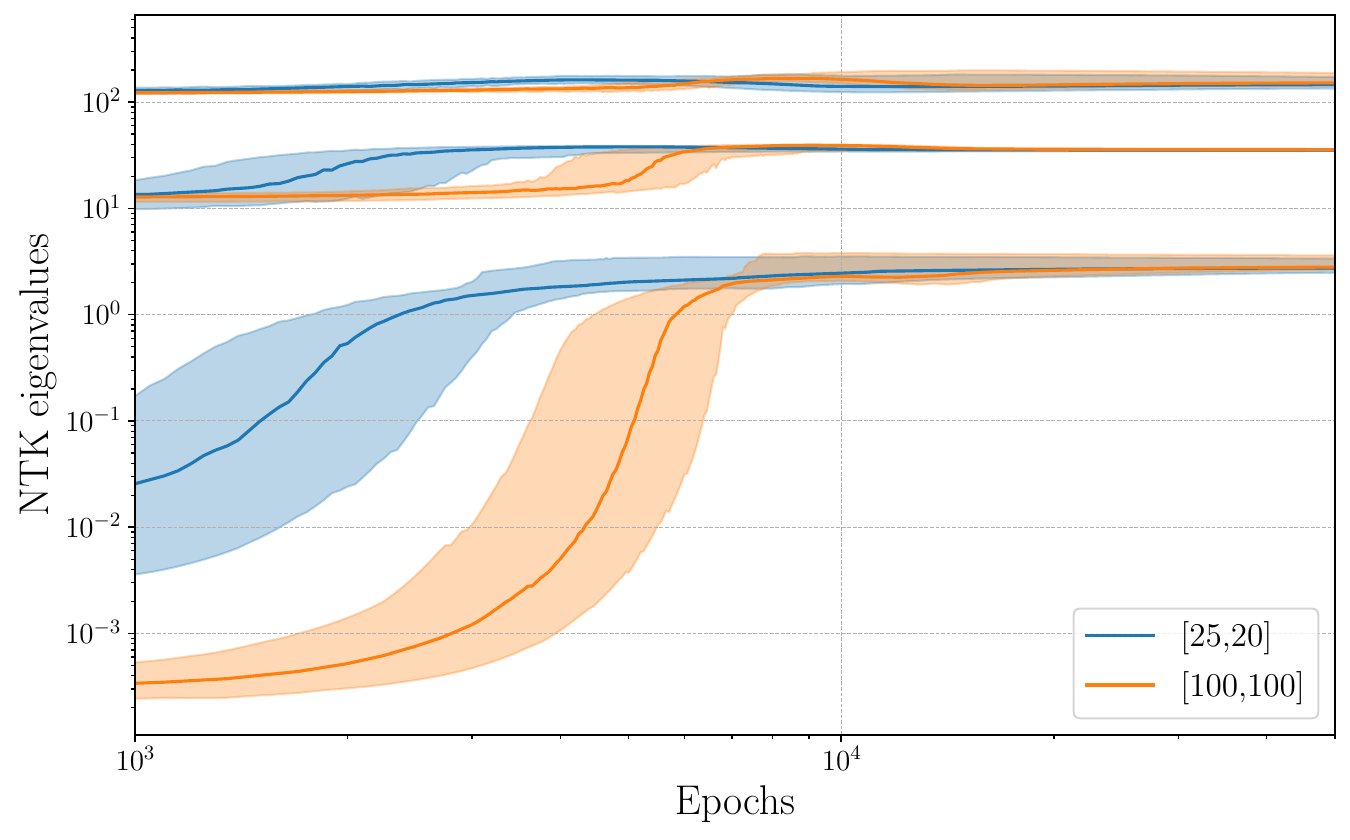}
  \caption{Comparison of the variation of the NTK during training (left) and the
  first three eigenvalues (right) for two different architectures with sizes
  $[28,20]$ and $[100,100]$ respectively. In both cases, L0 data is used. In the
  left plot, error bands represent the standard deviation over the ensemble of
  replicas. In the right plot, solid lines represent the median over the
  ensemble of networks, while solid bands correspond to 68\% confidence level.}
  \label{fig:NTKTimeDiffArch}
\end{figure}

It is interesting to consider what happens to the picture sketched so far as the
architecture of the neural networks is varied. In Fig.~\ref{fig:NTKTimeDiffArch}
we compare the time dependence of the Frobenius norm of the NTK and the
variation of the first three eigenvalues for two different architectures. The
smaller network is the $[28,20]$ used in the standard NNPDF analyses and in this
work, while the large one is a $[100,100]$ network, which is closer to the
infinite-width limit. For illustration purposes we focus in this plot on L1 data
and only three eigenvalues rather than the five we examined above. The
quantitative features are exactly the same for L0 and L2 data, and adding the
fourth and fifth eigenvalues does not add unexpected behaviours compared to what
we observe in Fig.~\ref{fig:NTKTime}. It is interesting to remark that the onset
of lazy training is slower for the larger network. This is to be expected if we
interpret the early stages of training as a phase where the network identifies
the learnable features in the space of functions that it can parametrize. For a
larger network, the space of parametrized functions is larger and the
identification of the physical features takes a larger number of epochs. 

\FloatBarrier
\section{Cut-off Tolerance of the NTK spectrum}
\label{sec:cutoff}

\begin{figure}[h!]
  \centering
  \includegraphics[width=0.6\textwidth]{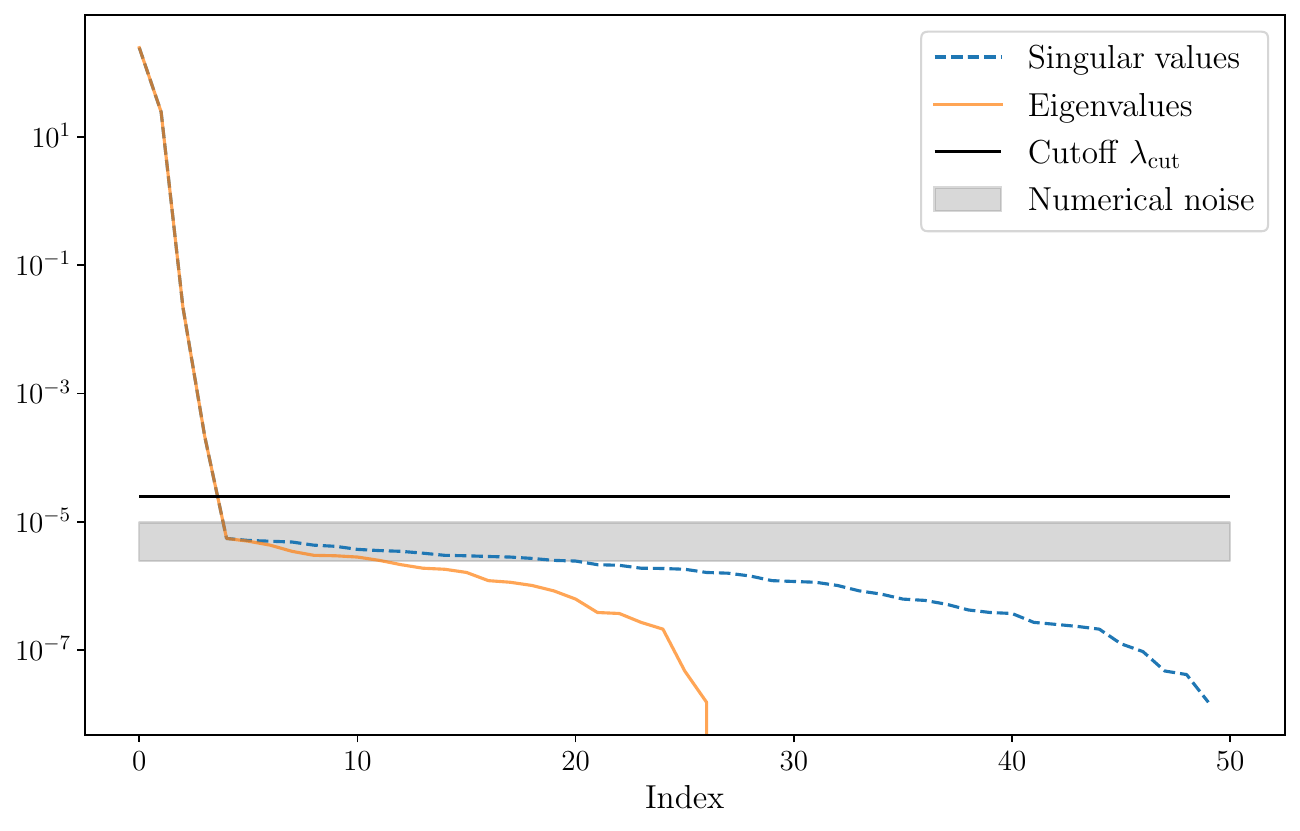}
  \caption{Visualisation of the cut-off tolerance for the NTK spectrum used in
    Section~\ref{sec:LazyTraining}. Eigenvalues (orange solid line) and singular values (blue dashed
    line) are shown, together with the cut-off tolerance (black) chosen as described
    in the text. We also show the corresponding numerical noise (grey shaded area)
    which occurs at small eigenvalues (or singular values).}
  \label{fig:cutoff}
\end{figure}

We determine the effective rank of the NTK by identifying eigenvalues that are
numerically significant. We classify eigenvalues $\lambda_i$ as numerically zero
if
\begin{equation}
  \lambda_i < \epsilon_{\rm tol} \cdot \lambda_{\rm max} \,,
\end{equation}
where $\epsilon_{\rm tol}$ is the relative tolerance and $\lambda_{\rm max}$ is
the largest eigenvalue of the NTK. Throughout this work, we use $\epsilon_{\rm
tol} = 10^{-7}$, which is close to machine precision for single-precision
floating point numbers. This choice is illustrated in Fig.~\ref{fig:cutoff},
where we show the NTK spectrum at initialization for the NNPDF-like architecture
discussed in the main text (orange solid line). Together with the eigenvalues,
we also show the singular values of the NTK (dashed blue line) to illustrate
that the two spectra are identical until numerical noise sets in. The cut-off
tolerance, indicated by the black horizontal line, is chosen to be slightly
above the numerical noise level (grey shaded area) to ensure that only
numerically significant eigenvalues are retained.

\section{Detailed derivation of the analytical solution}
\label{app:derivation}

The solution to Eq.~\eqref{eq:FlowPerp} can be written as the sum of the
solution of the homogeneous equation, $\hat{f}^{\perp}_{t,k}$, and a particular
solution of the full equation. The solution of the homogeneous equation is
\begin{align}
    \label{eq:HomoSoln}
    \hat{f}^{\perp}_{t,k} = \sum_{i=1}^{d^\perp} f^{\perp (i)}_{0} e^{-h^{(i)}t} w^{(i)}_k\, ,
\end{align}
where
\begin{align}
    \label{eq:InitialCi}
    f^{\perp (i)}_{0} = \sum_{k=1}^{d_\perp} w^{(i)}_k f^\perp_{0,k}\, ,
\end{align}
guarantees that the initial condition $\hat{f}^\perp_{t,k}=f^\perp_{0,k}$ is
satisfied. Similarly, if we define
\begin{align}
    \label{eq:BiDef}
    \Upsilon^{(i)} = \sum_{k=1}^{d_\perp} w^{(i)}_k B^\perp_{k}\, ,
\end{align}
then
\begin{align}
    \label{eq:PartSol}
    \check{f}^\perp_{t,k} = \sideset{}{'}\sum_{i} \frac{1}{h^{(i)}} \Upsilon^{(i)}
        \left(1 - e^{-h^{(i)}t}\right) w^{(i)}_k\, ,
\end{align}
where the sum only involves the non-zero modes of $H^\perp$, is a particular
solution of the inhomogeneous equation, which satisfies the boundary condition
$\check{f}^{\perp}_{0,k}=0$. Hence, the solution of the flow equation in the
subspace orthogonal to $\text{ker}\ \Theta$ is
\begin{align}
    f^\perp_{t,k}
    \label{eq:FlowSolution}
        &= \hat{f}^\perp_{t,k} + \check{f}^\perp_{t,k}
        \, .
\end{align}
Finally, collecting the parallel contribution, Eq.~\eqref{eq:FlowParallel}, and
the solution of the orthogonal component, Eq.~\eqref{eq:FlowSolution}, yields a
simple expression,
\begin{align}
    \label{eq:AnalyticSolApp}
    f_{t,\alpha}
        = U(t)_{\alpha\alpha'} f_{0,\alpha'} + V(t)_{\alpha I} Y_{I}\, .
\end{align}
The two evolution operators $U(t)$ and $V(t)$ are here summarised: 
\begin{align}
    \label{eq:UDef}
    U(t)_{\alpha\alpha'} = \hat{U}^\perp(t)_{\alpha\alpha'}
        + \check{U}^\perp(t)_{\alpha\alpha'} + U^\parallel_{\alpha\alpha'}\, ,
\end{align}
where
\begin{align}
    \hat{U}^\perp(t)_{\alpha\alpha'}
        = \sum_{k,k'\in\perp} \sqrt{\lambda^{(k)}} z^{(k)}_\alpha 
            \left[\sum_i w^{(i)}_{k} e^{-h^{(i)}t} w^{(i)}_{k'}\right]
            z^{(k')}_{\alpha'} \frac{1}{\sqrt{\lambda^{(k')}}}\, ,
\end{align}
and
\begin{align}
    U^\parallel_{\alpha\alpha'}
        = \sum_{k''\in\parallel} z^{(k)}_\alpha z^{(k)}_{\alpha'} \, .
\end{align}
The contributions from $\check{U}^\perp(t)$ and $V(t)$ are more easily expressed
by introducing the operator
\begin{align}
    \label{eq:MOperatorDef}
    \mathcal{M}(t)_{\alpha\alpha'} 
        = \sum_{k,k'\in\perp} \sqrt{\lambda^{(k)}} z^{(k)}_\alpha 
            \left[\sideset{}{'}\sum_{i} w^{(i)}_{k} \frac{1}{h^{(i)}}\, 
            \left( 1- e^{-h^{(i)}t}\right) w^{(i)}_{k'}\right]
            z^{(k')}_{\alpha'} \sqrt{\lambda^{(k')}}\,. 
\end{align}
Then, we can write
\begin{align}
    \label{eq:UperpCheck}
    \check{U}^\perp(t)
        = - \mathcal{M}(t)\; \FKtabT C_Y^{-1} \FKtab 
            \left[\sum_{k''\in\parallel} z^{(k'')} z^{(k'') T}\right]\, ,
\end{align}
and
\begin{align}
    \label{eq:VDef}
    V(t) = \mathcal{M}(t)\; \FKtabT C_Y^{-1}\, ,
\end{align}
where we note that the term in the bracket in Eq.~\eqref{eq:UperpCheck} is
simply the projector on the kernel of the NTK. The four terms that appear in the
analytical solution have a clear physical interpretation:
\begin{itemize}
    \item The first term $\hat{U}^\perp(t)$ suppresses the components of the
    initial condition that lie in the subspace orthogonal to the kernel of the
    NTK. These are the components that are learned by the network during
    training. While the trained solution still depends on its value at
    initialisation, that dependence is suppressed during training. This
    suppression is exponential in the training time, and the rates are given by
    the eigenvalues of $H^{\perp}$.
    \item The contribution from $U^\parallel$ yields the component of the
    initial condition that lies in the kernel of the NTK. As such, those
    components remain unchanged during training and are part of the trained
    field at all times $t$. 
    \item The two remaining contributions are best understood by combining them
    together,
    \begin{align}
        \label{eq:DataCorrectedInference}
        \check{U}^{\perp}(t) f_{0} + V(t) Y 
            = \mathcal{M}(t)\; \FKtabT C_Y^{-1} \left[Y - \FKtab f_{0}^{\parallel}\right]\, .
    \end{align}
    The parallel component of the initial condition $f_{0}^{\parallel}$ does not
    evolve during training, and therefore it yields a contribution $\FKtab
    f_{0}^{\parallel}$ to the theoretical prediction of the data points at all
    times $t$. This is taken into account by subtracting this contribution from
    the data, before the inference is performed.
\end{itemize}

\subsection{Crosschecks using L0 data}
\label{app:AnalyticalChecks}

The analytical solution enables rigorous validation of our implementation
through crosschecks with L0 data, where we have complete control over the data
generation process. In this case, the realization of the dataset is completely
determined by the input PDFs
\begin{equation}
    \label{eq:DataL0NoIndices}
    Y = \FKtab \fin\, .
\end{equation}
Note that using L0 data only affects the second term in
Eq.~\eqref{eq:AnalyticSol}.\footnote{To be more precise, since the analytical
solution requires the NTK to be frozen at a certain epoch $T_{\rm ref}$, the NTK
also depends on the data used in the training.} We can then rewrite the combined
term in Eq.~\eqref{eq:DataCorrectedInference} as follows
\begin{align}
  \label{eq:TrainingOnLevelZero}
  \check{U}^\perp(t) f_0 + V(t) Y 
    &= \mathcal{M}(t)\, \FKtabT C_{Y}^{-1} \FKtab\, 
      \left[\fin - f_{0}^\parallel\right]\, .
\end{align}
The subtraction taking place in the square brackets of
Eq.~\eqref{eq:TrainingOnLevelZero} suggests that the effective function that
the neural network actually sees is not the input function $\fin$ used to
generate the data, but rather the difference between $\fin$ and the component of
the initial function $f_0$ that lies in the subspace spanned by the kernel of
the NTK, \ie, $f_0^\parallel$. In other words, the parallel component
$f_0^\parallel$, which, we remind the reader, does not evolve during the analytic training,
acts as a constant ``bias'' in the training process, effectively shifting the 
input function seen by the neural network. Of course the actual magnitude of
this irreducible noise depends both on how $f_0$ and the kernel of the NTK are
distributed over the ensemble. 

\begin{figure}[t]
  \centering
  \includegraphics[width=\textwidth]{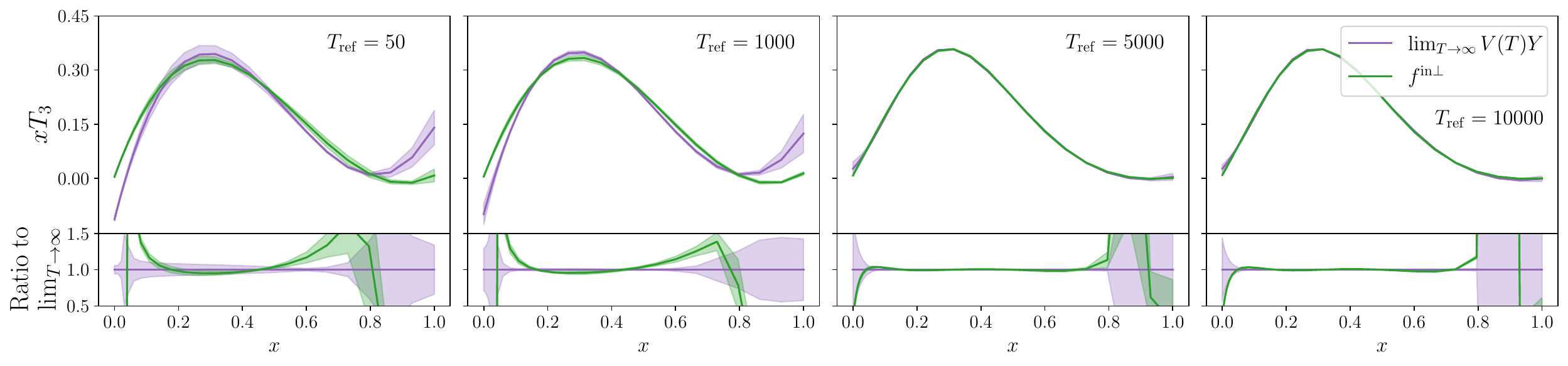}  
  \caption{Test of the $t\to\infty$ limit of the L0 training for different frozen
  NTKs. The green curve represents the projection of the input function $\fin$
  onto the subspace orthogonal to the kernel of the NTK at $t_{\rm ref}$, \ie,
  $\finperp$. The purple curve represents the contribution of the operator $V$,
  computed with the NTK at $T_{\rm ref}$, in the limit of infinite training
  time.}
  \label{fig:InfiniteTimeVterm}
\end{figure}

Note that the observation above remains true even in the limit of infinite
training, 
\begin{align}
    \label{eq:LevelZeroClosureInfiniteTraining}
    \lim_{t\to\infty} V(t) Y = \finperp + \mathcal{M}_{\infty} M \finpar\, ,
\end{align}
which shows that the $V$ component of the trained solution reproduces exactly
the component of the PDF that lies in the subspace orthogonal to the kernel of
$\Theta$. We compare the asymptotic behaviour of $V(t) Y$ and $\finperp$ in
Fig.~\ref{fig:InfiniteTimeVterm}, where we see that the analytical solution at
infinite training time reproduces the expected result, \ie, it coincides 
with $\finperp$, as long as $T_{\mathrm{ref}}> 1000$. The second term in 
the square bracket on the right-hand side of
Eq.~\eqref{eq:TrainingOnLevelZero} is the contribution from the parallel
component at initialisation that does not evolve in the training process. Given
that $f_0$ is almost normally distributed around zero, that term does not
contribute to the central value of the fitted PDF, \ie, to the average of the
trained solution over replicas. 

\begin{figure}[h!]
  \centering
  \includegraphics[width=0.95\textwidth]{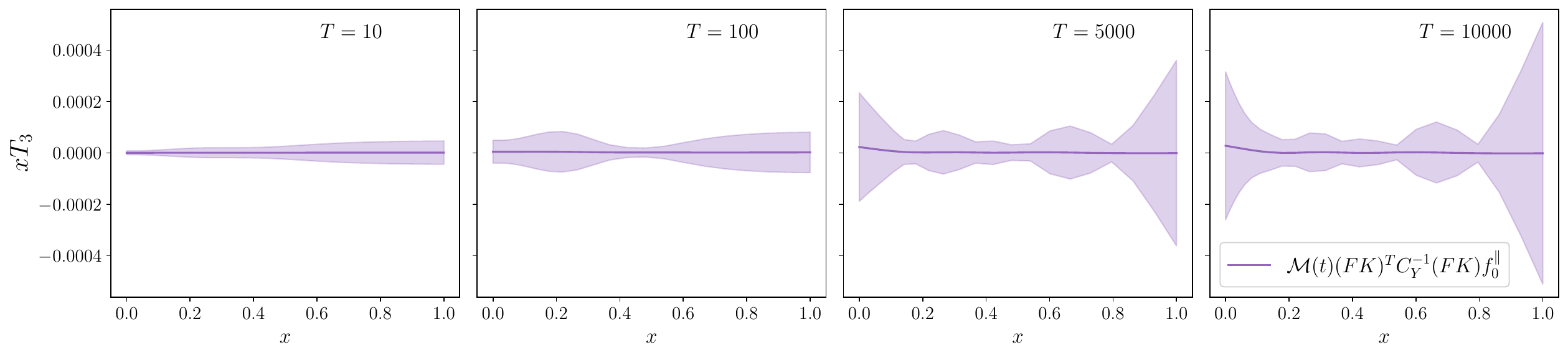} 
  \caption{Test of the average of the parallel contribution for different
  epochs. The reference epoch at which the frozen NTK is chosen is $T_{\rm ref}
  = 10000$. L2 data is used in the plot. Note that the scale on the vertical 
  axis is three orders of magnitude smaller than in Fig.~\ref{fig:InfiniteTimeVterm}.}
  \label{fig:AverageLevelZeroUcheck}
\end{figure}

The time evolution of 
\begin{align}
  \label{eq:AverageLevelZeroUcheck}
  \mathbb{E}\left[\mathcal{M}(t)\, \FKtabT C_{Y}^{-1} \FKtab\, 
    f_{0}^\parallel\right]\, ,
\end{align}
is shown in Fig.~\ref{fig:AverageLevelZeroUcheck}.

\subsection{Infinite Training Time}
In the limit of infinite training time, the evolution operators $U(t)$ and
$V(t)$ simplify and yield an elegant interpretation of the minimum of the cost
function. For large training times, we have
\begin{align}
    \label{eq:UhatInfty}
    \hat{U}^\perp_{\infty, \alpha\alpha'}
        &= \lim_{t\to\infty}\hat{U}^\perp(t)_{\alpha\alpha'} = 0\, \\
    \label{eq:MOperatorInfty}
    \mathcal{M}_{\infty, \alpha\alpha'} 
        &= \lim_{t\to\infty}\mathcal{M}(t)_{\alpha\alpha'} = \sum_{k,k'\in\perp} \sqrt{\lambda^{(k)}} z^{(k)}_\alpha 
        \left[\sideset{}{'}\sum_{i} w^{(i)}_{k} \frac{1}{h^{(i)}}\, 
        w^{(i)}_{k'}\right] z^{(k')}_{\alpha'} \sqrt{\lambda^{(k')}}\, ,
\end{align}
and explicit expressions for $\check{U}^\perp_{\infty}$ and $V_{\infty}$ are
obtained from $\mathcal{M}_{\infty}$. The term in the square bracket in
Eq.~\eqref{eq:MOperatorInfty} is the spectral decomposition of the pseudo-inverse
of $H^\perp$ in $d_\perp$-dimensional orthogonal subspace. So, the operator
$\mathcal{M}_{\infty}$ acts as follows on a field $f_{\alpha}$:
\begin{enumerate}
    \item The term on the right of the square bracket computes the coordinate
    $f_{k'}$ introduced in Eq.~\eqref{eq:OrthogonalComponents}. The $f_k$ are 
    the coordinates of the component $f^\perp$ that evolves during
    training, 
    \begin{align}
        \label{eq:RightOfTheBracket}
        f^\perp = \sum_{k\in\perp} \sqrt{\lambda^{(k)}} f_k\, z^{(k)}\,  .
    \end{align}
    \item The term in the square bracket applies the pseudo-inverse to the
    coordinates $f_k$, 
    \begin{align}
        \label{eq:ApplyPseudoInv}
        f'_k = \left(H^\perp\right)^+_{kk'} f_{k'}\, .
    \end{align}
    \item The final term on the left of the square bracket reconstructs the full
    field corresponding to the coordinates $f'_{k}$,
    \begin{align}
        \label{eq:LeftOfTheBracket}
        f^{'\perp} = \sum_{k\in\perp} \sqrt{\lambda^{(k)}} f'_{k}\, z^{(k)}\, .
    \end{align}
    
\end{enumerate}

As discussed at the end of Sect.~\ref{sec:AnlyticalLazySolution} it is convenient to combine the
contributions of $\check{U}^\perp_{\infty}$ and $V_{\infty}$,
\begin{align}
    \label{eq:DataCorrectedInferenceAtInfty}
    \check{U}^{\perp}_{\infty} f_{0} + V_{\infty} Y 
        = \mathcal{M}_{\infty}\; \FKtabT C_Y^{-1} \left[Y - \FKtab f_{0}^{\parallel}\right]\, .
\end{align}
The contribution to the observables from the parallel components of $f$ does not
change during training, therefore that contribution is subtracted from the data
and the orthogonal components of $f$ are adjusted to minimize the $\chi^2$ of
the corrected data. The minimum of the $\chi^2$ in the orthogonal subspace is
found applying $\mathcal{M}_{\infty}$, \ie, by projecting in the orthogonal
subspace, applying the pseudo-inverse and finally recomputing the full field as
detailed above.

\bibliographystyle{JHEP}
\bibliography{ntk.bib}

\end{document}